\documentclass[twocolumn]{28onr_art}
\usepackage{ulem}
\usepackage{graphicx}
\usepackage{epsfig}
\usepackage{multirow}
\usepackage{hyperref}
\usepackage{harvard}
\usepackage{fixltx2e}
\usepackage{color}

\pdfoutput = 1

\def\be{\begin{eqnarray}}
\def\ee{\end{eqnarray}}
\def\benl{\begin{eqnarray*}}
\def\eenl{\end{eqnarray*}}

\newcommand{\nwc}{\newcommand}
\nwc{\bm}{\boldmath}
\nwc{\m}{\mbox}
\nwc{\ubm}{\unboldmath}
\nwc{\bmU}{\m{\bm$U$\ubm}}
\nwc{\bmX}{\m{\bm$X$\ubm}}
\nwc{\bmu}{\m{\bm$u$\ubm}}
\nwc{\bmx}{\m{\bm$x$\ubm}}
\nwc{\bmz}{\m{\bm$z$\ubm}}
\nwc{\bmv}{\m{\bm$v$\ubm}}
\nwc{\bmw}{\m{\bm$w$\ubm}}
\nwc{\bmW}{\m{\bm$W$\ubm}}
\nwc{\bmn}{\m{\bm$n$\ubm}}
\nwc{\bmG}{\m{\bm$G$\ubm}}
\nwc{\bmF}{\m{\bm$F$\ubm}}
\nwc{\bmI}{\m{\bm$I$\ubm}}
\nwc{\bmN}{\m{\bm$N$\ubm}}
\nwc{\bmP}{\m{\bm$P$\ubm}}
\nwc{\bmcalP}{\m{\bm $\cal P$\ubm}}
\nwc{\bmV}{\m{\bm$V$\ubm}}
\nwc{\bmS}{\m{\bm$S$\ubm}}

\begin{document}

%
%

\title{A comparison of model-scale experimental measurements and computational predictions for a large transom-stern wave}

\author{David A. Drazen$^1$, Anne M. Fullerton$^1$, and Thomas C. Fu$^1$ {\\}  Kristine L.C. Beale$^2$, Thomas T. O'Shea$^2$, Kyle A. Brucker$^2$,{\\} Douglas G. Dommermuth$^2$, and Donald C. Wyatt$^2$ {\\} Shanti Bhushan$^3$, Pablo M. Carrica$^3$, and Fred Stern$^3$}

\affiliation{\small $^1$Naval Surface Warfare Center - Carderock, USA
\\$^2$Science Applications International Corporation, USA
\\$^3$University of Iowa, USA}

\maketitle

%
%

\section{ABSTRACT}

The flow field generated by a transom stern hullform is a complex,
broad-banded, three-dimensional system marked by a large breaking wave. This
unsteady multiphase turbulent flow feature is difficult to study experimentally
and simulate numerically. Recent model-scale experimental measurements and
numerical predictions of the wave-elevation topology behind a transom-sterned
hullform, Model 5673, are compared and assessed in this paper. The mean
height, surface roughness (RMS), and spectra of the breaking stern-waves were
measured by Light Detection And Ranging (LiDAR) and Quantitative Visualization
(QViz) sensors over a range of model speeds covering both wet- and dry-transom
operating conditions.  Numerical predictions for this data set from two Office
of Naval Research (ONR) supported naval-design codes, Numerical Flow Analysis
(NFA) and CFDship-Iowa-V.4, have been performed. Comparisons of experimental
data, including LiDAR and QViz measurements, to the numerical predictions for
wet-transom and dry transom conditions are presented and demonstrate the
current state of the art in the simulation of ship generated breaking waves.
This work is part of an ongoing collaborative effort as part of the ONR Ship
Wave Breaking and Bubble Wake program, to assess, validate, and improve the
capability of Computational Fluid Dynamics (CFD).

%
%

\section{INTRODUCTION}

The flow field generated by a transom stern hullform is a complex system
dependent upon a number of  variables, including the transom height and shape,
buttock slope, the wave system generated upstream, the boundary layer flow
around the hull, etc., and has been the subject of a number of studies. While
earlier work has focused on the surface wave profile \cite[for
example]{Maki07}, detailed laboratory scale measurements of the flow field have
been
performed by Lasheras \cite{Rodriguez08}, who focused on
identifying the general flow topology.  More recently, the development of advanced
instrumentation has allowed for measurement of the breaking transom wave of a
full-scale ship to be made \cite{Fu06a}.  In an
associated effort, numerical predictions of the full-scale stern wake and
comparisons of the mean height, surface roughness (RMS), and spectra of the
breaking stern-waves were made \cite{Wyatt08}. Although these initial
comparisons of the numerical predictions and full-scale measurements showed
generally good agreement, differences remained that were difficult to associate
with the underlying phenomenology due to the complexities of in-situ
collection.

\begin{figure} [ht!] \centering
\includegraphics[width=.9\columnwidth]{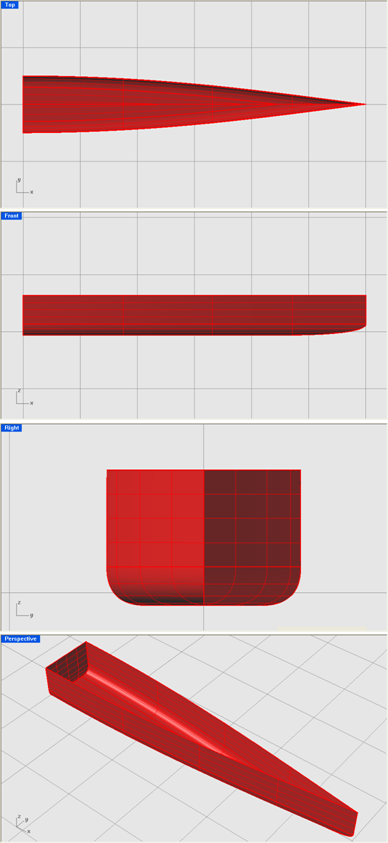}
\caption{\label{fig:5673}Model 5673. Rendered at scale. Grid line separation is
1.524 m (5').} \end{figure}

To obtain quantitative breaking wave data from a large-scale transom stern hull
form, measurements of the free-surface elevation in stern region of a large
transom model were made in June of 2007 \cite{Fu09} and October/November
of 2008 \cite{Fu10}. This large-scale laboratory experiment provides a
more canonical transom wave for study, removing the effects of the propellers,
appendages, and ambient conditions. For this experiment a large, specially
designed transom stern model was developed and tested in the Naval Systems Warfare Center - Carderock (NSWCCD) Carriage 2
Towing Tank. The hullform used, Model 5673, was purpose-built for these
experiments. It was designed to provide a minimal bow-wave disturbance and
maximum stern-wave disturbance. Its dimensions were determined by the maximum
practical size allowable in the Carriage 2 facility. Her hull
construction was fiberglass. A schematic of Model 5673 is shown in Figure \ref{fig:5673},
and the vessel geometry is detailed in Table \ref{tab:details}. The experiment and data
reduction and analysis are described in the Experimental Measurements section
of this paper.

%
%

%
%

\begin{table} \caption {\label{tab:details} Model 5673 Details }
\begin{center} \begin{tabular}{|c|c|} \hline \hline Length Overall, $L_o$  &
9.144 m (30 ft) \\ \hline Waterline Length   &   9.144 m (30 ft)    \\ \hline
Extreme Beam       &   1.524 m (5 ft)     \\ \hline Bow Draft          &
Fixed              \\ \hline Stern Draft, ${T_{max}}$  &   Variable      \\
\hline Construction & Fiberglass                 \\ \hline Displacement       &
771.1 kg (1700 lb) \\ \hline \hline \end{tabular} \end{center} \end{table}

The model-scale measurement data is compared to predictions from two Computational Fluid Dynamics (CFD) codes
currently under development by the Office of Naval Research (ONR), Numerical Flow Analysis (NFA) and
CFDship-Iowa-V.4. NFA is a Cartesian grid formulation of the Navier-Stokes
equations utilizing a cut-cell technique to impose the hull boundary conditions
\cite{dommermuth07,dommermuth08,nfa1}. CFDship-Iowa-V.4 is an unsteady
Reynolds-Averaged Navier-Stokes (URANS)/detached eddy simulation (DES) code
that uses a single-phase level-set method, advanced iterative solvers,
conservative formulations, and the dynamic overset grid approach for free
surface flows \cite{Bhushan07}. The two CFD techniques
are compared in separate but complimentary sections in the Numerical
Predictions portion of this paper.

%
%

\section{EXPERIMENTAL MEASUREMENTS}

\begin{figure} \centering
\includegraphics[width=1.0\columnwidth]{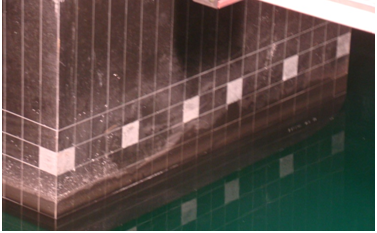}
\caption{\label{transom_sub} The initial submergence of the transom in 2007 and
2008 was 0.305 m} \end{figure}

\begin{figure} \centering
\includegraphics[width=1.0\columnwidth]{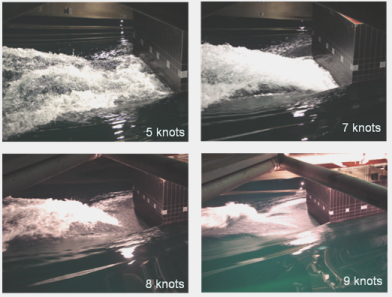}
\caption{\label{wake_at_speeds} A series of still images collected at each
speed tested. Only results from speeds of 3.60 m/s (7 knots) and 4.12 m/s (8
knots) will be discussed.} \end{figure}

The initial experiment, conducted in 2007, was designed to enable the
examination of the transom-stern wake transition from fully wet to fully dry
and focused on obtaining mean wave field data. Table \ref{trim_and_draft} lists
the test conditions. Four speeds [2.57, 3.60, 4.12, and 4.63 m/s (5,7,8, and 9
knots)] were tested. During the experiment the transition from a wet to dry
transom was observed to occur between 3.60 and 4.12 m/s (7 and 8 knots). The
follow on phase, conducted in 2008, focused on collecting a more detailed data
set at the Froude numbers which straddle the wet/dry transom condition. The
model was statically ballasted to the same waterline in 2007 and 2008, with a
transom submergence of 0.305 m, as shown in Figure \ref{transom_sub}. A view of
the transom stern wake at each of the Froude numbers tested is given in Figure
\ref{wake_at_speeds}. A variety of instrumentation was deployed to characterize
the transom wake field. A summary of the experimental work will be described
herein, further specifics can be found in \citeasnoun{Fu09} and
\citeasnoun{Fu10}. The discussion in this paper will focus on the surface
elevation measurement results from runs at 3.60 (7 knots) and 4.12 m/s (8
knots) due to the large volume of data collected at these speeds.

%
%
\begin{table*} \caption{\label{trim_and_draft} Calculated trim angle and draft
for the 2007 and 2008 data. The Froude numbers based on ship length and draft
are $F_{r{_L}}$ and $F_{r_{T_{max}}}$ respectively. $R_{e{_L}}$ is the Reynolds
number based on ship length.} \begin{center}
\setlength{\tabcolsep}{5pt} \hspace*{-.25in}
\begin{tabular}{|c|c|c|c|c|c|c|c|c|c|}
\hline \hline
Speed & Speed & Test & $F_{r{_L}}$ & $R_{e{_L}}$ & Trim  & T$_{FP}$ & T$_{AP}$ & $F_{r_{T_{max}}}$ & Transom\\
 m/s  & knots & Yr.  &             &             & (deg) & (m)      & (m)      &                   & Condition \\
\hline \hline
2.57 & 5 & '07 & 0.27 & $2.35\cdot 10^{7}$ & 0.19 & 0.296 & 0.323 & 1.43 & Wet \\\hline
3.60 & 7 & '07 & 0.38 & $3.29\cdot10^{7}$  & 0.51 & 0.284 & 0.366 & 1.9  & Wet \\ \hline
3.60 & 7 & '08 & 0.38 & $3.29\cdot10^{7}$ & 0.48 & 0.287 & 0.363 & 1.91 & Wet \\ \hline
4.12 & 8 & '07 & 0.43 & $3.77\cdot10^{7}$ & 0.73 & 0.274 & 0.366 & 2.1  & Dry \\ \hline
4.12 & 8 & '08 & 0.43 & $3.77\cdot10^{7}$ & 0.67 & 0.277 & 0.384 & 2.12 & Dry \\ \hline
4.63 & 9 & '07 & 0.49 & $4.23\cdot10^{7}$ & 0.78 & 0.274 & 0.393 & 2.34 & Dry \\
\hline \hline
\end{tabular}
\hspace*{0.0in} \end{center} \end{table*}

%

%
%

\subsubsection{Light Detection and Ranging (LiDAR)}

The free-surface deformation was measured using a scanning Light Detection And
Ranging (LiDAR) system. The LiDAR system used at NSWCCD, Riegl LMS-Q140-80i
\cite{Fu06b,Fu09b}, has a range accuracy of $\pm$2.54 cm for highly reflective
surfaces. The instrument scans an angular region of $\pm$40$^{\circ}$ at a
maximum line sample rate of 40 Hz with a maximum laser pulse frequency of 30
kHz. As infrared radiation is absorbed by water, only a small fraction of the
incident energy is scattered back to the instrument.

The LiDAR system was mounted to a traverse which moved in a direction parallel
to the centerline of the model, on a pan and tilt unit which allowed for remote
control of the LiDAR's position during testing.  In 2007 the traverse was
located 3.87 m above the mean waterline and 0.25 m starboard of centerline. In
2008 the traverse was at an elevation of 4.82 m and was 0.143 m port of
centerline.

\begin{figure} [t] \centering
\includegraphics[width=1.0\columnwidth]{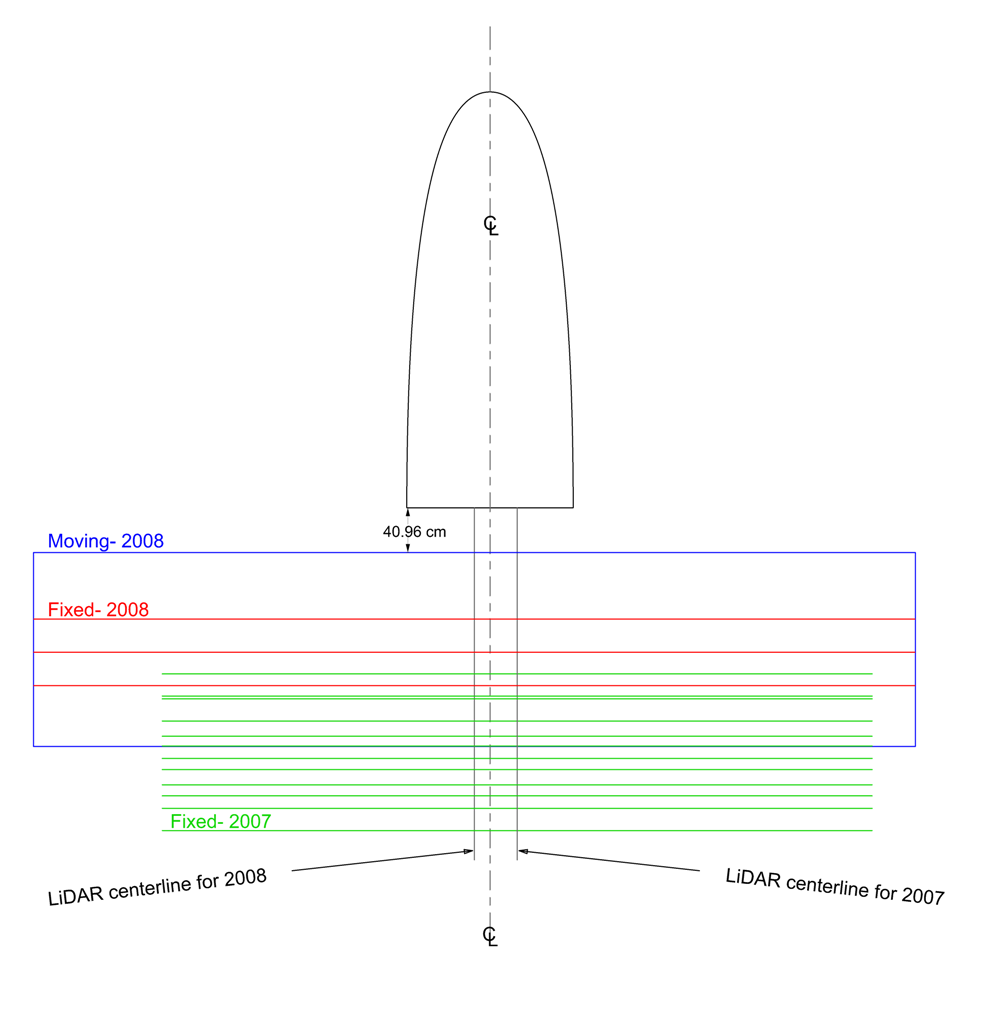}
\caption{\label{lidar_setup} Schematic showing the location of the LiDAR
measurements in 2007 and 2008. The field of view for the 2008 fixed and moving
LiDAR measurements are shown as red and blue lines. The locations of the fixed
LiDAR data from 2007 are shown in green. Not all measurement locations aft of
the transom were shown for 2007 due to the close spacing between lines. All
scan line widths are theoretical maximums. See Table \ref{table_2} for specific
distances. } \end{figure}

The primary goal of the LiDAR system's measurements during the 2007 test period
was to capture the gross properties of the transom stern wake. Data was
collected at a number of discrete locations aft of the
transom at a line sampling rate of 20 Hz.  During the 2008 test period the
focus was on capturing the statistical properties (mean and standard deviation)
of the transom stern wake with the LiDAR.  A set of three fixed locations aft
of the transom were selected and multiple runs were made at a line sampling
rate of 20 and 40 Hz. This data will be referred to as the fixed LiDAR data.
Additional tests were conducted in 2008 where the LiDAR was driven down the
traverse at a constant speed of 3.70 cm/s with a sampling rate of 20 Hz,
yielding a map of the surface elevation with an inter-line spacing of 2 mm. The
surface map extended from 0.422 m to 2.19 m aft of the stern in one run. This
data will be referred to as the moving LiDAR data set. A summary of the
discrete locations measured by the LiDAR during both testing phases is given in
Table \ref{table_2} and is also shown in Figure \ref{lidar_setup}.

%
%

%
%
\begin{table} [t] \caption{\label{table_2} Fixed locations aft of the transom
where stationary LiDAR measurements were taken during 2007 and 2008. See Figure
\ref{lidar_setup} for a schematic of the LiDAR coverage area. } \begin{center}
\begin{tabular} {|c|c|c|c|c|c|c|} \hline \hline Speed (knots) & 5 & 7 & 7 & 8 & 8
& 9\\ \hline Test Year & '07 & '07 & '08 & '07 & '08 & '07 \\ \hline 1.019 m &
&     & X &     & X &  \\ \hline 1.324 m &     &     & X &     & X &  \\ \hline
1.519 m & X  & X  &    & X  &    &  X\\ \hline 1.629 m &     &     & X &     &
X &  \\ \hline 1.723 m & X  & X  &    &  X &    &  X\\ \hline 1.748 m & X  & X
&    &  X &    &  X\\ \hline 1.951 m & X  & X  &    &  X &    &  X\\ \hline
1.964 m &     & X  &    &     &    &  \\ \hline 2.091 m &     &     &    &
&    &  X\\ \hline 2.180 m &  X &  X &    &  X &    &  X\\ \hline 2.193 m &  X
&     &    &     &    &  X\\ \hline 2.294 m &     &  X &    &     &    &  \\
\hline 2.396 m &  X &  X &    &  X &    &  X\\ \hline 2.536 m &     &  X &    &
X &    &  \\ \hline 2.637 m &  X &  X &    &  X &    &  X\\ \hline 2.752 m &
&  X &    &  X &    &  \\\hline 2.955 m &     &     &    &     &    &  X\\
\hline \hline \end{tabular} \end{center} \end{table}

%
%
\begin{table} \caption{\label{lidar_data_volume} Total amount of data used
to generate the mean and standard deviations shown in Figure
\ref{lidar_fixed_7kn} and \ref{lidar_fixed_8kn}. These represent concatenations
of approximately three to six separate runs.} \begin{center}
\begin{tabular}{|c|c|c|c|c|} \hline \hline Speed & Speed &
\multicolumn{3}{c}{$x$ Position}\vline \\ m/s  & (knots) & 1.019 m & 1.324 m &
1.629 m \\ \hline \hline 3.60  & 7 & 202 s & 321 s & 318 s          \\ \hline 4.12  &
8 & 137 s & 192 s & 119 s          \\ \hline \hline \end{tabular} \end{center}
\end{table}

The LiDAR returns a measurement of the distance to the
free-surface and the data was first corrected to yield elevation above the mean
water level (MWL). The MWL was computed each day during the testing. For each
measurement point the elevation, spatial location along the line, elevation
above the MWL, and return signal strength (i.e., amplitude) was returned.  Any
point where the amplitude of the return signal was 0 was ignored during
subsequent processing by setting it to be a NaN (Not a Number). As the LiDAR
has a constant angular step between measurement points, the spacing between
adjacent measurements increases away from the centerline. In order to correct
for this, the data was binned and the average of all points within the bin
returned. For the 2007 data the bin spacing was 2 cm and 5 cm for the 2008
data.

\begin{figure} \centering
\includegraphics[width=1.0\columnwidth]{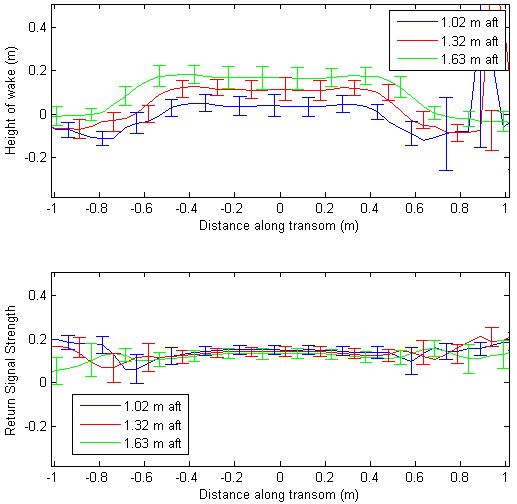}
\caption{\label{lidar_fixed_7kn} LiDAR measurements at three fixed locations
aft of the transom for a speed of 3.60 m/s (7 knots). The data was collected in
2008 and was binned with a bin size of 5 cm. The mean and standard deviation of
the wake elevation are shown in the top figure. The mean and standard deviation
of the return signal strength is shown in the bottom figure.} \end{figure}

\begin{figure} [t] \centering
\includegraphics[width=1.0\columnwidth]{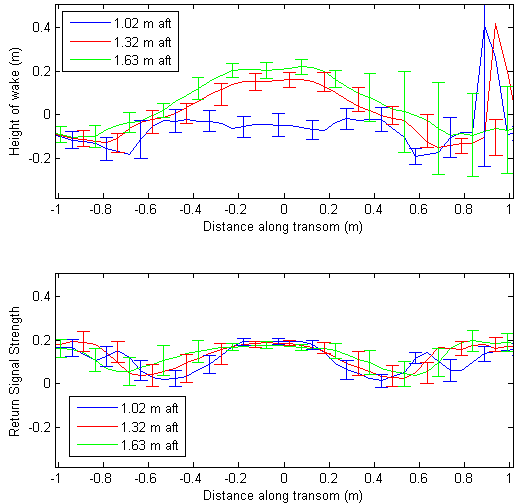}
\caption{\label{lidar_fixed_8kn} LiDAR measurements at three fixed locations
aft of the transom for a speed of 4.12 m/s (8 knots). The data was collected in
2008 and was binned with a bin size of 5cm . The mean and standard deviation of
the wake elevation are shown in the top figure. The mean and standard deviation
of the return signal strength is shown in the bottom figure.} \end{figure}

The fixed LiDAR measurements of the mean and standard deviation of the transom
wake at 3.60 m/s and 4.12 m/s (7 and 8 knots) from 2008 are shown in Figure
\ref{lidar_fixed_7kn} and Figure \ref{lidar_fixed_8kn}. The mean and standard
deviation of the signal strength is also shown, as the standard deviation could
be affected by a large variability in the measured signal strength.  The large
elevation spike seen at $\approx$89 cm is caused by the LiDAR reflecting
off of the traverse supporting the Quantitative Visualization (QViz) system.  The large RMS values between
$Y=0.4-0.7$ are likely due to the QViZ camera
system moving within the field of view of the LiDAR. At a speed of 4.12 (8 knots)
the standard deviation of the elevation is seen to increase along the shoulder
consistent with the reduction in signal strength as one moves away from the
transom centerline.

%
%

%
%

The results shown in Figures \ref{lidar_fixed_7kn} and \ref{lidar_fixed_8kn}
were computed from data collected at sample rates of both 20 and 40 Hz, see
Table \ref{lidar_data_volume} for the total amount of data collected in 2008.
The sampling rate was varied in order to obtain a mixture of improved spatial
and temporal resolution.

\begin{figure}  \begin{center}
\includegraphics[width=1.0\columnwidth]{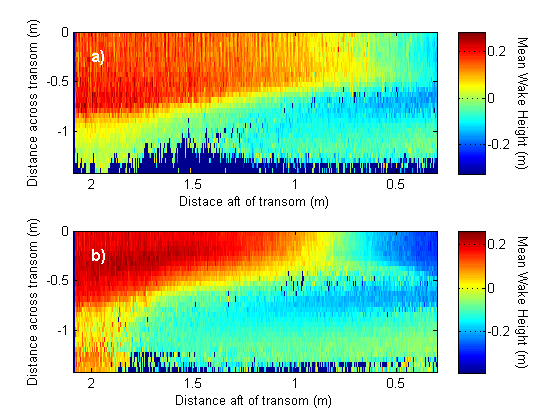}
\end{center} \caption{\label{lidar_2008_78kn} a) Surface map
of transom wake elevation generated from mean of three LiDAR runs collected in 2008
at a speed of 3.60 m/s (7 knots). b) Surface map of transom wake elevation
generated from mean of five LiDAR runs collected in 2008 at a speed of 4.12 m/s (8
knots).} \end{figure}

Collection of the moving LiDAR data was a secondary goal of the test
and therefore the volume of data is much less when compared to the fixed LiDAR
case. There are three individual runs at a
speed of 3.60 m/s (7 knots) and five individual runs at a speed of 4.12 m/s (8
knots). Each run was started individually and therefore each run had to be
aligned during post-processing. After each individual run was re-aligned, the
mean elevation from the moving LiDAR was compared against the fixed LiDAR data
to insure that no mis-alignment had occurred. The LiDAR results from 2008 is
shown in Figure \ref{lidar_2008_78kn}. RMS values are not shown for the moving
case due to the small sample size.

\begin{figure} \centering
\includegraphics[width=1.0\columnwidth]{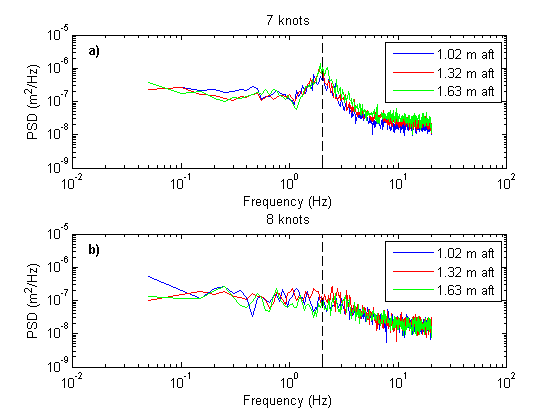}
\caption{\label{lidar_spectra_2008} Frequency power spectra of LiDAR
measurements at three fixed locations aft of the transom for a speed of a) 3.60
m/s (7 knots) and b) 4.12 m/s (8 knots). The data was collected in
2008 and three adjacent measurements along the centerline were averaged after
computing the spectrum.  The vertical dashed line is the location of the
spectral peak seen in the 3.60 m/s (7 knots) data.} \end{figure}

Frequency spectra were computed along the centerline at the three fixed aft
locations measured in 2008. They are shown in Figure \ref{lidar_spectra_2008}.
Peaks in the spectra near 2 Hz are evident in the 3.60 m/s (7 knots) dataset
(shown by a dashed line) for all distances aft of the transom. This phenomena
is thought to be related to the shedding of vortices off the transom and was
also seen by \citeasnoun{Wyatt08}. The peak is seen to vanish at a speed of 4.12
m/s (8 knots) when the transom becomes fully ventilated and would be
consistent with a process driven in part by vortex shedding. \citeasnoun{Wyatt08}
also describe this result as a possible generation mechanism and suggest that it could
also be due to waves which interact after having propagated off of the transom
corner.  A scaling of the Strouhal number between the model and full-scale Athena
measurements predicts a peak between 1.7-2.4 Hz based on a transom submergence
of $\approx 18.5$ cm at 3.60 m/s (7 knots), consistent with the vortex shedding
hypothesis.

%

\subsubsection{Quantitative Visualization (QViz)}

\begin{figure} \centering
\includegraphics[width=1.0\columnwidth]{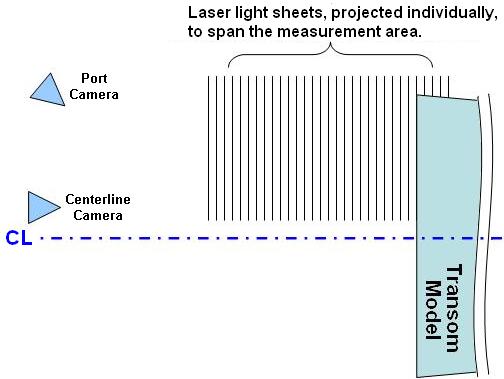}
\caption{Schematic representation of the QViz system:  two cameras captured
images of the intersection between the wave field generated by the transom and
systematic laser light sheets, projected as transverse lines, from a
cylindrical lens.} \label{qviz_schematic} \end{figure}

Free-surface elevation measurements of the breaking wave were obtained by an
optical laser sheet QViz system \cite{Furey02}.
These measurements represent an effort to capture the statistical properties of
the near-hull wake region, where accessibility is difficult by alternate
sensors, and to evaluate the dynamics across the wake shoulder into the
breaking region. Shown schematically in Figure \ref{qviz_schematic}, the QViz
system collected digital images of the intersection of the breaking transom
wave with a laser light sheet, projected parallel to the transom edge, to
generate lateral free-surface profiles for a systematic set of axial positions.
Subsequent image processing has been completed to transform the resulting wave
shape from pixels in the image plane into world coordinates.

\begin{figure} [t] \centering
\includegraphics[width=1\columnwidth]{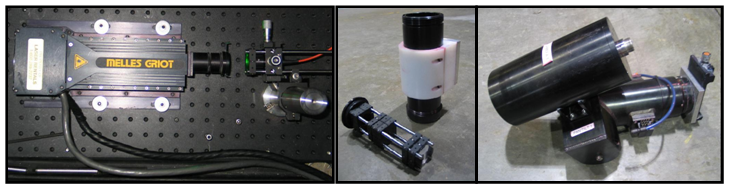}
\caption{Major QViz system components: laser as directed into a fiber-optic
cable (left), cylindrical lens assembly and housing (center), and a camera
housed in a motorized pan and tilt unit (right).} \label{qviz_parts}
\end{figure}

\begin{figure} \centering
\includegraphics[width=1.0\columnwidth]{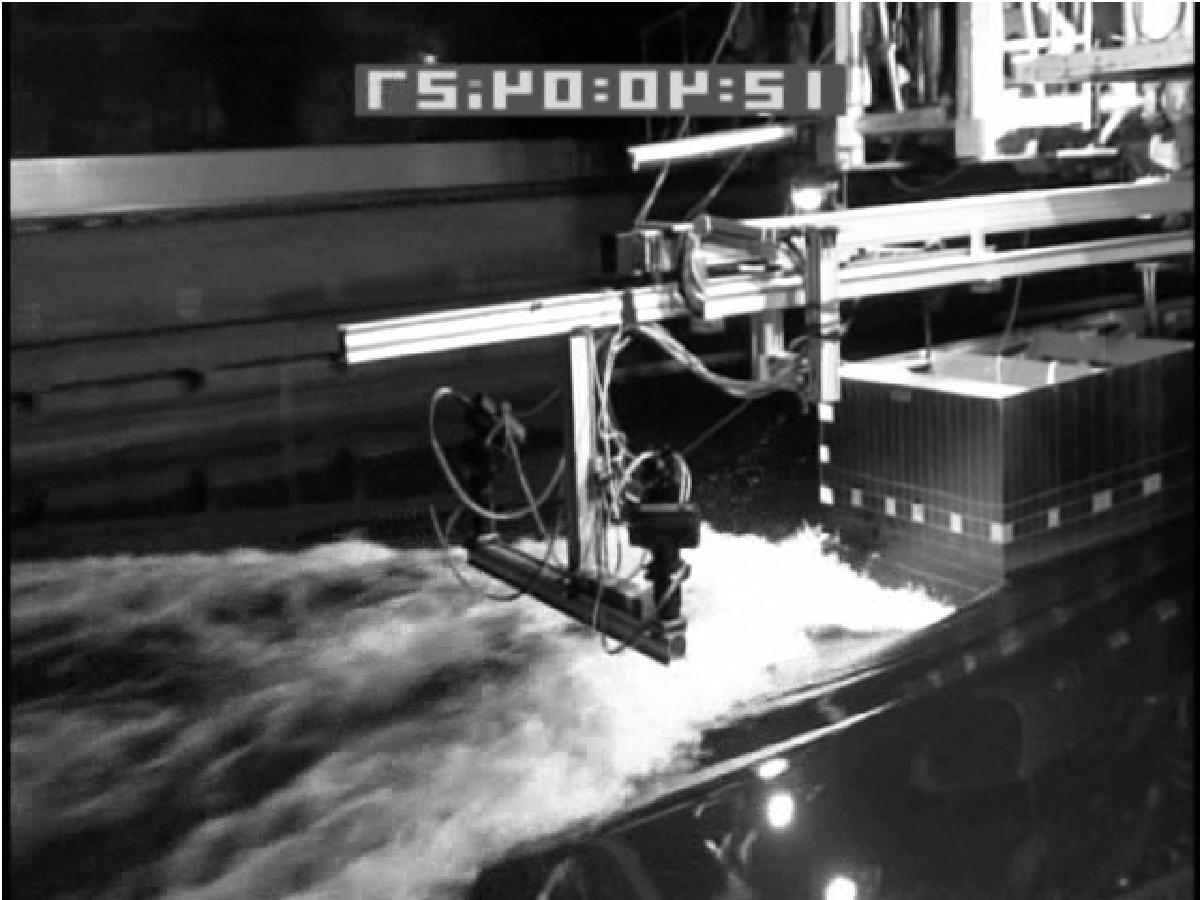}
\caption{Photograph of the QViz transverse as mounted during the 2008 Transom
Experiment.} \label{7kts_qviz_traverse} \end{figure}

The major components of the Qviz system deployed during this experiment are
depicted in Figure~\ref{qviz_parts}. A laser beam produced by a 3 watt laser
with 532 nm wavelength was fed through a fiber-optic cable to an
enclosed housing containing a cylindrical lens.  The beam was then converted by
the lens into a light sheet which was projected perpendicular to the disturbed
free-surface. Digital images of the resulting deformed light at the
intersection were acquired by two progressive scan non-interlaced video cameras
(JAI model CV-A11) operating at 30 frames per second, and lenses (Computar
model H6Z0812M) fitted with 532 nm filters to reduce ambient light noise. The
cameras were housed by Applied Microvideo pan and tilt units mounted to an
optical rail system, along with the lens housing. The rail assembly was moved,
in fixed increments, longitudinally aft of the model by a motorized traverse,
shown in Figure~\ref{7kts_qviz_traverse}.

The QViz system successfully acquired cross-sections of the flow at 3.6 m/s (7
knots) spanning roughly 0.1270 to 0.7874 meters to port of the model centerline
for axial locations spaced every 0.0254 meters from 0.0254 to 0.6604 meters aft
of the transom and every 0.0508 meters from 0.6604 to 1.2192 meters aft.  The
measurements of the flow at 4.12 m/s (8 knots) spanned roughly from centerline to 0.8128
meters to port, and axially from 0.0254 to 0.1524 meters at 0.0254 meter
increments, and at 0.0508 meter increments from 0.1524 to 1.2192 meters aft of
the transom.  The resolution of the images for this experiment was on the order
of 1 millimeter.

The images were each processed using a matched filtering approach to
define the edge acquired at each instant.  This algorithm and its
capabilities are described in detail in~\citeasnoun{Beale10}.  For
processing of the 2008 data, a single impulse
response was used to determine the matched filter output for each
image obtained from both cameras and for both model speeds.  The
selected impulse response was generated by pre-processing each run
to determine the edge detector with the maximum ratio of half-height
to half-width, and averaging each of these for all runs.  The
resulting matched filter output was thresholded using a value of
three times the standard deviation as a false alarm cutoff.

\begin{figure}[t]
\centering
\begin{tabular}{c}
    (a) \includegraphics[trim=20 180 5 200,clip,width=1.0\columnwidth]{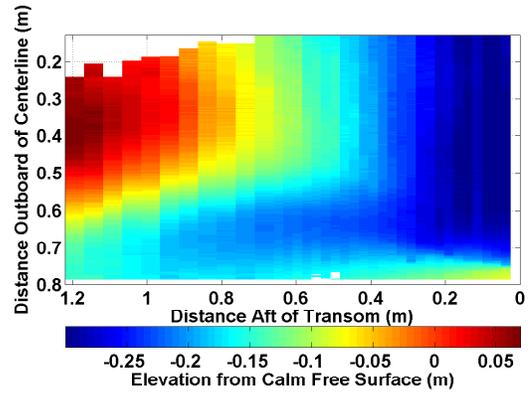} \\
    (b) \includegraphics[trim=20 180 5 200,clip,width=1.0\columnwidth]{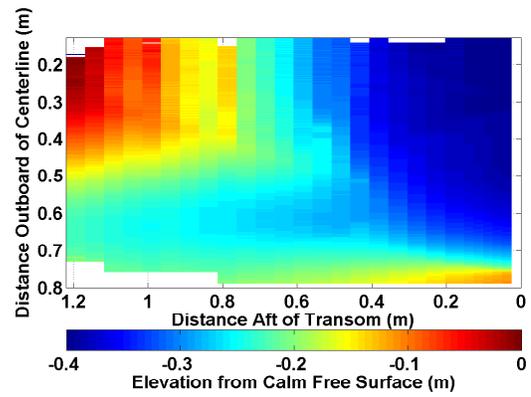} \\ \\
\end{tabular}
\caption{Transom wake elevations generated from QViz measurements,
for model speeds of (a) 3.6 m/s (7 knots) and (b) 4.12 m/s (8
knots).} \label{qviz_mean_surfaces}
\end{figure}

The complex free-surface elevation and its roughness have been
determined by binning the calibrated edge detection results of both
cameras. The QViz elevation fields are shown in
Figure~\ref{qviz_mean_surfaces} for model speeds of 3.6 m/s (7
knots) and 4.12 m/s (8 knots).

The frequency content, as measured by the QViz technique for
individual time series, have also been examined. Previous versions
of this technique have been successfully employed to obtain full-scale
field measurements of ship generated waves
\cite{Furey02,Rice04}, as well as laboratory breaking bow waves
\cite**{Karion03}. The production of meaningful spectral content,
however, is a new capability resulting from the matched filtering
processing scheme.

\subsubsection{Senix Ultrasonic Sensors}

The Senix Corporation's ToughSonic\textregistered \ Ultrasonic sensor is a non-contact, acoustic instrument
for measuring distances through air. A number of these were deployed during the
2007 and 2008 phases of the testing for the collection of longitudinal wavecut
data. In 2008 a single Senix sensor was used to supplement measurements of the
transom wake elevation. The sensor was mounted to the QViz traverse and
approximately 10 seconds of data were collected at 20 Hz at a number of  static
locations aft of the stern.   The traverse was also moved longitudinally at a
steady speed in order to obtain a profile of the wake. The sensor was located 8
mm starboard of  the model centerline, and was traversed from the stern to
1.134 m aft of the stern.

\begin{figure} \centering
\includegraphics[width=1.0\columnwidth]{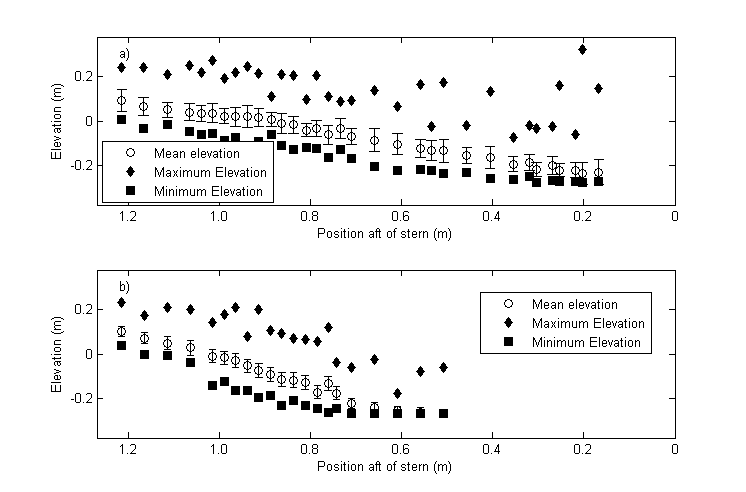}
\caption{\label{centerline_sonic} Centerline ultrasonic measurements at 3.60
m/s (7 knots) (a) and 4.12 m/s (8 knots) (b). Positions aft are referenced to
stern at 0 cm. There is no data forward of 0.508 cm for the 8-knot case as the
water level was outside of the calibrated range.} \end{figure}

The data was processed to yield maximum and minimum elevations as well as  mean and
standard deviations of the wake elevations. Multiple measurements at the same
fixed location were averaged to obtain a single value. The data from the moving
ultrasonic sensor was binned into 2.5 cm increments and then averaged for each bin
reported. No data was reported for the 4.12 m/s (8 knots) case within 0.508 m of
the transom as the water level was outside the calibrated range of the
instrument.  The results for 3.60 m/s (7 knots) and 4.12 m/s (8 knots) are shown
in Figure \ref{centerline_sonic}.

\subsubsection{Void Fraction Measurements}

A set of six impedance void fraction probes were used to measure the fraction of air entrained in
 the transom stern wake. Data was collected in both 2007 and 2008, with reduced noise in the 2008
measurements due to improved electronics \cite{Waniewski99}. The probes consist of two concentric
stainless steel electrodes separated by insulation. The outer  electrode is grounded and a sinusoidal voltage of
$\pm$2.5 V with an excitation of 500 kHz is applied to the inner electrode.  The impedance across the two
electrodes increases with increased void fraction and is mainly resistive for excitation frequencies below
the megahertz level. When an air bubble is pierced by a probe, the current between the two electrodes decreases
and voltage output of the probe is a large negative spike. The sampling rate of the probes was set at 20 kHz and was
determined based on the limitations of the data acquisition system.  See \citeasnoun{Fu09} and \citeasnoun{Fu10} for further details on
the setup of the probes.

\begin{figure} [t] \centering
\includegraphics[width=1.0\columnwidth]{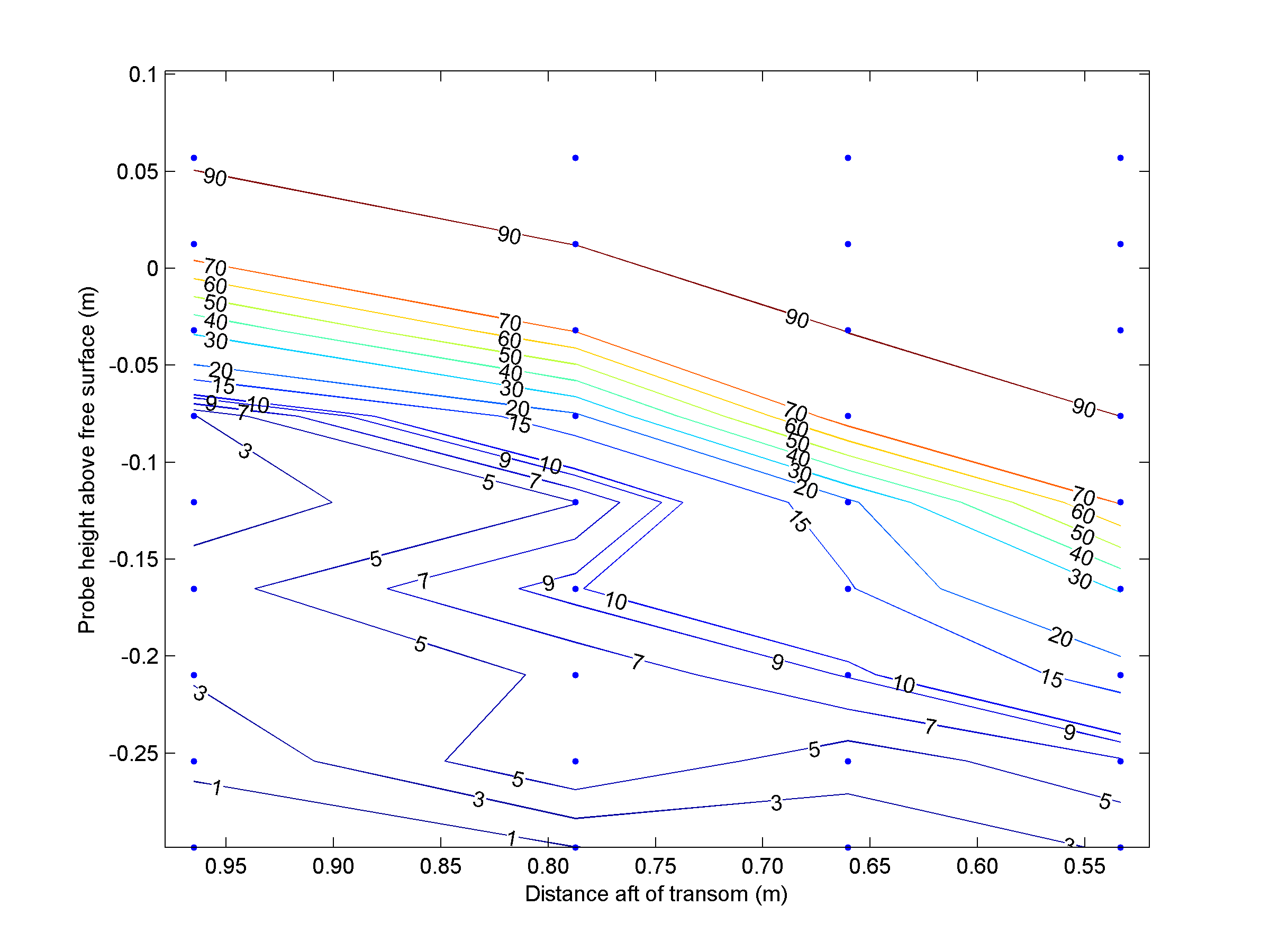}
\caption{\label{fig:VF_7kn} Contour plot of void fraction measurements taken in 20087 at a speed of 3.60 m/s (7 knots). The measurement
locations are denoted by blue dots.}
\end{figure}

\begin{figure} [h] \centering
\includegraphics[width=1.0\columnwidth]{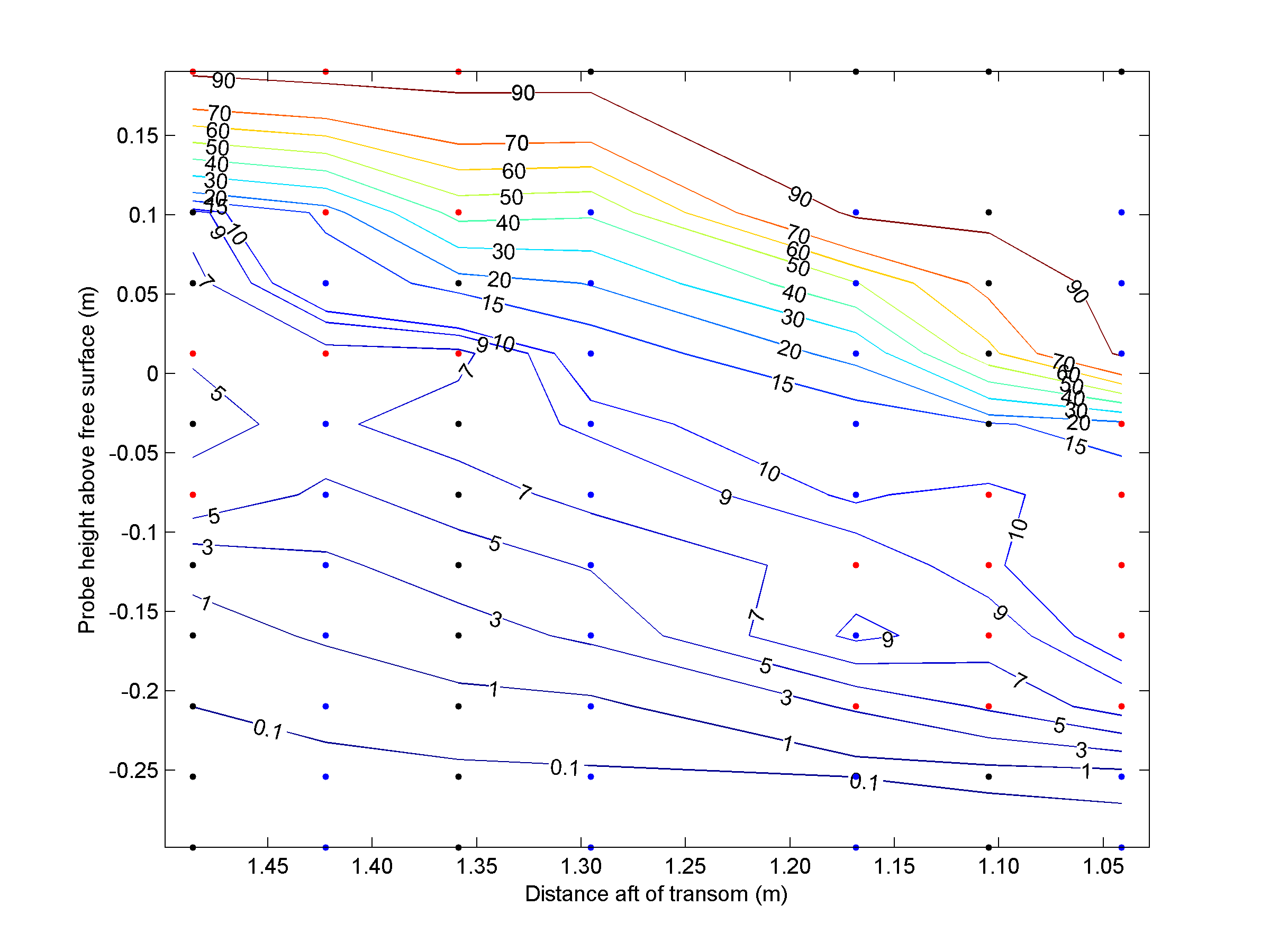}
\caption{\label{fig:VF_8kn} Contour plot of void fraction measurements taken in 2007 and 2008 at a speed of 4.12 m/s (8 knots).
The blue dots represent the 2007 data, the red dots 2008 data, and the black dots are the results of linear interpolation between 2007 and 2008 results. }
\end{figure}

The probes were mounted in a brass strut with a vertical separation of 8.9 cm.
The vertical position of the strut relative to the mean water level could be
adjusted and data was collected at three locations, yielding a final vertical
resolution of 4.4 cm. In 2007, data was collected at  3.60 m/s (7 knots) at
four longitudinal locations aft of the transom: 0.53 m, 0.66 m, 0.79 m, and
0.91 m.  At 4.12 m/s (8 knots), data was collected at 1.04 m, 1.17 m, 1.29 m,
and 1.42 m aft of the transom.  Between two to three runs were collected at
each position for both speeds and the results were then averaged. Error
analysis can be found in Appendix A of \citeasnoun {Fu09}. Data was collected
in 2008 at seven longitudinal locations aft of the transom (1.04 m, 1.10 m,
1.17 m, 1.29 m, 1.36 m, 1.42 m, and 1.48 m) for 4.12 m/s (8 knots) only. Due to
issues with noise when all six probes were operating, only two or three probes
were used at a time during the 2008 test period.  Results for 3.60 m/s (7
knots) from 2007 and 4.12 m/s (8 knots) from 2007 and 2008 are in Figures
\ref{fig:VF_7kn} and \ref{fig:VF_8kn}, respectively.  The spacing of contour
lines was chosen based on the largest error for a given void fraction range.
The contour plots agree qualitatively well with observations made from images
of the transom wake.  The spikes in Figure \ref{fig:VF_8kn} at (1.1 m, -0.076
m) and (1.17 m, -0.159 m)  are thought to be the result of the crude merging of
the 2007 and 2008 datasets. Comparison of results from 2008 and 2007 seem to
indicate that the data point at (1.36 m, 0.013 m) might be an outlier, possibly
due to noise in the system.

\subsubsection{Other Measurements}
The velocity field within the wake was measured using a Nortek Acoustic Wave
and Current (AWAC) profiler and SonTek\textregistered \ Acoustic Doppler Velocimeter (ADV\textregistered) (YSI Incorporated). The
AWAC profiler measured velocity and acoustic backscatter in the water column
while the ADV measured water velocity only. An estimate of the free-surface
deformation was also available from the AWAC profiler see
\citeasnoun{Fullerton08}. Defocused Digital Particle Image Velocimetry (DDPIV),
\cite{Pereira00,Jeon03}, was used to measure the bubble size distribution at
a fixed depth aft of the transom. Video measurements of the flow field aft of
the transom were made with both standard and high-speed video cameras. For the
high-speed video data, a fluorescent dye was injected at the transom as a
tracer.

\subsection{Experimental Measurements Discussion}

\begin{figure} \centering
\includegraphics[width=1.0\columnwidth]{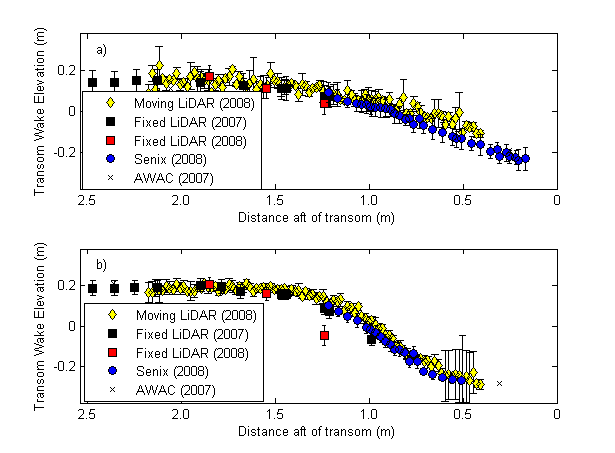}
\caption{\label{elevation_compare} A comparison between measurements of the
transom wake elevation from a variety of instruments used in 2007 and 2008. The
symbols are the mean values and the errorbars represent the standard deviation
of the measured data.}
\end{figure}

All of the instrumentation used to quantify the free-surface is non-intrusive.
The main difficulty in interpreting the data is in determining which surface
each instrument is actually measuring. A comparison between the LiDAR
measurements, centerline ultrasonic data, and AWAC data from both 2007 and 2008
is shown in Figure \ref{elevation_compare}.

All measurements agree qualitatively well up to $\approx$150 cm aft of the
transom.  Beyond this point, the LiDAR data begins to diverge from the
ultrasonic and AWAC results. This is likely caused by a reduction in the
returned signal strength as the number of scatters at the water surface
decreases, i.e., less surface roughness. The good qualitative agreement
indicates that all instruments appear to be measuring the same surface. Which
surface this corresponds to is unknown and remains an important question to be
answered.

When using infrared instrumentation, registration of the exact measurement
location relative to a fixed origin is often difficult. An initial 0.41 m
offset was applied to the LiDAR data presented here. Comparisons of 2008 LiDAR
measurements to other measurements that are not shown, including 2007 LiDAR
measurements and 2007 and 2008 longitudinal wave cuts, suggest that the 2008
LiDAR measurements are not registered properly in the longitudinal direction.
At one point during the 2008 test period, the forward attachment point on the
model came loose and the model moved from its original position.  An attempt
was made to place it back in its original position, but this could account for
some of the identified offset. Work is currently underway in an attempt to
identify the magnitude of the offset and any possible causes.

%
%

\section{NUMERICAL PREDICTONS} \label{sec:numerics}

Initial comparisons of the experimental measurements are made to
predictions from NFA and CFDship-Iowa-V.4 in the following sub-sections.
Comparison of the 2008 LiDAR data with both numerical predictions and
additional experimental measurements indicates that the 2008 LiDAR data is
likely not registered properly in the longitudinal direction. A brief discussion is
given in the Experimental Measurements portion of the paper above. Unless otherwise
specified, no offsets have been applied to the LiDAR or QViz datasets. Some
differences in the apparent agreement or disagreement between the various
simulations and the measurements could be attributed to a mis-registration
error.

    %
    %

\subsection{NFA Predictions and Assessments}

The objective of the numerical predictions is to assess the capability of NFA
to predict unsteady transom flows of model-scale and full-scale ships.

\subsubsection{Computational Method}

\begin{table*} [h!]
\begin{center} 
\begin{tabular}{|c|c|} \hline
Grid Cells  & Sub domains \\ \hline 
$N_x \times N_y \times N_z$     & $n_i \times n_j \times n_k$   \\ \hline 
$832 \times 384 \times 192=61,341,696$    & $13 \times 6 \times 6 = 468$  \\ \hline 
$1664 \times 786 \times 384=490,733,568$   & $13 \times 6 \times 6=468$ \\ \hline
$2688 \times 1024 \times 384=1,056,964,608$ & $21 \times 8 \times 6=1008$  \\ \hline 
\end{tabular}
\caption{\label{nfa_details} Details of NFA Numerical Simulations.} 
\end{center}
\end{table*}

The NFA code provides turnkey capabilities to model breaking waves around a
ship, including both plunging and spilling breaking waves, the formation of
spray, and the entrainment of air.   A description of NFA and its current
capabilities can be found in \citeasnoun{dommermuth07};
\citeasnoun{dommermuth08}; and \citeasnoun{nfa1}.  NFA solves the Navier-Stokes
equations utilizing a Cartesian-grid formulation.   The flow in the air and
water is modeled, and as a result, NFA can directly model air entrainment and the
generation of droplets.  The interface capturing of the free surface uses a
second-order accurate, volume-of-fluid technique.  A cut-cell method is used to
enforce no-flux boundary conditions on the hull.  A boundary-layer model has
been developed \cite{nfa3}, but it is not used in these numerical simulations, and as a result, the 
tangential velocities are free to slip over the hull.  NFA uses an implicit sub-grid scale (SGS)
model that is built into the treatment of the convective terms in the momenum
equations \cite{nfa1}.  A surface representation of the ship hull is all that
is required as input in terms of hull geometry.  The numerical scheme is
implemented on parallel computers using Fortran 90 and Message Passing
Interface (MPI).  Relative to methods that use a body-fitted grid, the potential advantages of NFA's approach
are significantly simplified gridding requirements and greatly improved
numerical stability due to the highly structured grid.

\subsubsection{Domain, Grids, Boundary and Simulation Conditions}

\citeasnoun{dommermuth08} compare NFA predictions to the results of 2007
measurements.   The comparisons include drag, free-surface contours in the
transom region, and longitudinal wave cuts.   For the 2008 comparisons, the NFA
results are focused on the transom region, including mean  elevations, free-surface spectral content, transverse cuts, and air entrainment.   Other 2008 comparisons include predictions of drag and longitudinal wave cuts.   Three grid
resolutions have been performed for the 2008 data set.   The highest grid
resolution for the 2008 experiments uses about 45 times more grid points than
the comparisons to the 2007 experiments.   Most of the increased grid
resolution has been concentrated in the transom region.  A plane of symmetry is
eliminated on the centerline, and the length of the domain has been extended by
2.5 ship lengths astern and 0.5 ship lengths ahead.  NFA comparisons to the
2008 experiments show better agreement with experiments than the comparisons to
the 2007 experiments that are reported in \citeasnoun{dommermuth08} due to improvements in the 
theory and numerical implementation, and increased grid resolution.

NFA predictions of the free-surface elevations near a transom-stern model
moving with constant forward speed are compared to laboratory measurements from
the model experiment producing full-scale breaking.  Table \ref{trim_and_draft} provides details
of the transom-stern model tests, including the length of the model, the depth
of the transom, the speed of the model, and the Froude and Reynolds numbers.   

Numerical simulations for the 7- and 8-knot cases at the 2008 trim are performed,
corresponding respectively to Froude numbers $F_r=0.38\;$ and $\;0.43$ and Reynolds numbers $Re =2.9\times10^7$ and $Re=3.3\times10^7$.   For 3.60 m/s (7 knots), the
transom is partially wet, and for 4.12 m/s (8 knots), the transom is dry.   All
length and velocity scales are respectively normalized by the model's length
($L_o$) and speed ($U_o$) .  

Table \ref{nfa_details} provides details of the numerical simulations.   The number of grid cells along the $x$, $y$, and $z$-axes are respectively denoted by $N_x, N_y,\;$ and $\;N_z$.    The number of sub-domains and processors along the $x$, $y$, and $z$-axes are respectively denoted by
$n_i, n_j,\;$ and $\;n_k$.   The coarsest resolution simulation uses about 61
million grid cells, and the highest resolution simulation uses over 1 billion
grid cells.  The two highest grid resolutions have twice as much grid
resolution along each direction as the coarsest simulation.   The main
difference between the two highest grid resolutions is improved grid refinement
near the bow and in the lateral direction.   The length, width, depth, and
height of the computational domains are respectively 6.0, 1.6983, 0.66667, 0.5
ship lengths ($L_o$).   These dimensions match the cross section of the NSWCCD
towing tank.  The computational domain extends 4 ship lengths behind the
transom and 1 ship length ahead of the bow.    The fore perpendicular and
transom are respectively located at $x=0$ and $x=-1$.   The $z-$axis is
positive up with the mean waterline located at $z=0$.   A plane of symmetry is
not used on the centerline of the hull because small-scale turbulent structures
are adversely affected.    

Grid stretching is employed in all directions. Details of the grid-stretching algorithm are provided in
\citeasnoun{dommermuth06}.  For the highest resolution case, the smallest grid
spacing is 0.0005 near the hull and mean waterline, and the largest grid
spacing is 0.01  in the far field.     The numerical simulations are slowly
ramped up to full speed.   The period of adjustment is $T_o=0.5$
\cite{dommermuth06}.   Mass conservation is ensured using the regridding
algorithm that is implemented by \citeasnoun{dommermuth06}.  Density-weighted
velocity smoothing is used every 400 time steps using a 3-point filter (1/4,
1/2, 1/4)  \cite{nfa1}.  The non-dimensional time step is $\Delta t=0.00025$.

The simulations are run for 30,000 time steps, corresponding to 7.5 ship
lengths, on the SGI\textregistered \ Altix\textregistered \ ICE (Silicon Graphics, Inc.) at the U.S. Army Engineer Research and
Development Center (ERDC).  The data sets are so large that only time steps 20,0000 through 30,000 are saved every
40 time steps for the purposes of post processing. The 1.06 billion cell simulation takes about 90
hours of wall-clock time to run 30,000 time steps using 1008 processors.  The
wall-clock time can be cut in half by doubling the number of processors because
NFA scales linearly.	  Alternatively, increasing the number of processors to over 8,000 will enable
numerical simulations of breaking ship waves and tsunamis with over 10 billion grid cells within the next year.

\subsubsection{Prediction Assessments}

\begin{figure*} 
\begin{center}
\begin{tabular}{llll} 
(a) & & (b) & \vspace{-15pt} \\ 
& \includegraphics[width=0.4\linewidth]{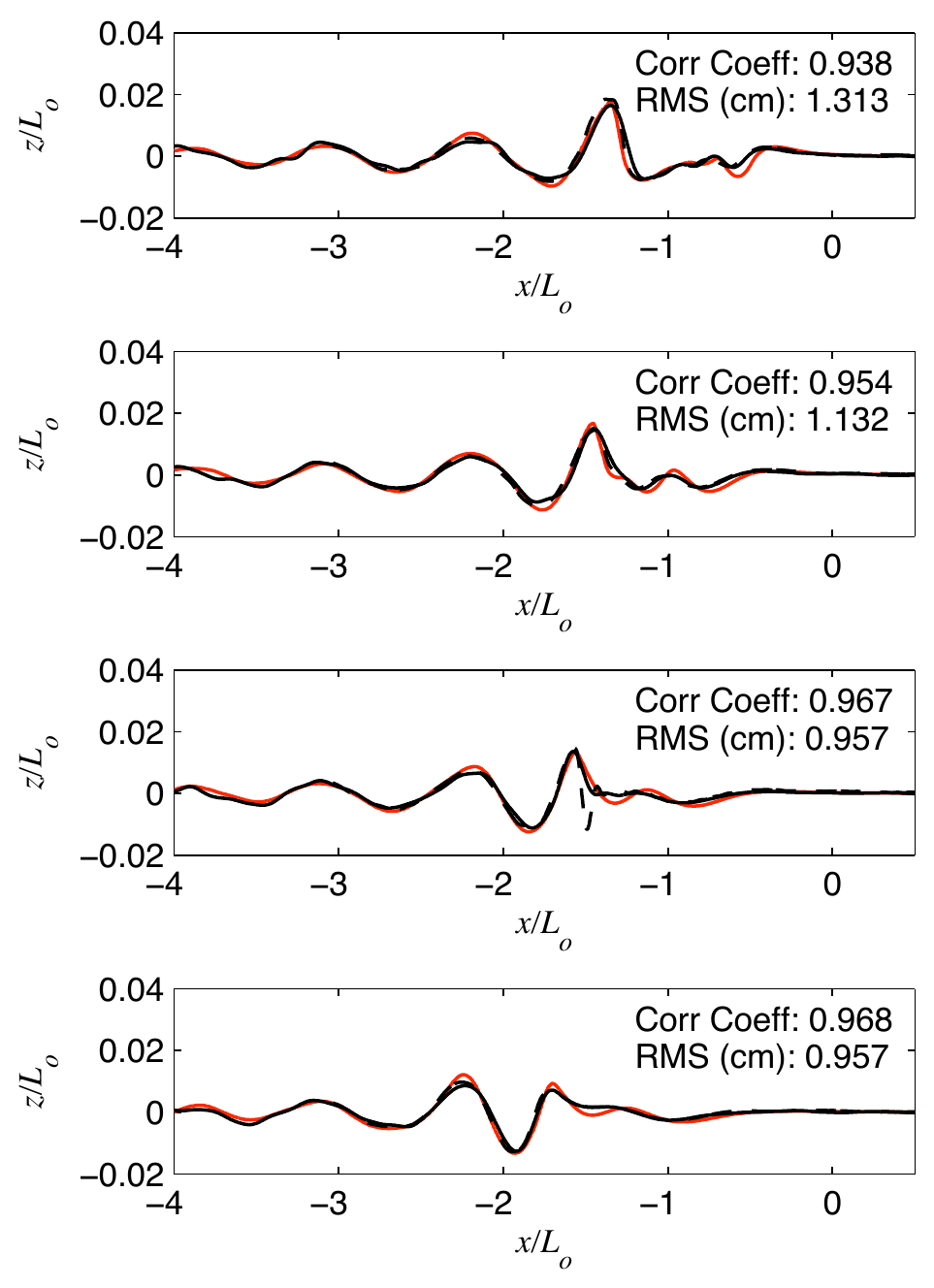} 
& & \includegraphics[width=0.4\linewidth]{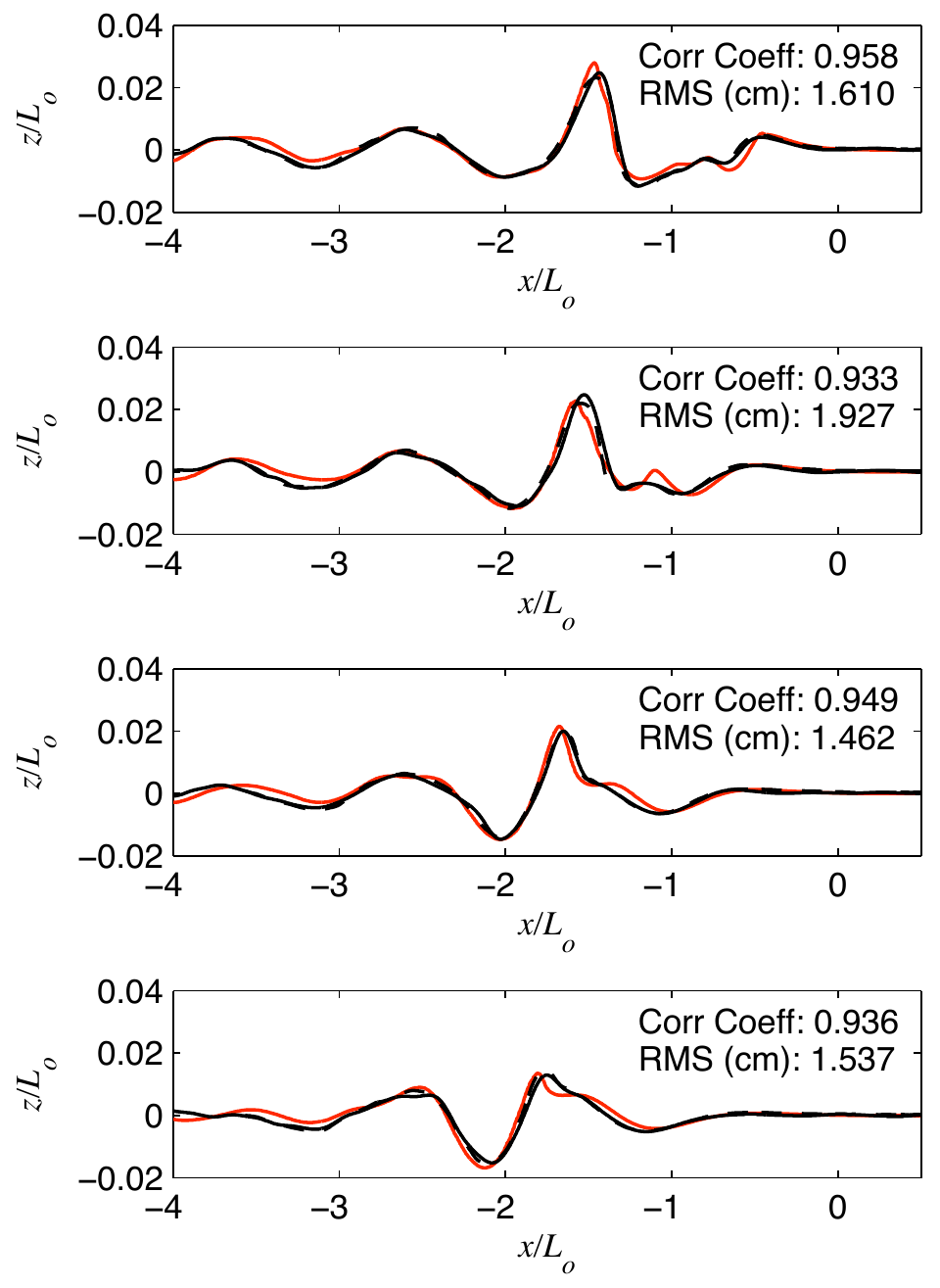}
\end{tabular} 
\end{center} 
\caption{\label{nfa_wavecut} Transom-stern model wave cuts.  (a) 3.60 m/s (7 knots)  and (b) 4.12 m/s (8 knots). NSWCCD measurements (2008: black lines, 2007: black dashed lines) are compared to NFA predictions (red lines).  For each speed and from top to bottom, the longitudinal cuts are located at $y/L_o$=0.14375, 0.22847, 0.3125, and 0.39514. $x/L_o=0$ corresponds to the bow and $x/L_o=-1$, the transom stern.  The correlation coefficients are between NFA predictions and 2007 measurements. } 
\end{figure*}

Figure \ref{nfa_wavecut} compares measured wave cuts using sonic probes from
the
2007 and 2008 experiments to predicted wavecuts using NFA for 3.60 m/s (7 knots) and 4.12 m/s (8
knots) speeds and four transverse locations.   The NFA predictions have been
time-averaged over the last 10,000 time steps for the 1.06 billion cell
simulations.   There is  good agreement between measurements and predictions.
The greatest  error occurs for the cusp line wave, where NFA predictions
have a lower trough than measurements. The portions of the NFA algorithm that
affect the prediction of the cusp line wave are under consideration.
The agreement between NFA predictions and experimental measurements in the
stern region where wave breaking occurs is very good.   The correlation
coefficients between the predictions and the measurements average 0.95 for both
3.60 m/s and 4.12 m/s.   The correlation coefficients between the 2007 and 2008
measurements average 0.97 for 3.60 m/s and 0.99 for 4.12 m/s.   The RMS error
between predictions and measurements average 1.0cm for 3.60 m/s and  1.6cm for
4.12 m/s.   The RMS errors between 2007 and 2008 experiments are 0.86cm. 

\begin{figure*}
\begin{center}
\begin{tabular}{c} 
\includegraphics[width=0.8\linewidth]{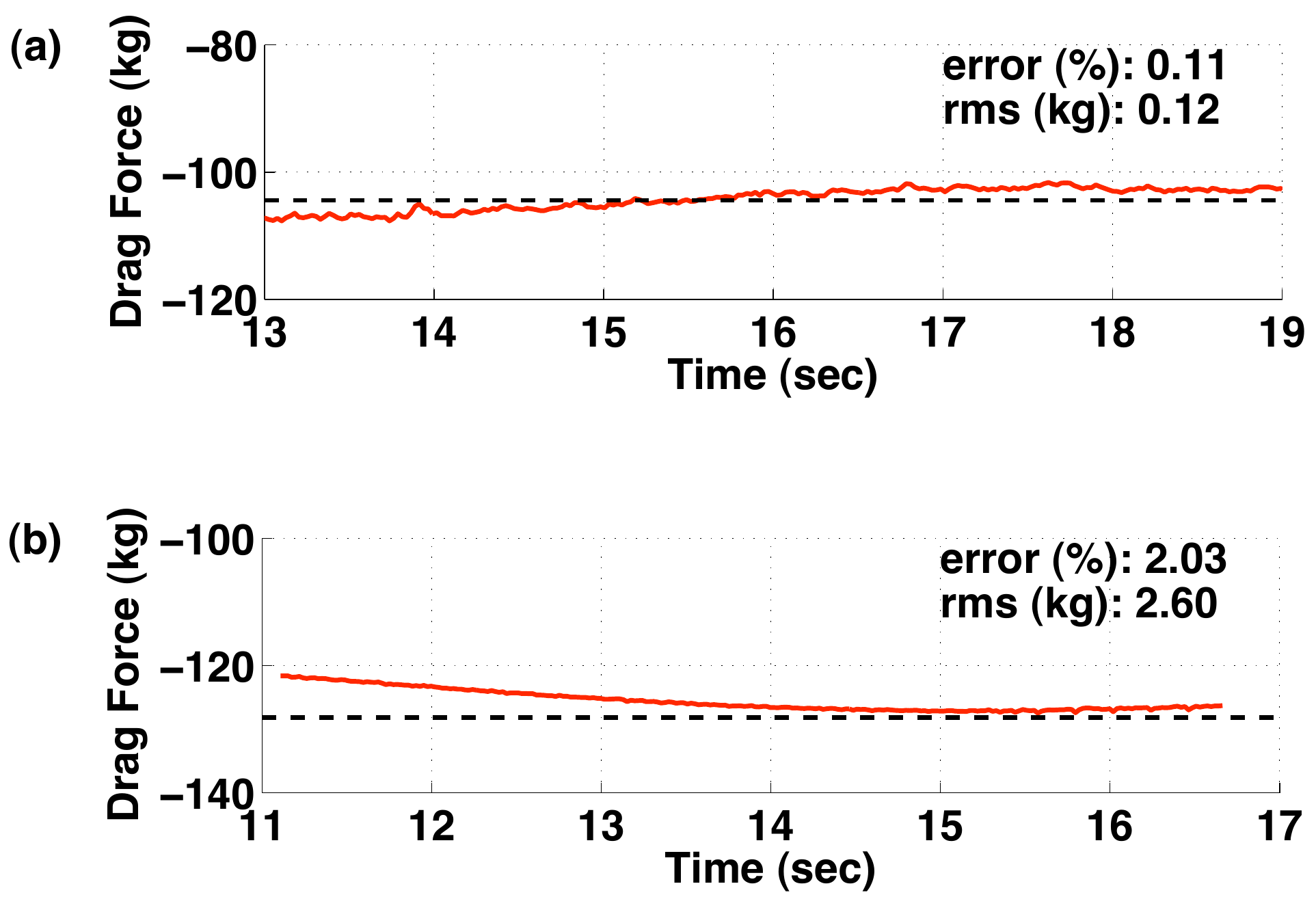}
\end{tabular} 
\end{center} 
\caption{\label{nfa_drag} Drag predictions compared to experimental measurements.  (a) 3.60 m/s (7 knots)  and (b) 4.12 m/s (8 knots). NSWCCD  measurements (black dashed lines) are compared to NFA predictions (red solid lines). } 
\end{figure*}

Figure~\ref{nfa_drag} compares drag predictions to measurements.  The viscous
and wavemaking portions of the drag are calculated using the procedures
outlined by \citeasnoun{dommermuth08}.  The NFA results are plotted as a
function of time late in the simulations to show the convergence to steady
state.  The numerical results have a low-frequency oscillation.   As shown by
\citeasnoun{Wehausen64}, unsteady oscillations can occur in the wave
resistance, and by implication the surface elevations, due to starting
transients. There are also starting transients in the buildup of separation and
the boundary layer on the hull, but the viscous time constants are
significantly shorter than the wave resistance. The oscillations in the wave
resistance occur at a period  equal to $T=8\pi U/g$, where $U$ is the speed of
the ship and $g$ is the acceleration of gravity.  For the 3.60 m/s (7 knots)
and 4.12 m/s (8 knots) cases, the periods of oscillation are equal to 9.2 and
10.6 seconds.   The time records of the drag are not long enough for full
periods of oscillation, but for the 3.60 m/s (7 knots) case, the half period of
oscillation that is shown agrees with theory.  The relative errors in the mean
drag and RMS errors are calculated over the duration of the time record that is
shown in the plot.   Overall, there is very good agreement between numerical
predictions and experimental measurements. 

\begin{figure*} [ht!]
\begin{center} 
\begin{tabular}{llll} (a) & & (b) & \vspace{-15pt} \\ 
& \includegraphics[width=0.4\linewidth]{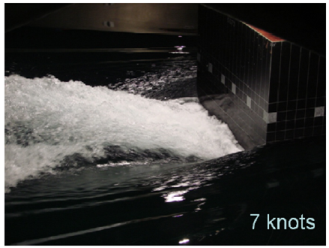} 
& & \includegraphics[width=0.4\linewidth]{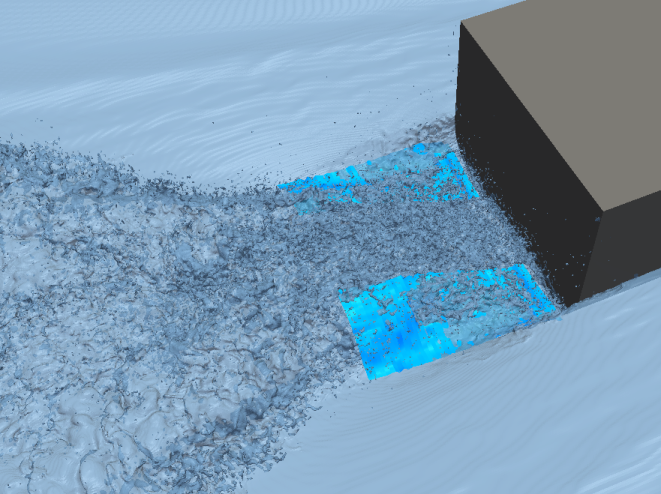} \\
(c) & & (d) & \vspace{-15pt} \\ 
& \includegraphics[width=0.4\linewidth]{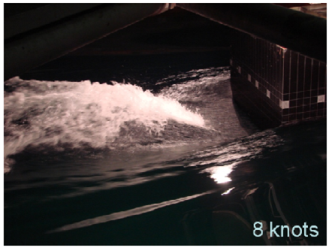} 
& & \includegraphics[width=0.4\linewidth]{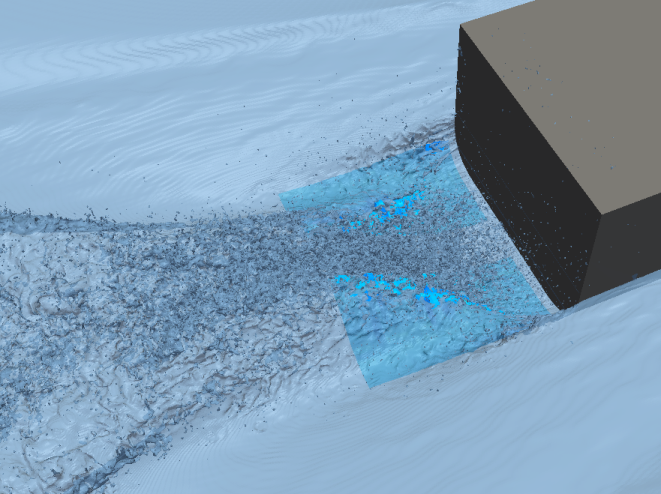}
\end{tabular} 
\end{center} 
\caption{\label{nfa_perspective} Perspective views
of transom.  (a) 3.60 m/s (7 knots), NSWCCD.  (c) 4.12 m/s (8 knots), NSWCCD.  (b)
3.60 m/s (7 knots), NFA.  (d) 4.12 m/s (8 knots), NFA. The darker blue surface
represents the QViz data. Both the QViz and NFA results are translucent to
allow for comparison between the two results.} 
\end{figure*}

Figure \ref{nfa_perspective} compares perspective views of laboratory and NFA
results for 3.60 m/s (7 knots) and 4.12 m/s (8 knots).   Instantaneous NFA predictions at t=5
(dark blue) are overlaid time-averaged QViz measurements (light blue).  The 0.5
isosurface of the volume fractions are shown for the NFA predictions. The NFA
results are translucent to show the entrainment of air due to
free-surface turbulence.  Regions of air entrainment are denoted by a slightly
darker shade of blue.  The transom is partially wet for 3.60 m/s (7 knots) and
fully dry for 4.12 m/s (8 knots).    A glassy region is  evident behind the
transom at the 4.12 m/s (8 knots) speed.  Significant air entrainment occurs for
the 3.60 m/s (7 knot) case at the transom, in the rooster-tail region, and along the edges
of the breaking stern wave.    For the 4.12 m/s (8 knot) case, air entrainment first
occurs on the forward face of the rooster tail and along the edges of the
breaking stern wave.    For both the 3.60 m/s (7 knots) and 4.12 m/s (8 knots) speeds, the measured mean profile of the free-surface elevation agrees well with instantaneous
predictions.  Animations of NFA results are available at
\href{http://www.youtube.com/waveanimations}{http://www.youtube.com/waveanimations}.

\begin{figure*}
\begin{center}
\begin{tabular}{llll} (a) & & (b) & \vspace{-15pt} \\ 
& \includegraphics[width=0.4\linewidth]{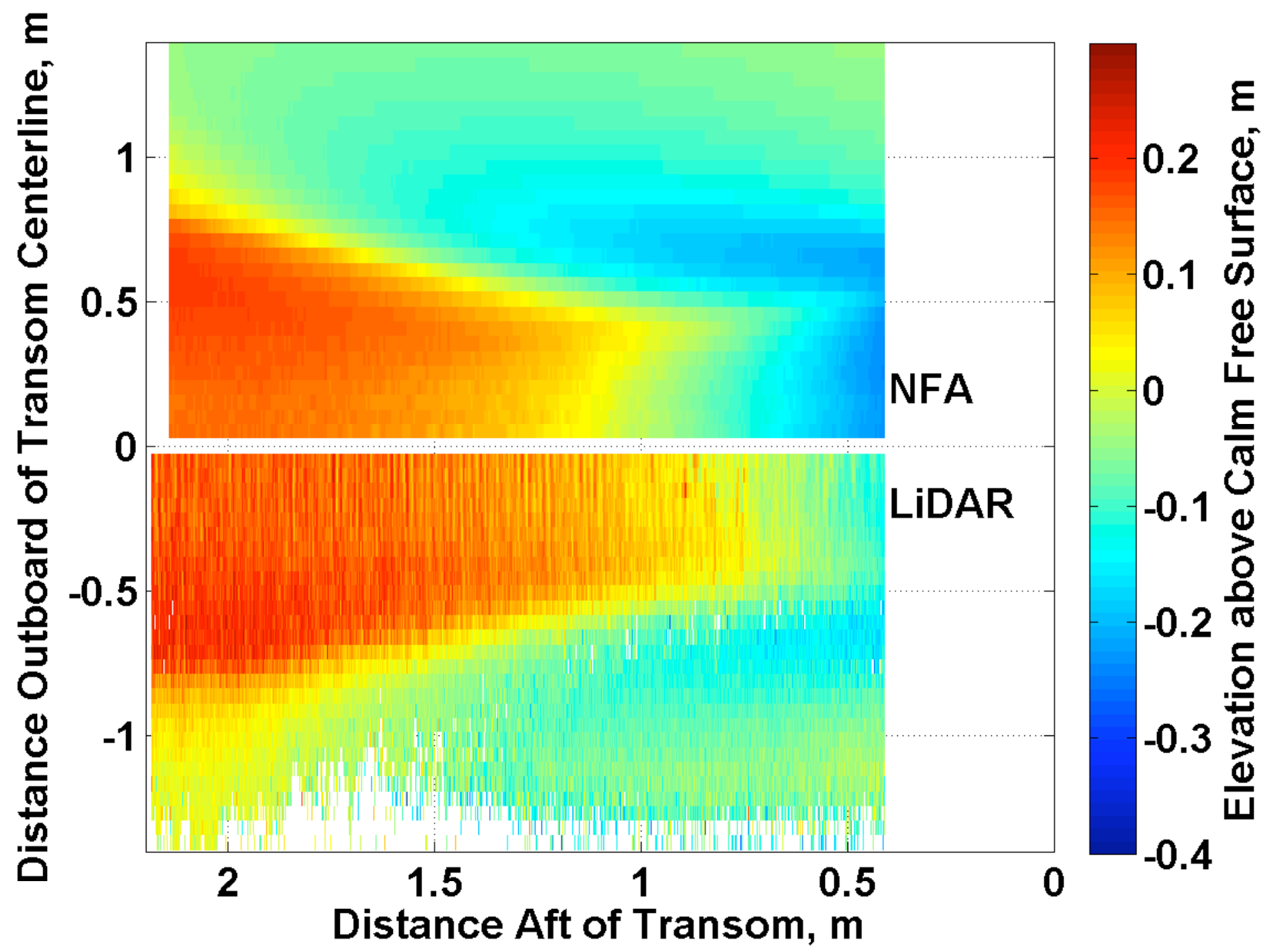} 
& & \includegraphics[width=0.4\linewidth]{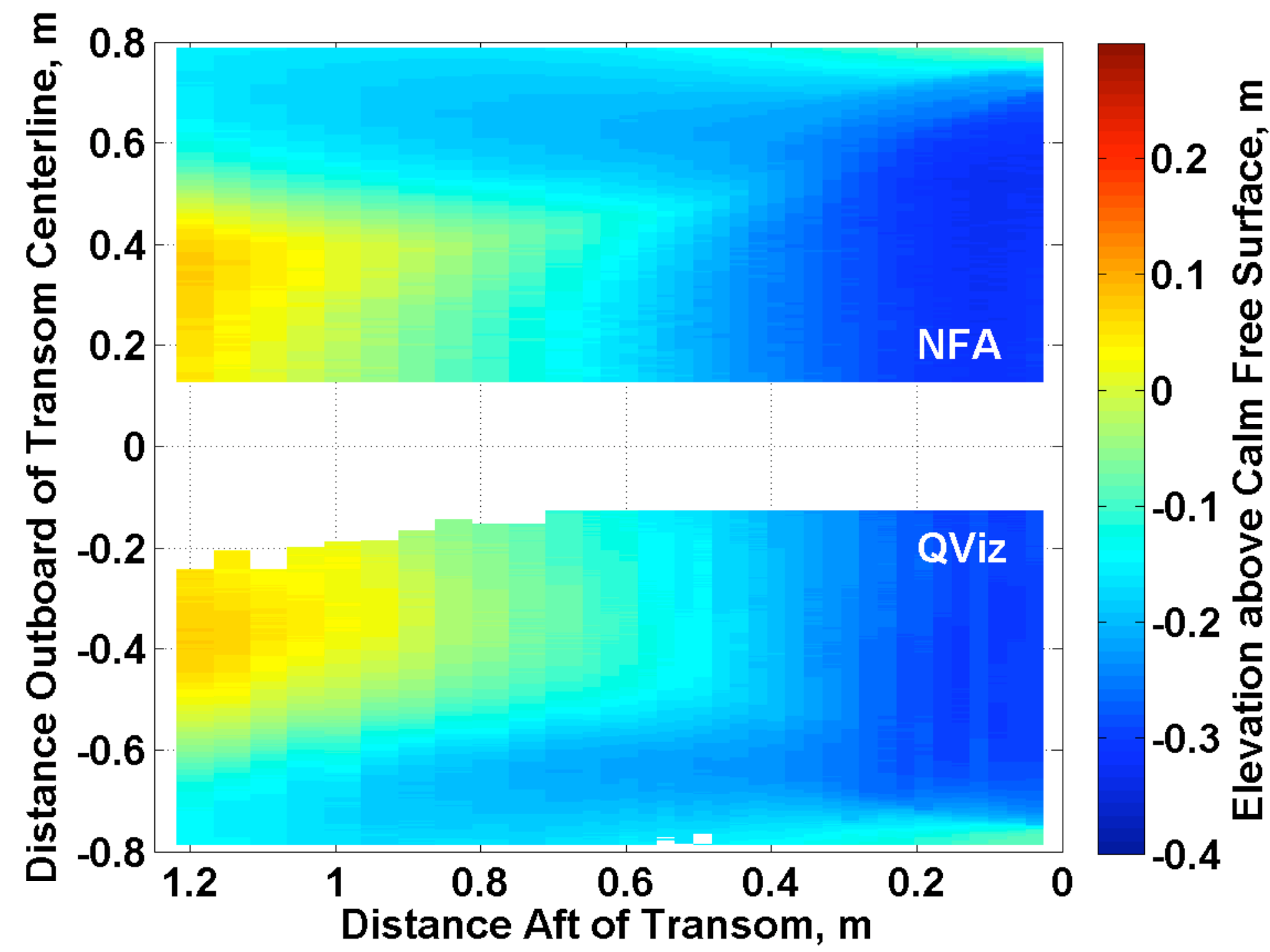} \\ 
(c) & & (d) & \vspace{-15pt} \\
& \includegraphics[width=0.4\linewidth]{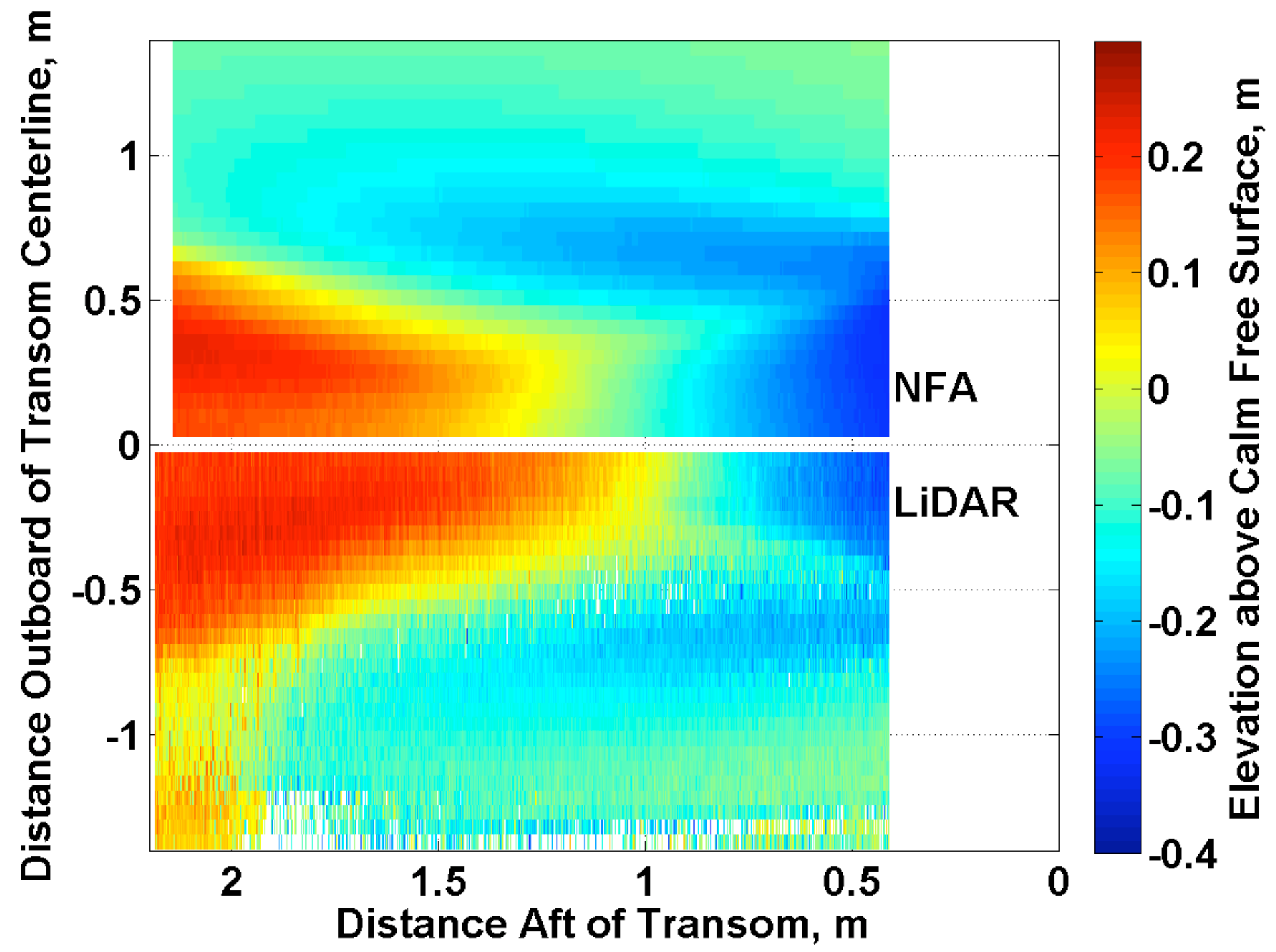} 
& & \includegraphics[width=0.4\linewidth]{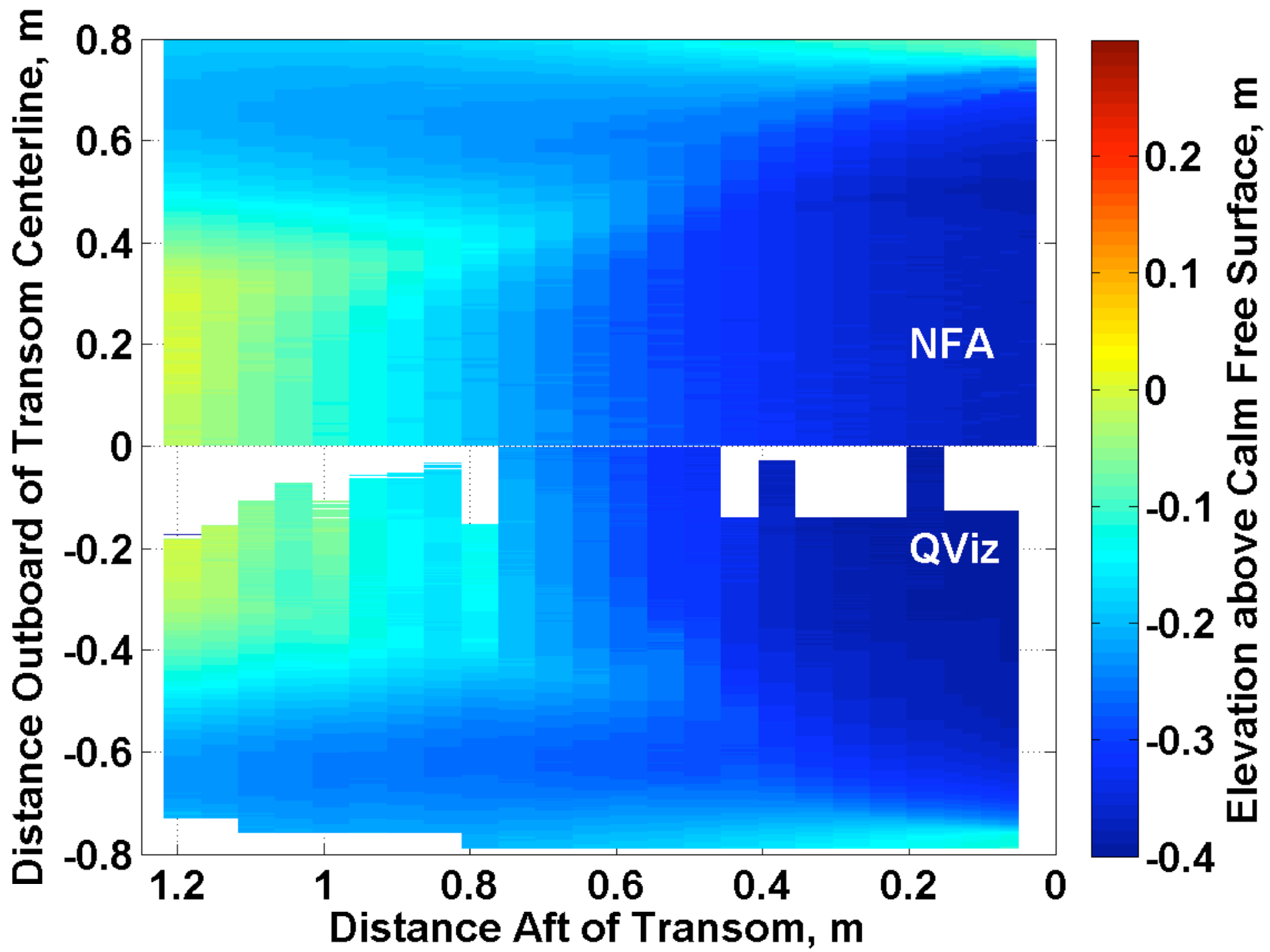}
\end{tabular} 
\end{center} \caption{\label{nfa_contours} Free-surface contours.
NFA predictions are above and measurements are below.  (a) NFA versus LiDAR,
3.60 m/s (7 knots).  (b) NFA versus QViz, 3.60 m/s (7 knots).  (c) NFA versus
LiDAR, 4.12 m/s (8 knots). (d) NFA versus QViz, 4.12 m/s (8 knots). All NFA results are top-down processing.}
\end{figure*}

Figure \ref{nfa_contours} compares contours of QViz and LiDAR measurements to
NFA predictions for the free-surface elevation in the transom region.   Since QViz and LiDAR look down on the free surface,  the NFA results are processed in a similar manner.    Based on volume-fraction data, a time series of free-surface heights is calculated at particular $x$ and
$y$ locations. The algorithm starts at the top of the domain and moves downward
searching for transitions from air to water.   Spray droplets of one and two grid
cells are filtered out, and the highest resulting free-surface intersection is
calculated.  For future reference, this type of processing of the data is denoted as top-down. The correlation coefficients for Figure \ref{nfa_contours} (a) \& (c), the 3.60 m/s
(7 knots) and 4.12 m/s (8 knots) comparisons to LiDAR data, are respectively 0.882 and 0.905.
For Figure \ref{nfa_contours} (b) \& (d), the 3.60 m/s (7 knots) and 4.12 m/s (8 knots)
comparisons to QViz data, the correlation coefficients are respectively 0.969
and 0.986.   As discussed in the introduction, an unresolved registration issue is the likely cause of the lower correlation coefficients for the LiDAR measurements.

\begin{figure*}
 \begin{center} 
\begin{tabular}{llllll} (a) & & (b) & & (c) & \vspace{-18pt} \\ 
& \includegraphics[width=0.25\linewidth]{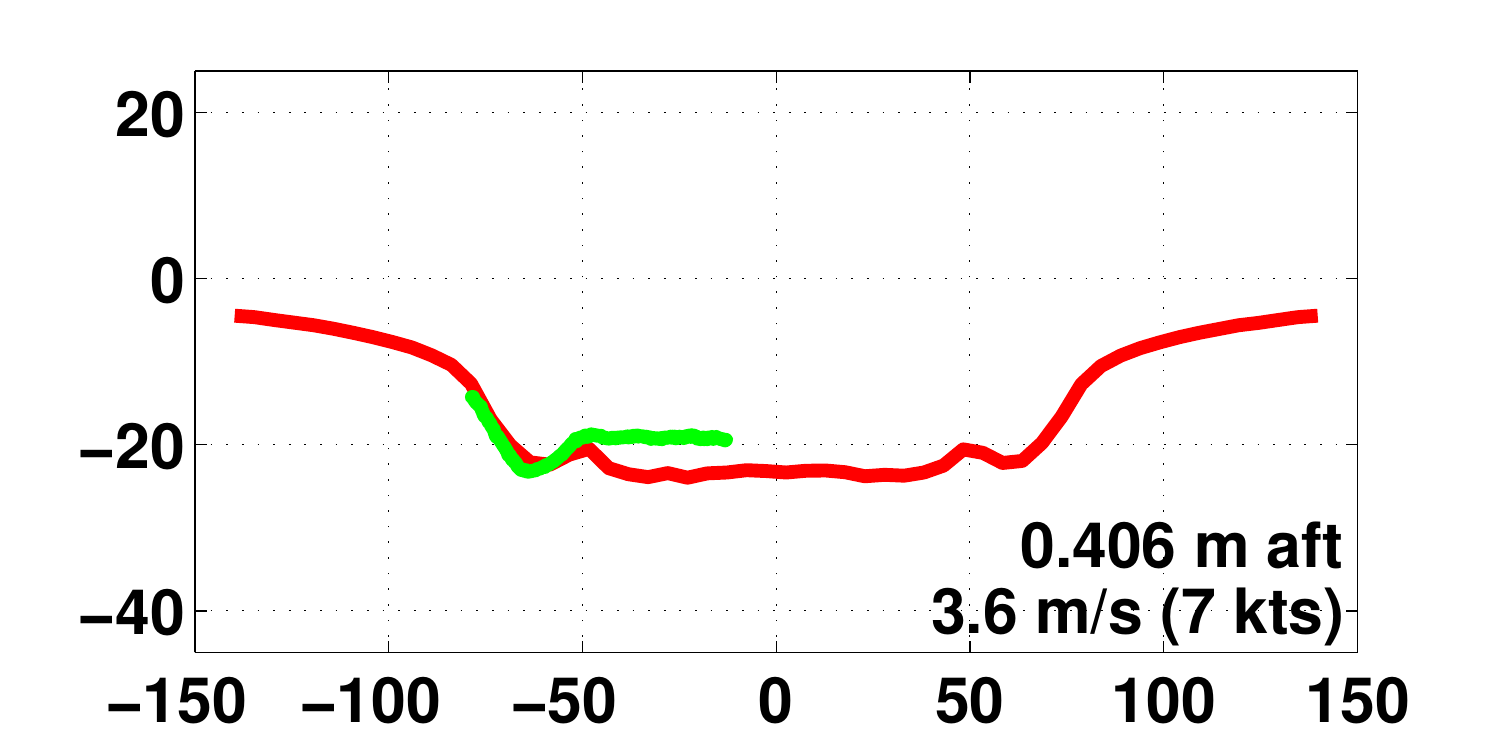} 
& &  \includegraphics[width=0.25\linewidth]{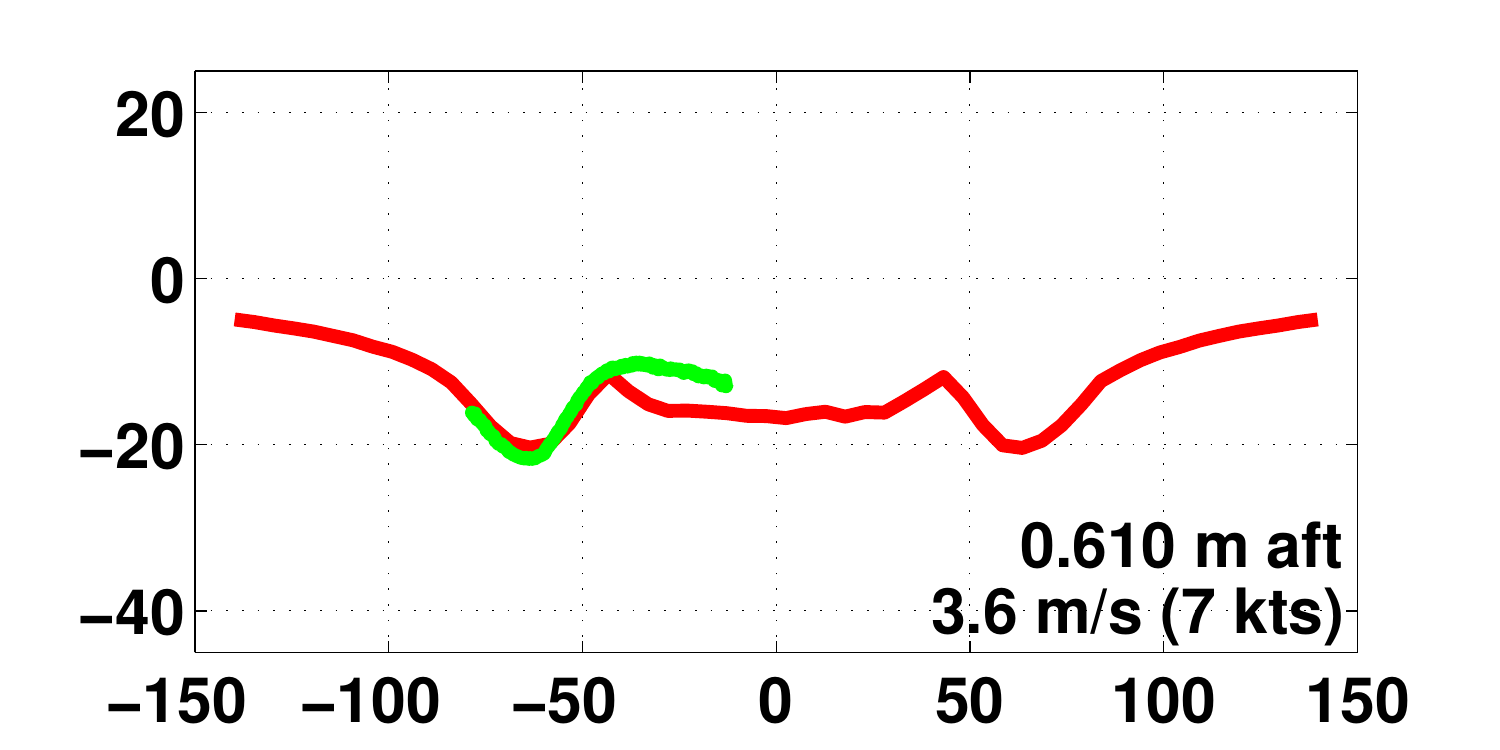}
& &  \includegraphics[width=0.25\linewidth]{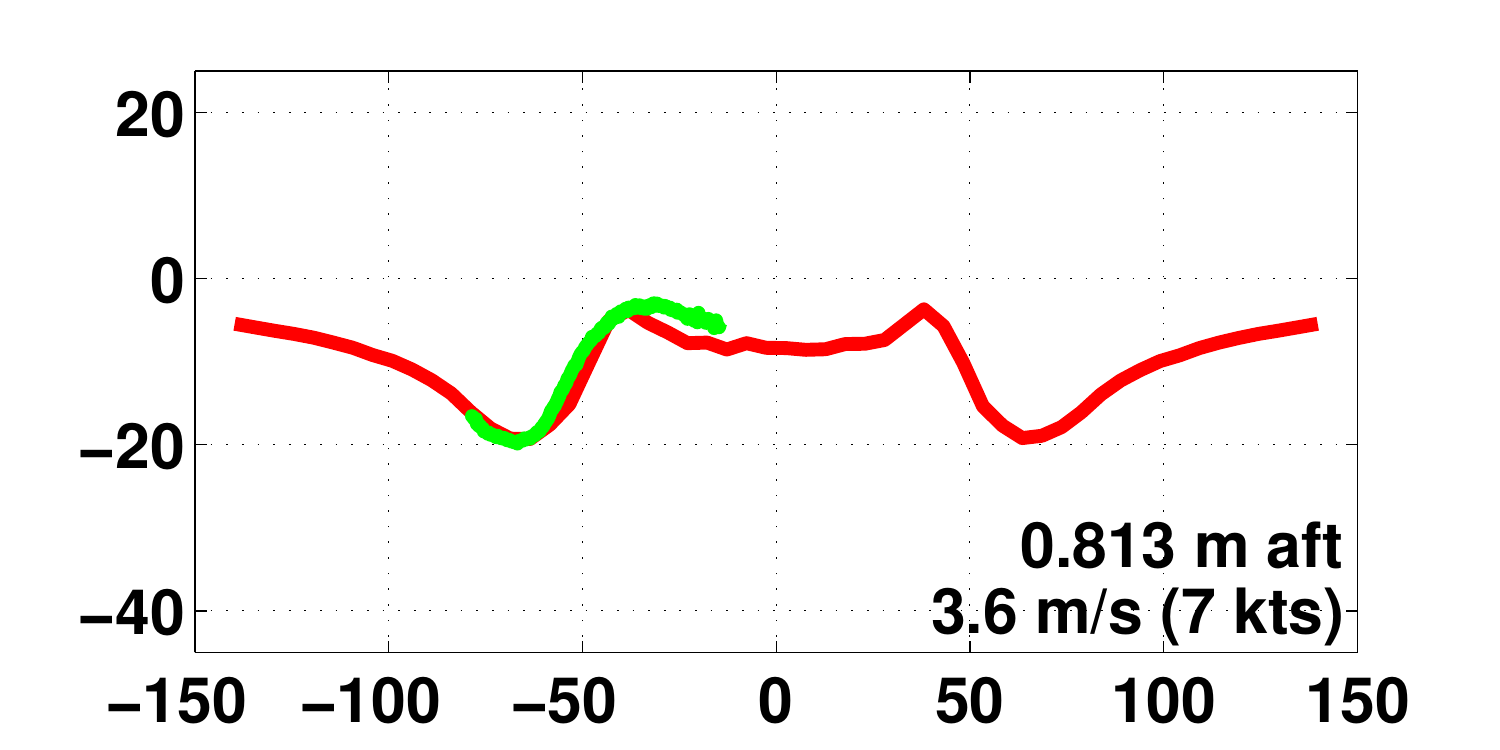} \\
(d) & & (e) & & (f) & \vspace{-18pt} \\ 
& \includegraphics[width=0.25\linewidth]{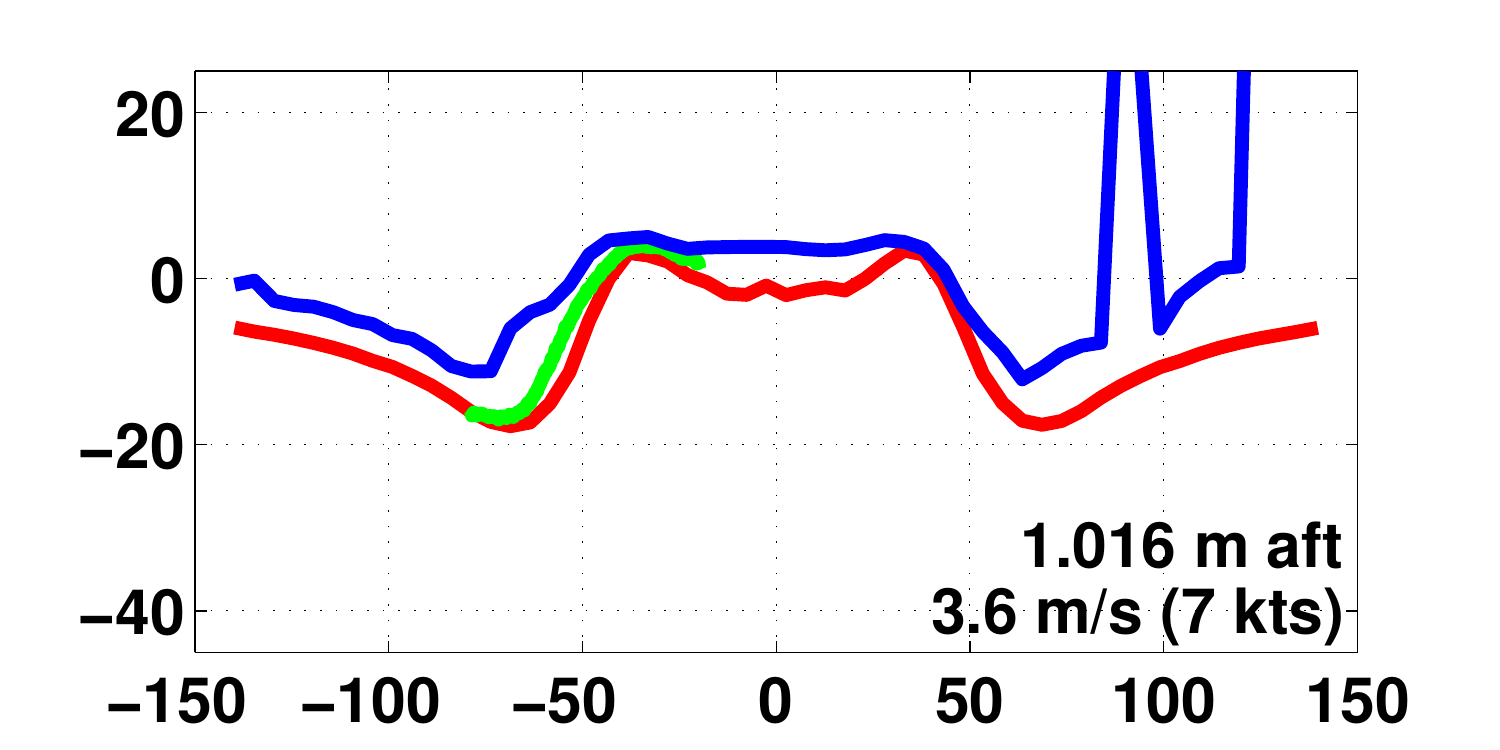}
& & \includegraphics[width=0.25\linewidth]{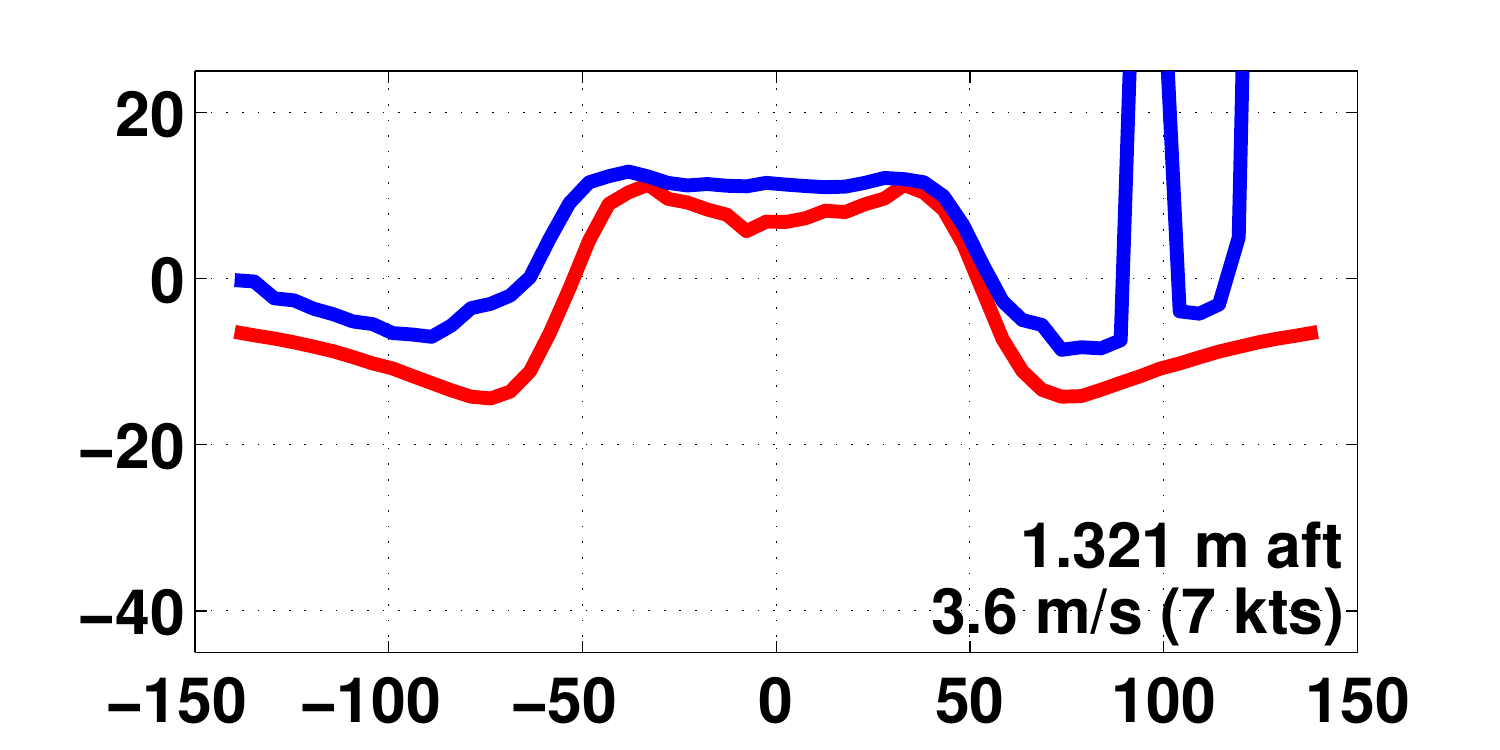}
& & \includegraphics[width=0.25\linewidth]{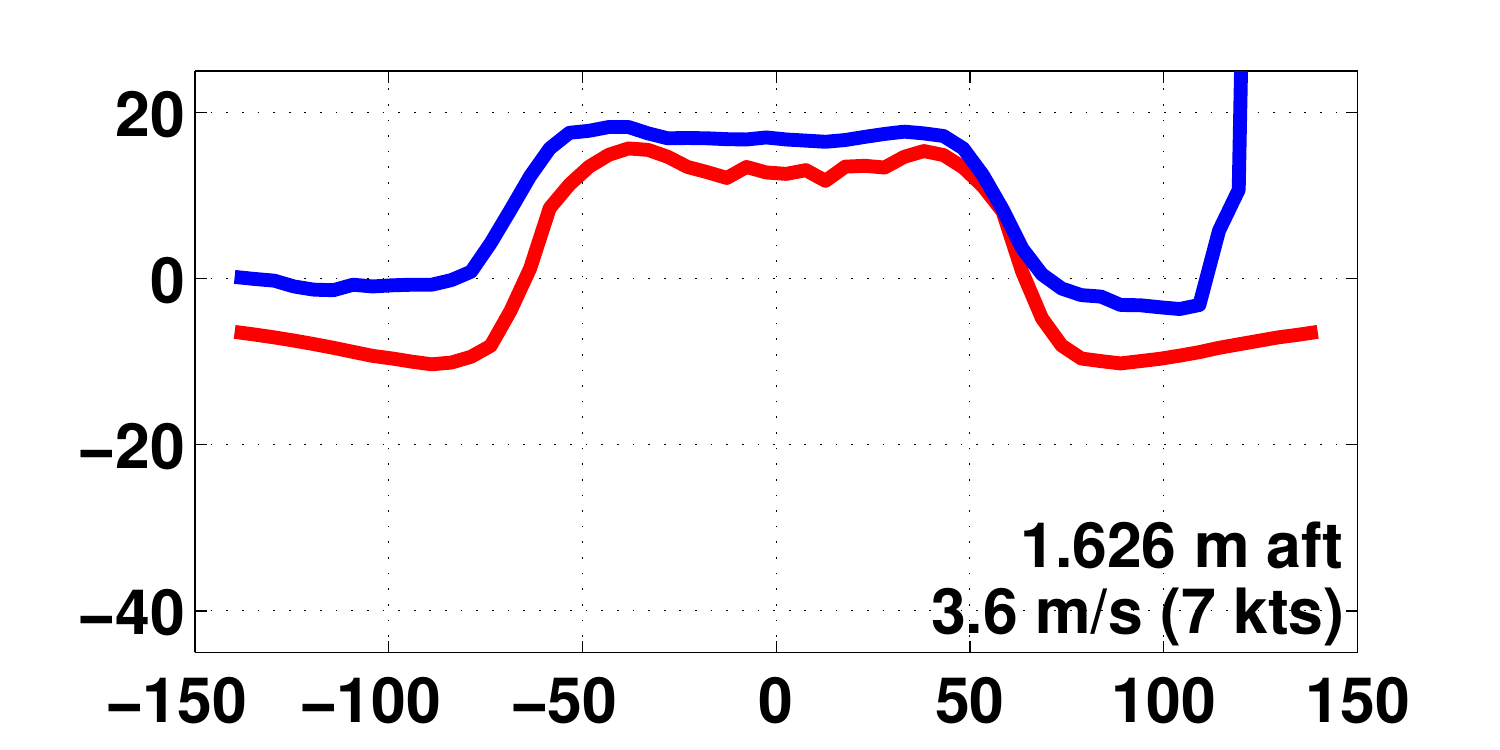} \\
(g) & & (h) & & (i) & \vspace{-18pt} \\ 
& \includegraphics[width=0.25\linewidth]{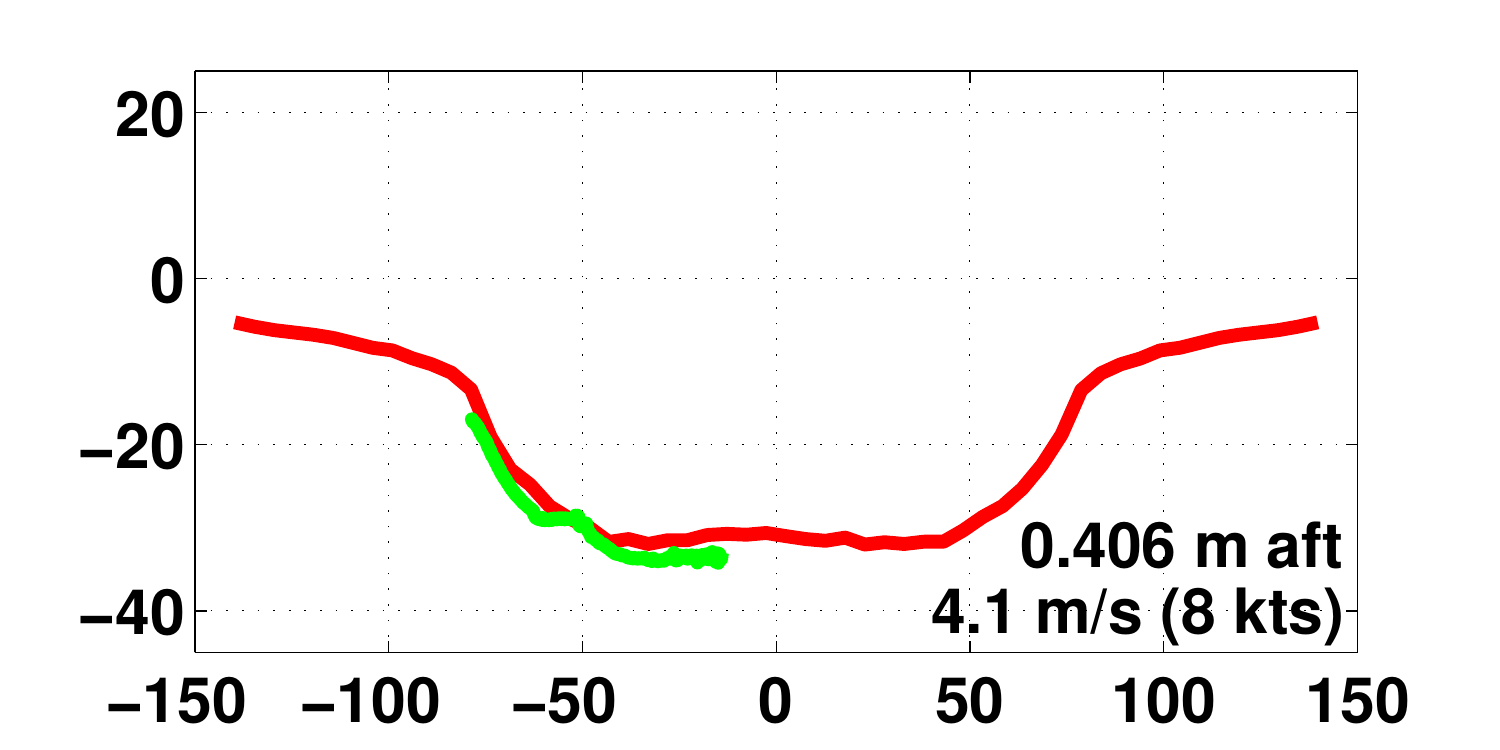}
& & \includegraphics[width=0.25\linewidth]{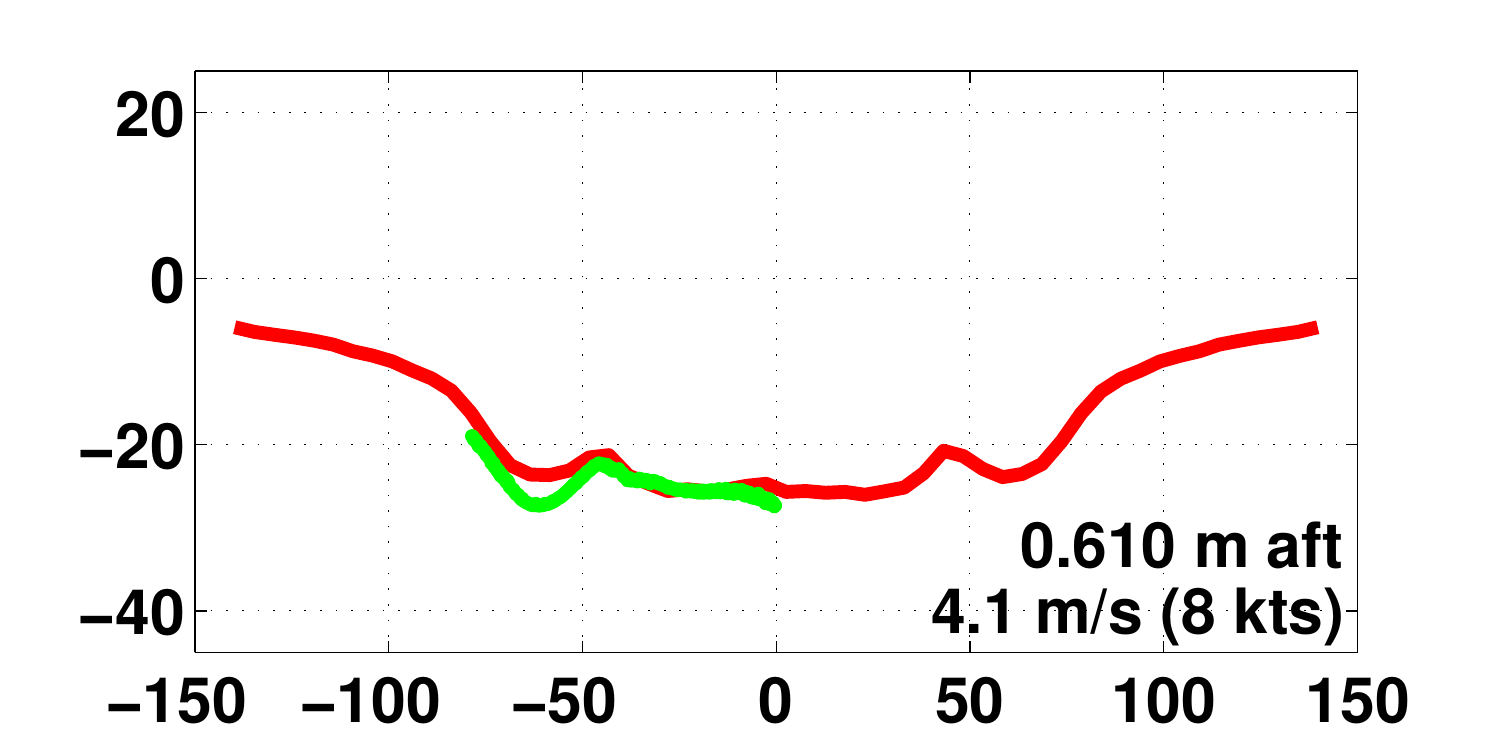}
& & \includegraphics[width=0.25\linewidth]{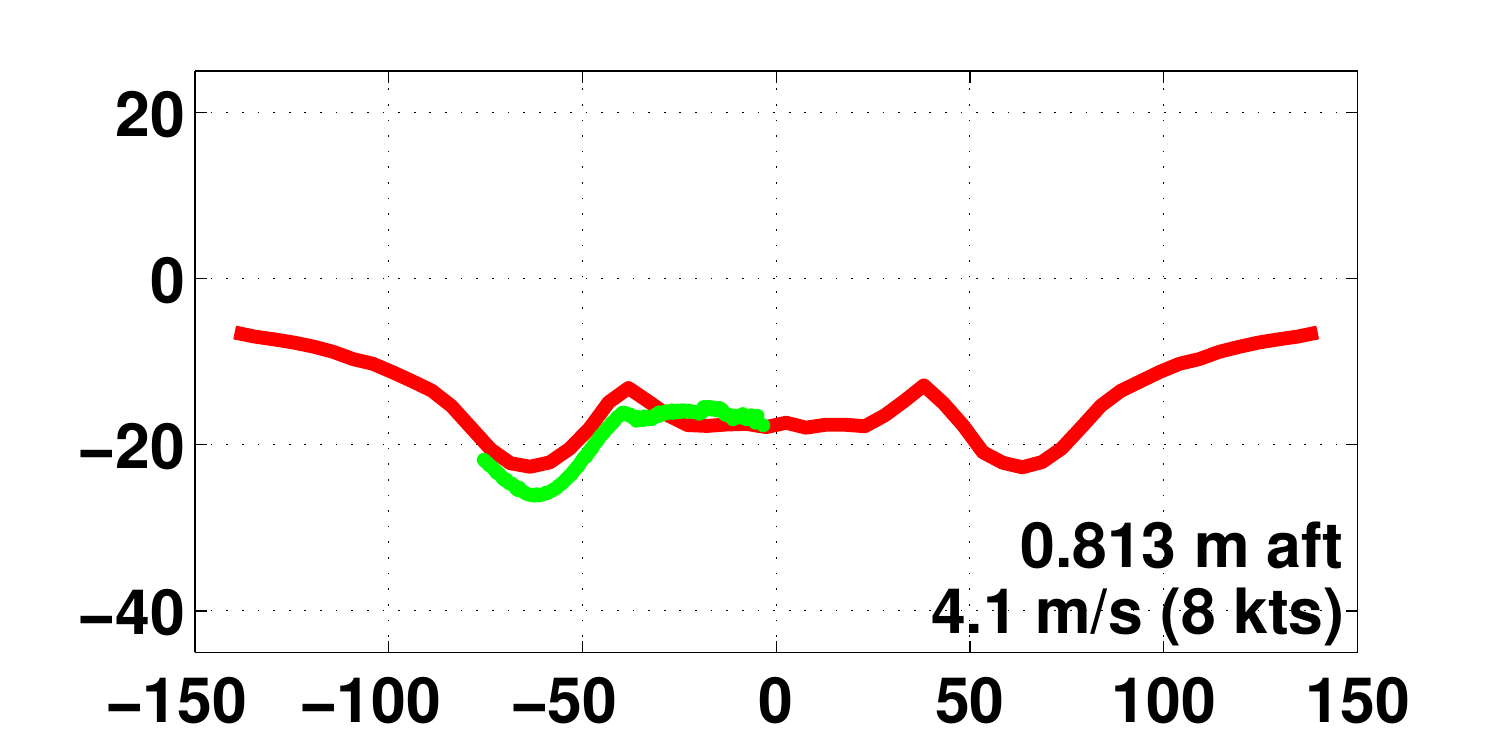} \\ 
(j) & & (k) & & (l) & \vspace{-18pt} \\ 
& \includegraphics[width=0.25\linewidth]{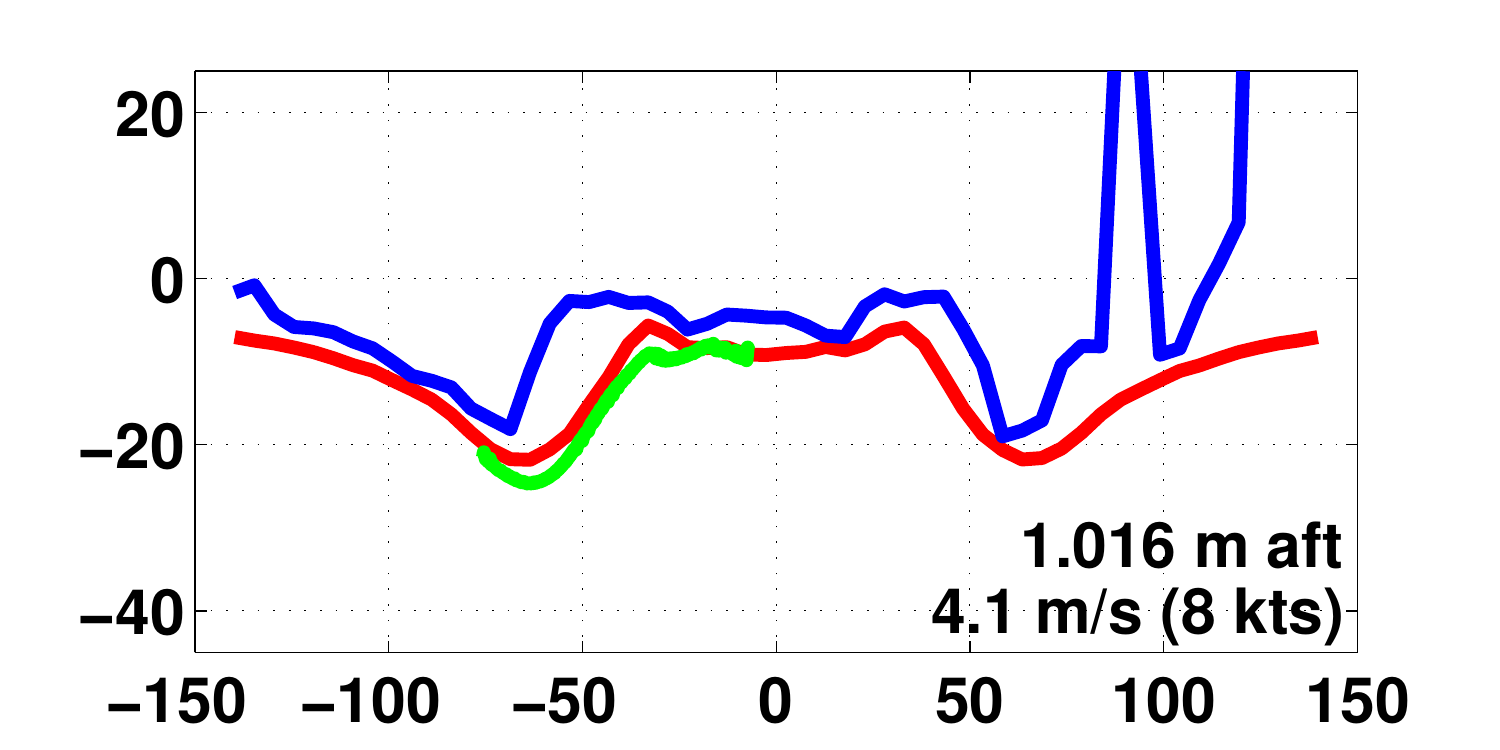}
& & \includegraphics[width=0.25\linewidth]{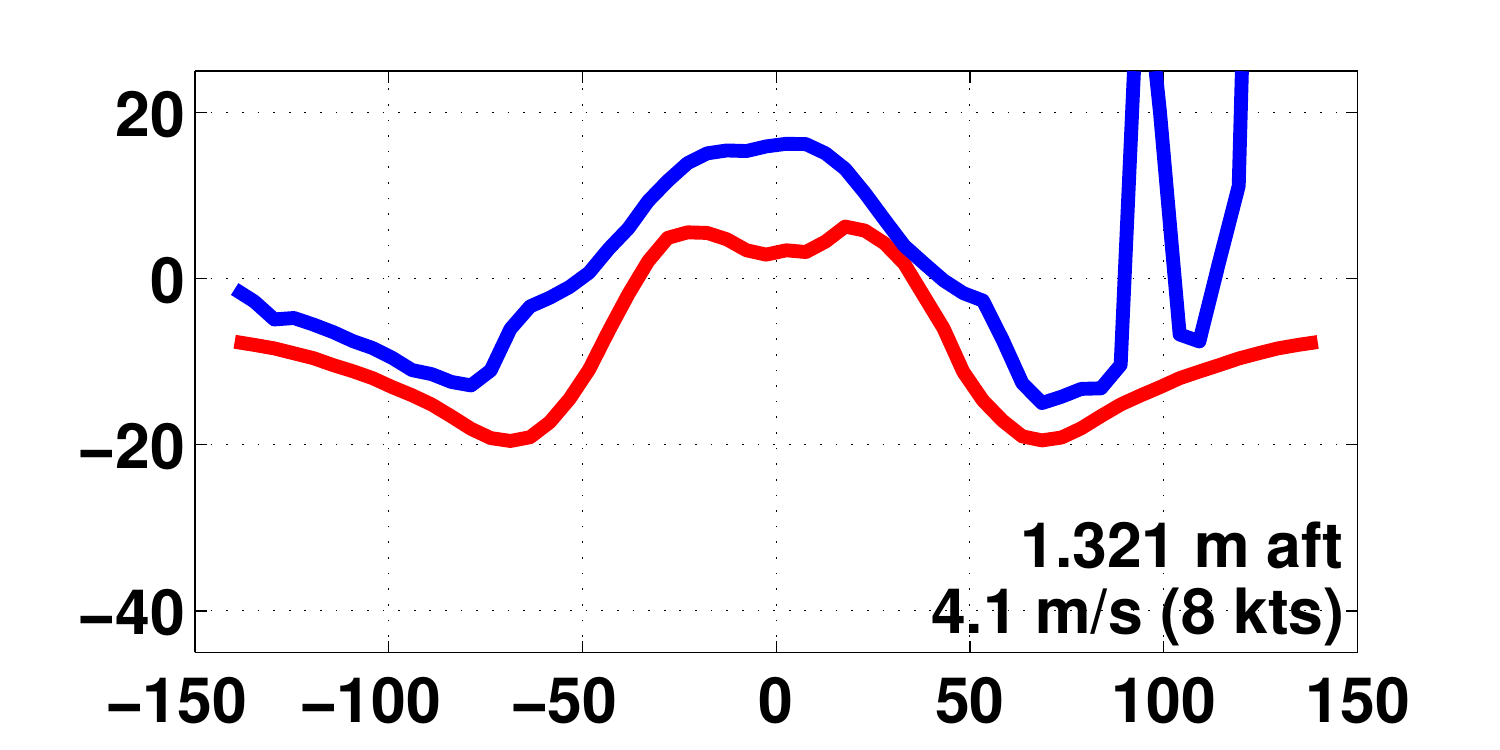}
& & \includegraphics[width=0.25\linewidth]{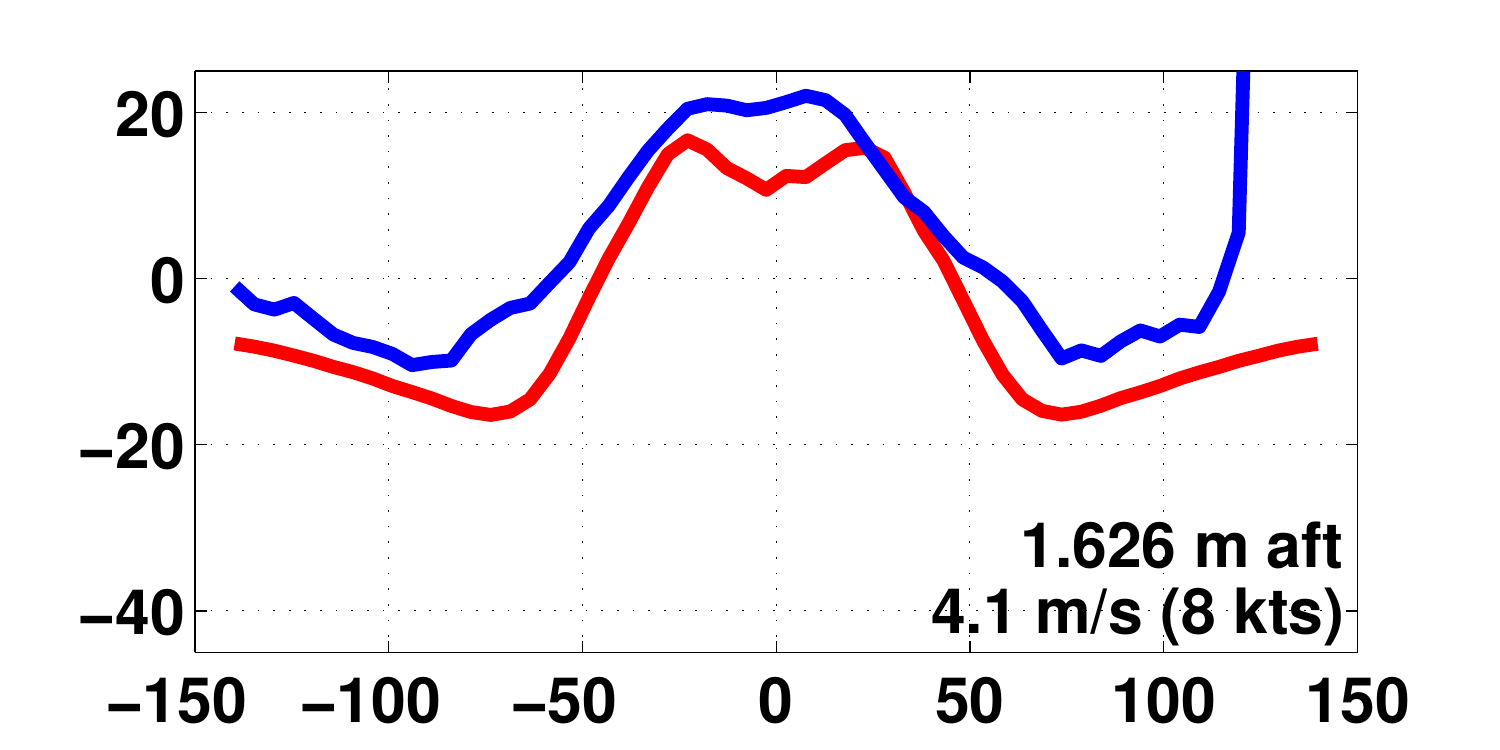}
\end{tabular}
\end{center} \caption{\label{nfa_transverse_cuts} Free-surface
transverse cuts.  $x$-axis denotes distance from centerline in cm.  The $y$-axis
represents the elevation from the calm free surface in cm.  Red, green, and
blue lines denote NFA, QViz, and LiDAR, respectively .  The large signal at $\approx$90 cm is due to
the QViz system being within the LiDAR's field of view. See the experimental
measurement section for further details.  All NFA results are top-down processing.} 
\end{figure*}

Figure \ref{nfa_transverse_cuts} compares transverse cuts of the free-surface
elevation.   NFA, QViz, and LiDAR data are shown for various distances aft of
the transom.   In general, the agreement between NFA and QViz is excellent,
whereas comparisons between NFA and LiDAR are poor.    In the one region where
all three transverse cuts overlap, NFA and QViz agree very well, but 2008 LiDAR
measurements appear to be offset, see the Experimental Measurements section for a
description of the offset. Aside from affecting the position of the data, the
effect of improper registration upon the processing of the measurements is
unknown. The LiDAR data also have some artifacts on the port side
because the QViz traverse and instrumentation often prevented the LiDAR from
measuring the free surface.

\begin{figure*}
\begin{center}
\begin{tabular}{llll} (a) & & (b) \vspace{-12pt} \\ 
& \includegraphics[trim = 10mm 60mm 10mm 60mm, clip,width=0.37\linewidth]{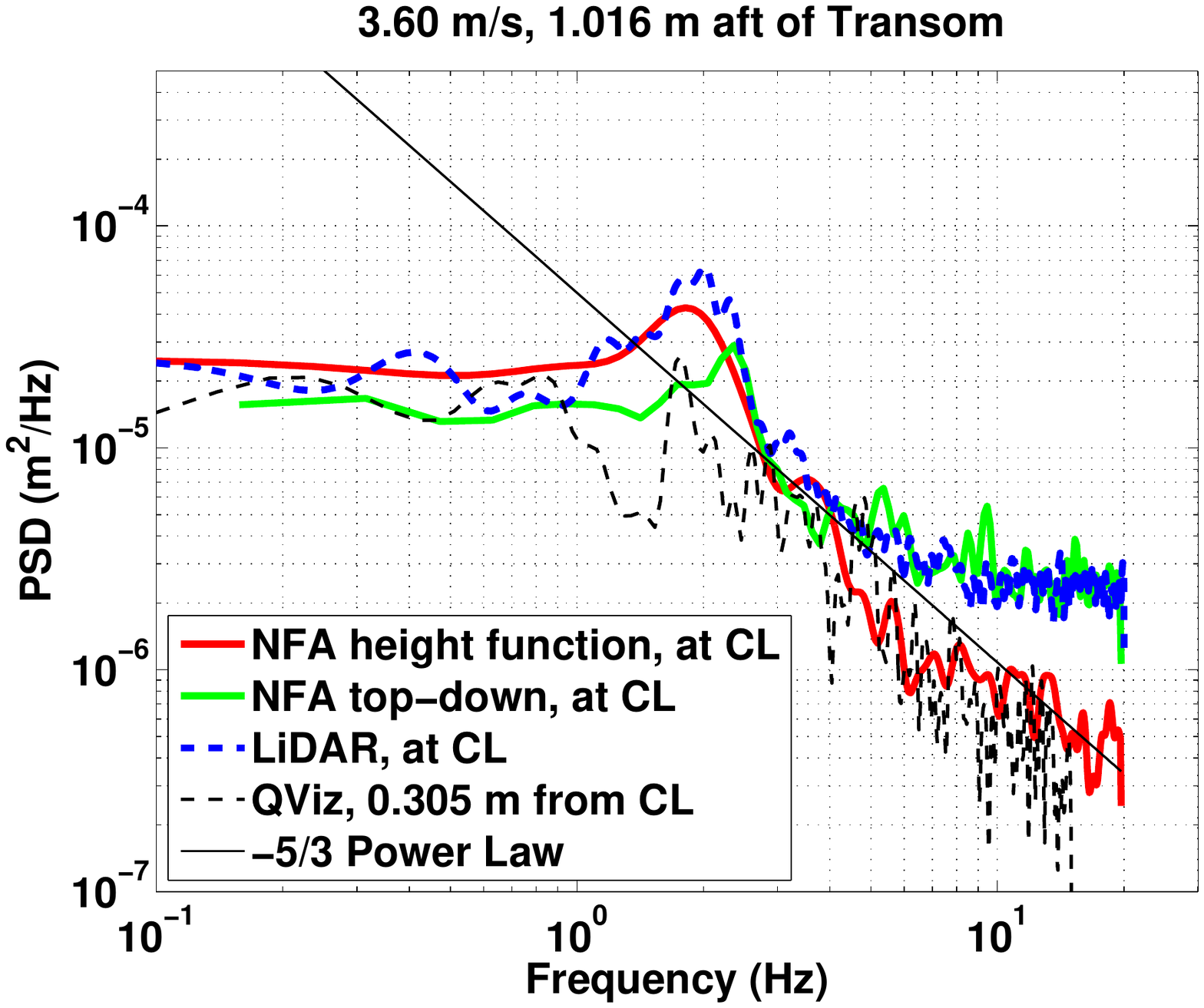}
& & \includegraphics[trim = 10mm 60mm 10mm 60mm, clip,width=0.37\linewidth]{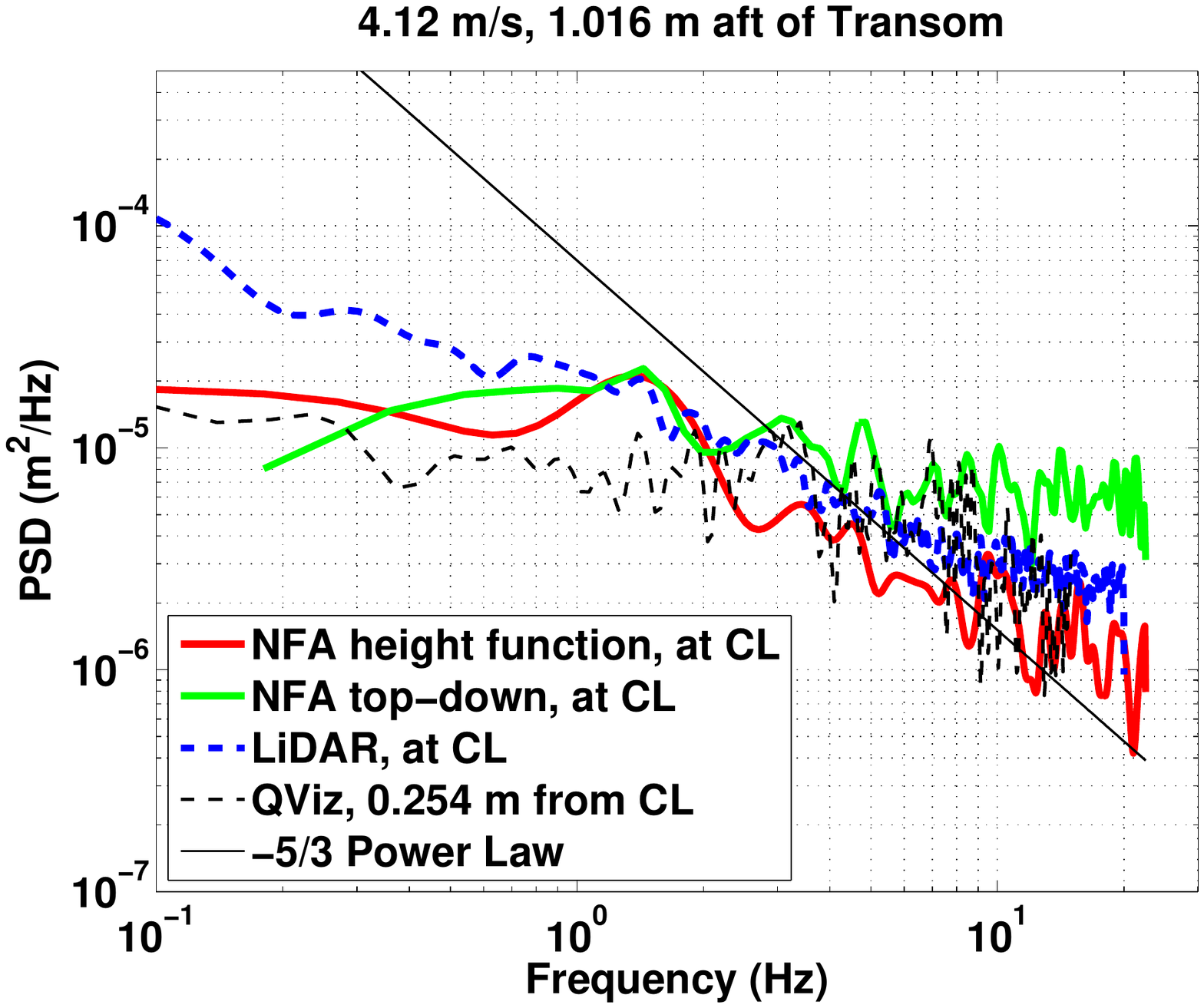} \\ 
& & & \vspace{-12pt} \\ 
& \includegraphics[trim = 10mm 60mm 10mm 60mm, clip,width=0.37\linewidth]{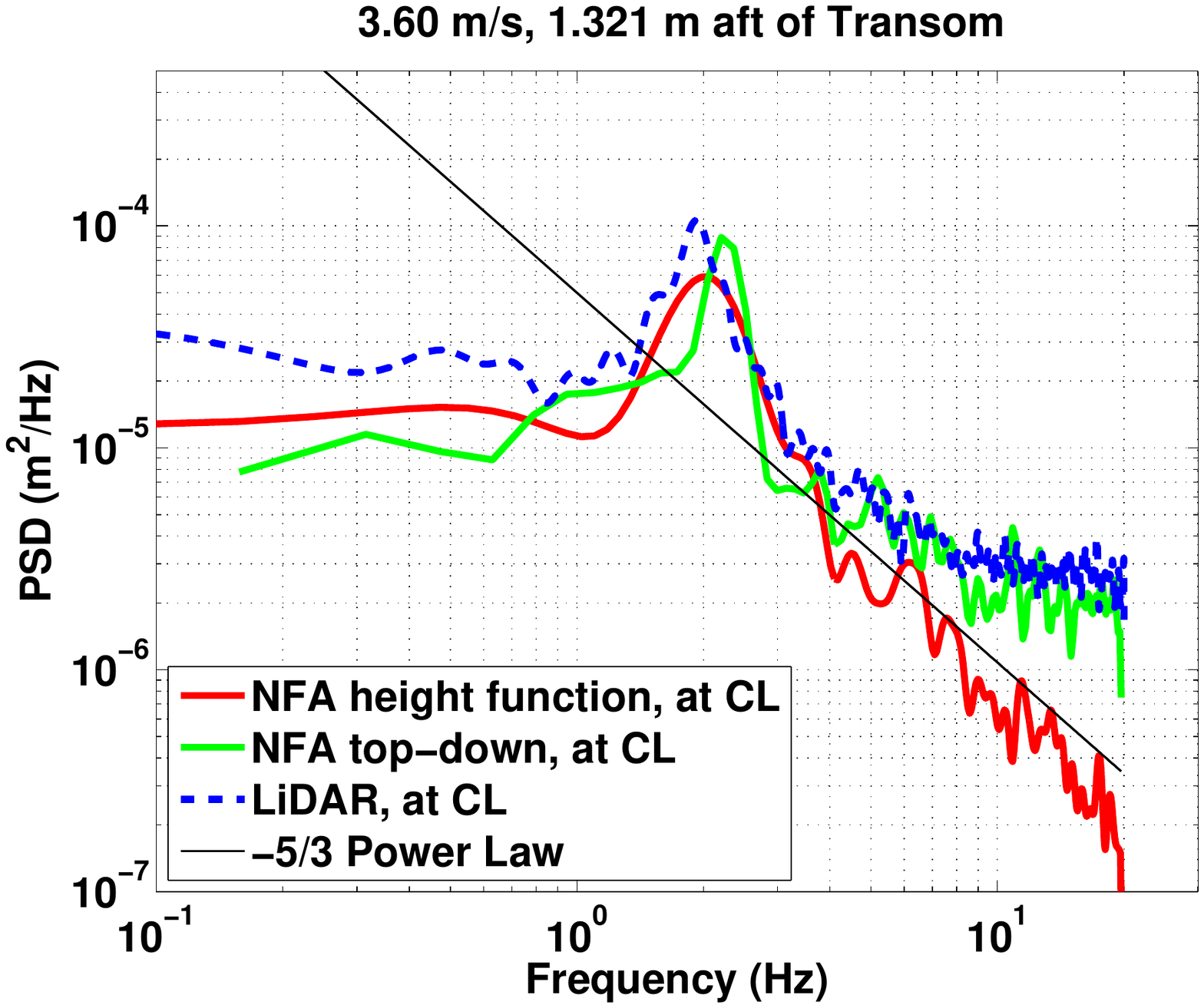}
& & \includegraphics[trim = 10mm 60mm 10mm 60mm, clip,width=0.37\linewidth]{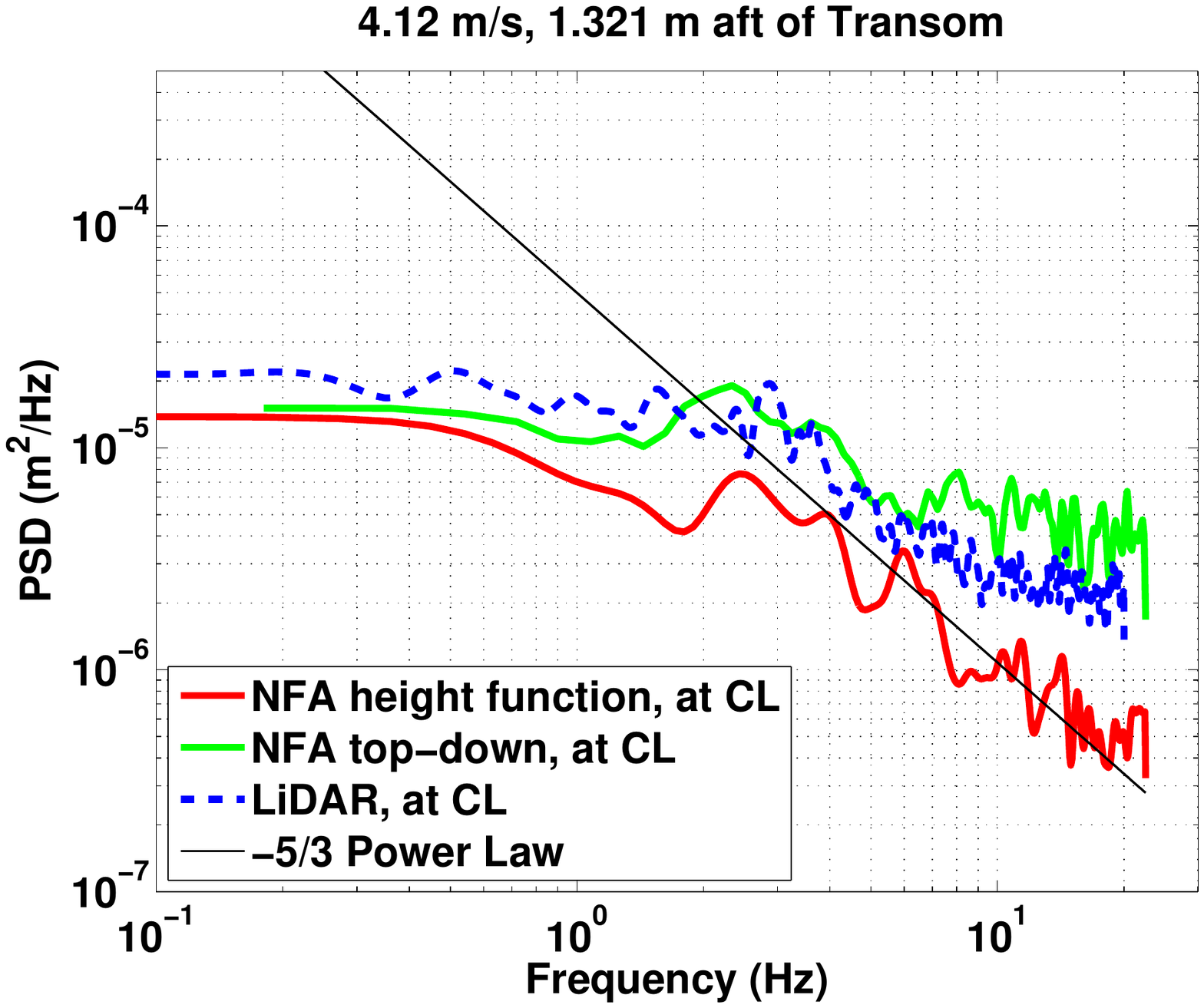} \\ 
& & & \vspace{-12pt} \\ 
& \includegraphics[trim = 10mm 60mm 10mm 60mm, clip,width=0.37\linewidth]{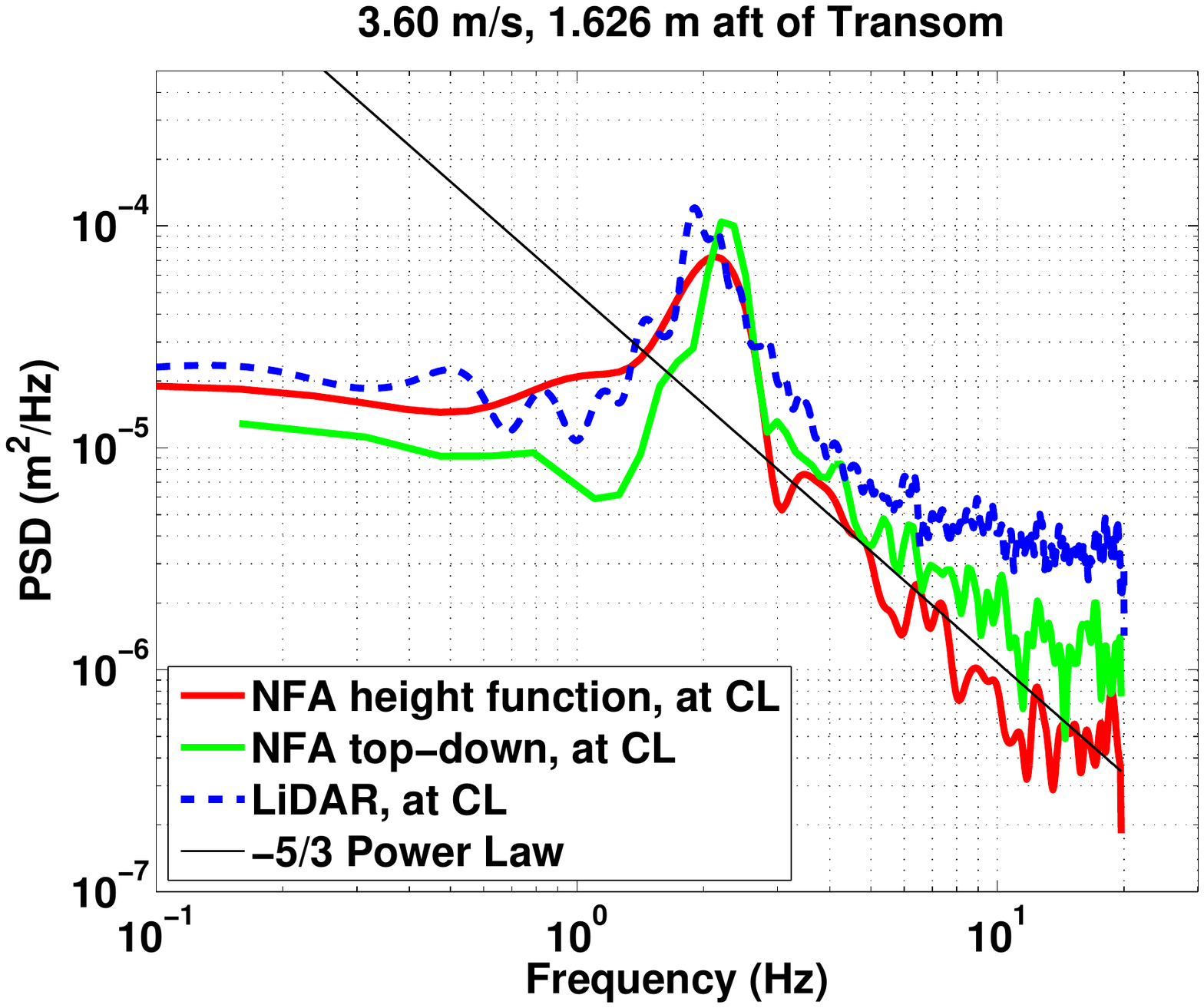}
& & \includegraphics[trim = 10mm 60mm 10mm 60mm, clip,width=0.37\linewidth]{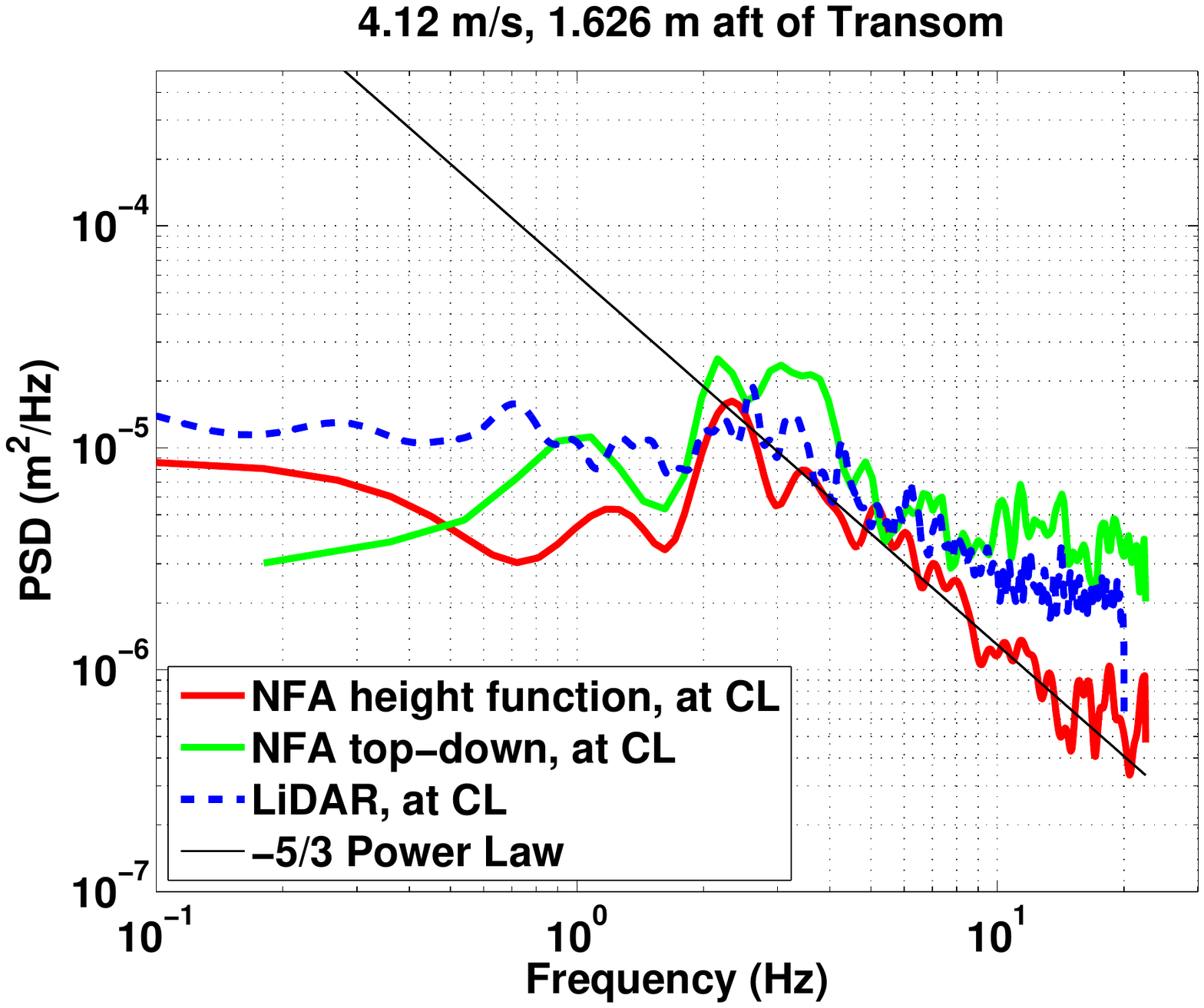} \\
& & & \vspace{-12pt} \\ 
& \includegraphics[trim = 10mm 60mm 10mm 60mm, clip,width=0.37\linewidth]{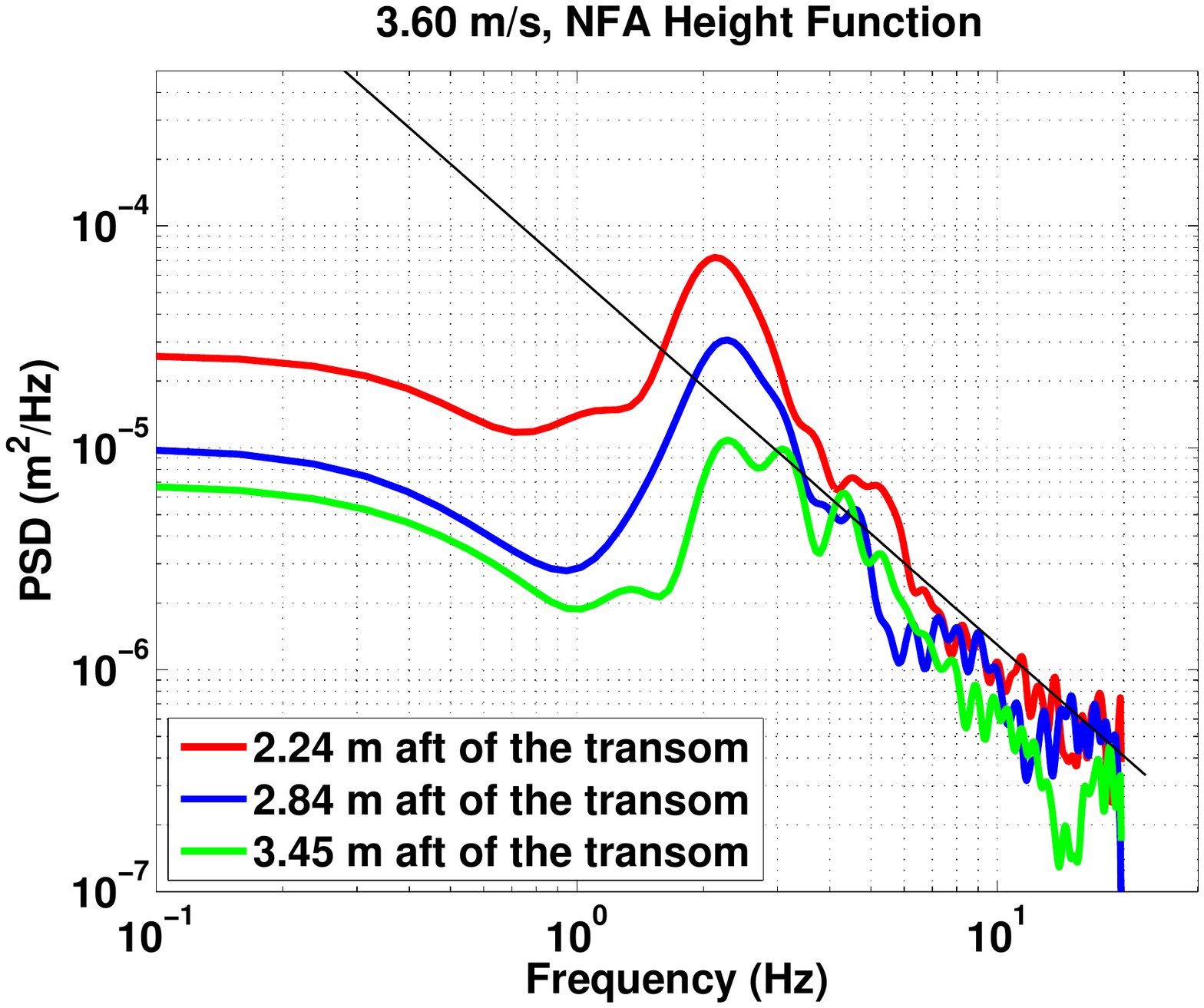} 
& & \includegraphics[trim = 10mm 60mm 10mm 60mm, clip,width=0.37\linewidth]{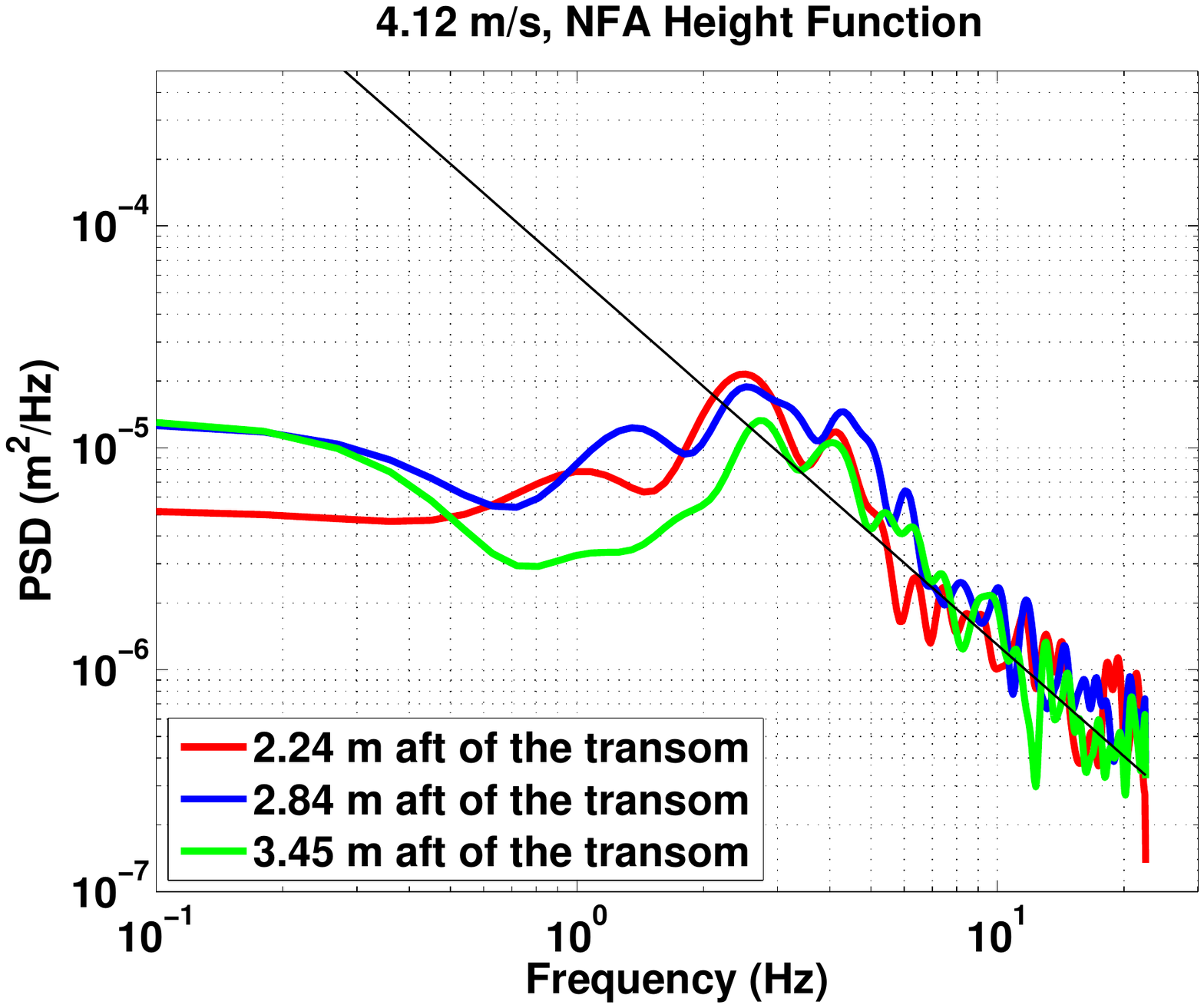}
\end{tabular}
\end{center} 
\caption{\label{nfa_spectra} Free-surface spectra.   NFA versus measurements.
(a) 3.60 m/s (7 knots).  (b) 4.12 m/s (8 knots).} 
\end{figure*}

Figure \ref{nfa_spectra} compares free-surface spectra in the transom region
between measurements using QViz and LiDAR and predictions using NFA.
Calculating free-surface spectra in the transom region is difficult because the
position of the free surface is hard to define due to the presence of droplets,
bubbles, and foam.   As a complement to top-down processing, we have also
developed a new processing technique that is meant to reduce noise by
minimizing the influence of spray and thin sheets. We define the free surface
in terms of height functions expressed in terms of the volume fraction.   The
calculation is performed in three steps.   First, we define a height function
integrating the volume fraction from the bottom of the computational domain to
the top.  This gives the total amount of water in the column $H_1$:
\begin{equation}
H_1(x,y,t)=\int_{-D}^\mathcal{H} dz \alpha(x,y,z,t) \; ,
\end{equation}
where $D$ is the water depth, $\mathcal H$ is the height of the domain, and $\alpha$ is the volume fraction. Then the air pockets that are trapped beneath $H_1$ are added back to provide a water column without bubbles.
\begin{equation}
H_2(x,y,t)=H_1(x,y,t)+\int_{-D}^{H_1(x,y,t)} dz (1-\alpha(x,y,z,t)) \; .
\end{equation}
Finally, the droplets above $H_2$ are subtracted out to provide a water column without bubbles and without droplets.
\begin{equation}
H_3(x,y,t)=H_2(x,y,t)-\int_{H_2(x,y,t)}^\mathcal{H} dz \alpha(x,y,z,t) \; .
\end{equation}

The NFA spectral analyses that are shown in Figure~\ref{nfa_spectra} are
performed using height functions ($H_3(x,y,t)$) and top-down processing. A
cosine taper is applied to 5\% of each end of each time series.   An FFT of
each time series is then applied to obtain the power spectral density plot. 

\begin{figure*} [h!] 
\begin{center}
\begin{tabular}{llll} 
(a) & & (b) & \vspace{-15pt} \\ 
& \includegraphics[trim = 12mm 12mm 12mm 15mm, clip,width=0.4\linewidth]{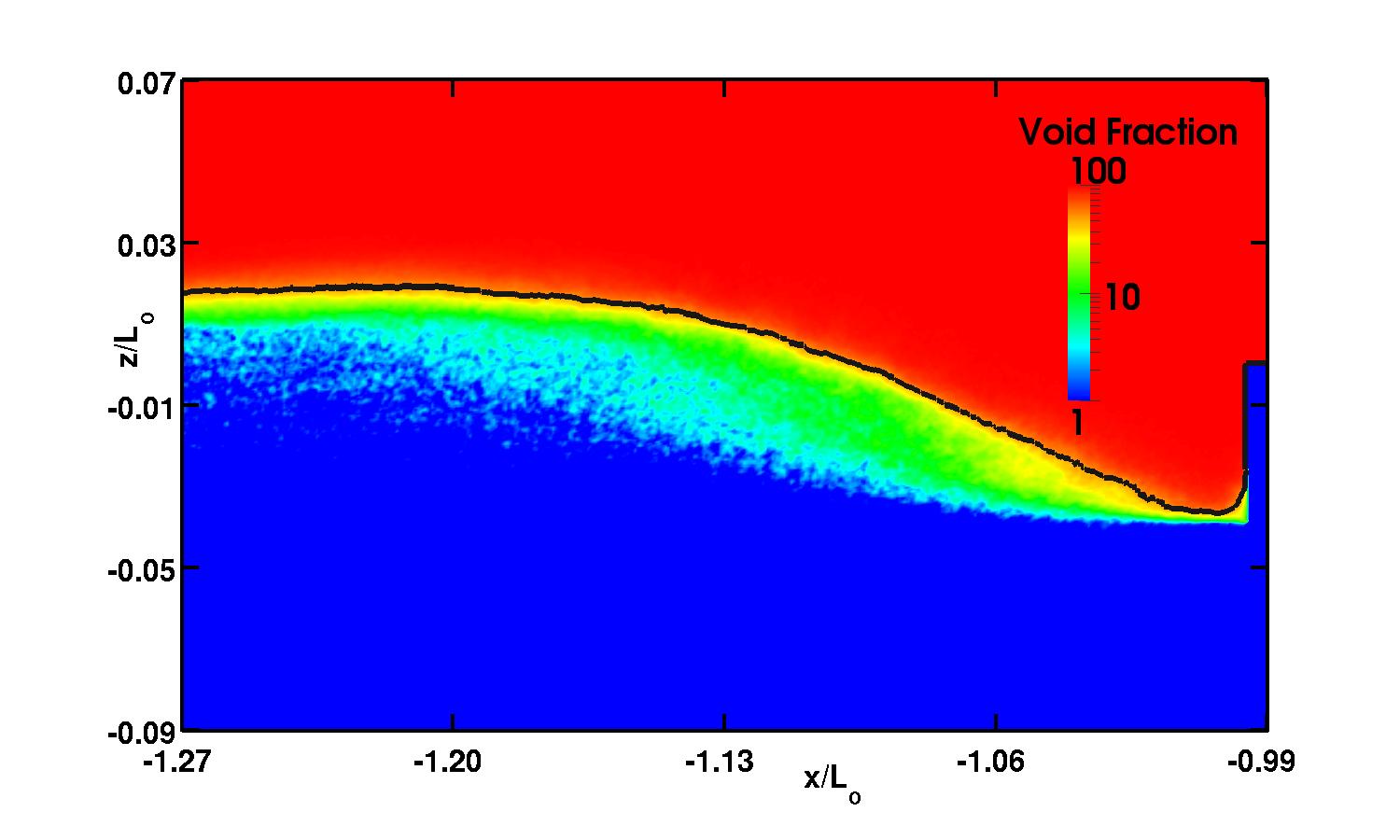} 
& & \includegraphics[trim = 12mm 12mm 12mm 15mm, clip,width=0.4\linewidth]{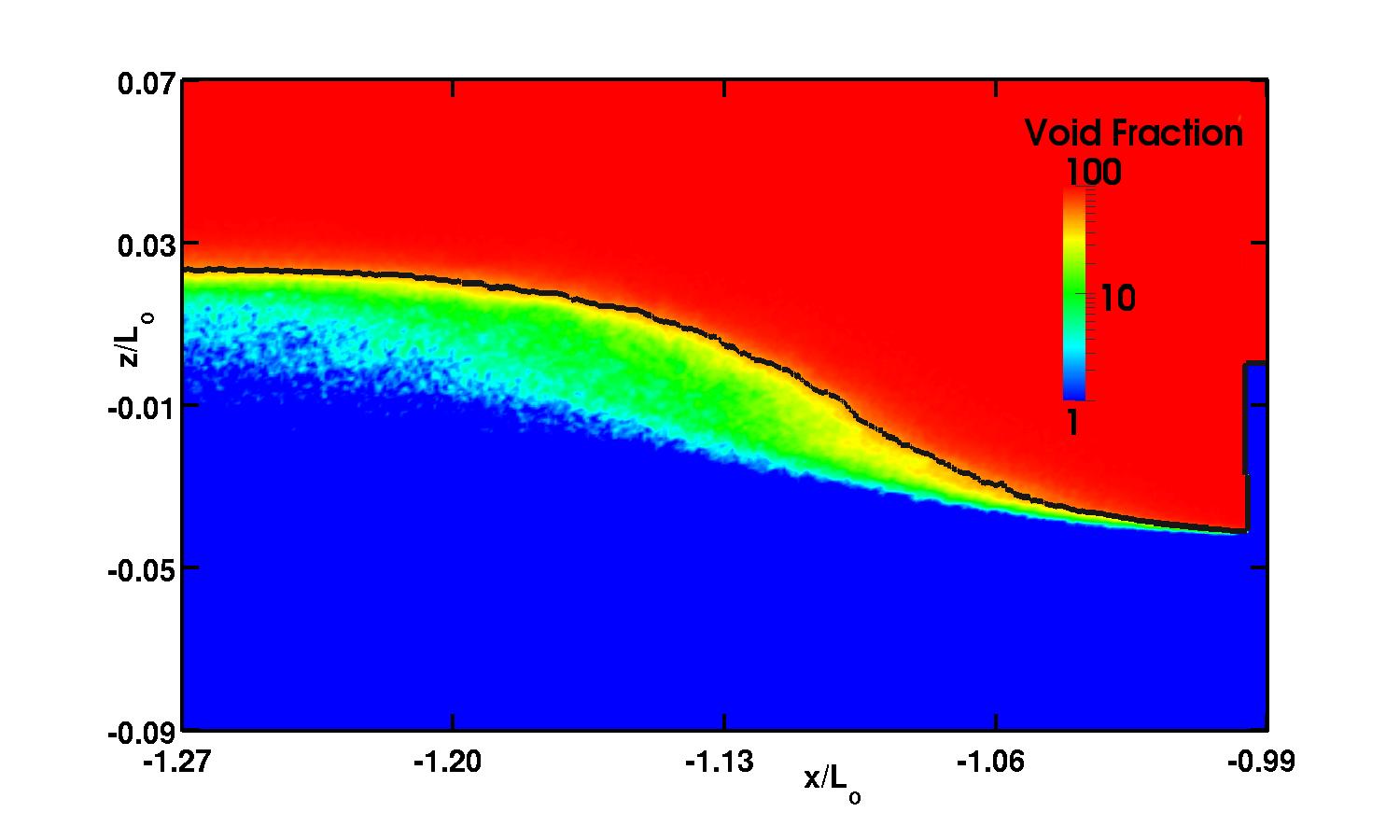} \\
(c) & & (d) & \vspace{-15pt} \\ 
& \includegraphics[trim = 12mm 12mm 12mm 15mm, clip,width=0.4\linewidth]{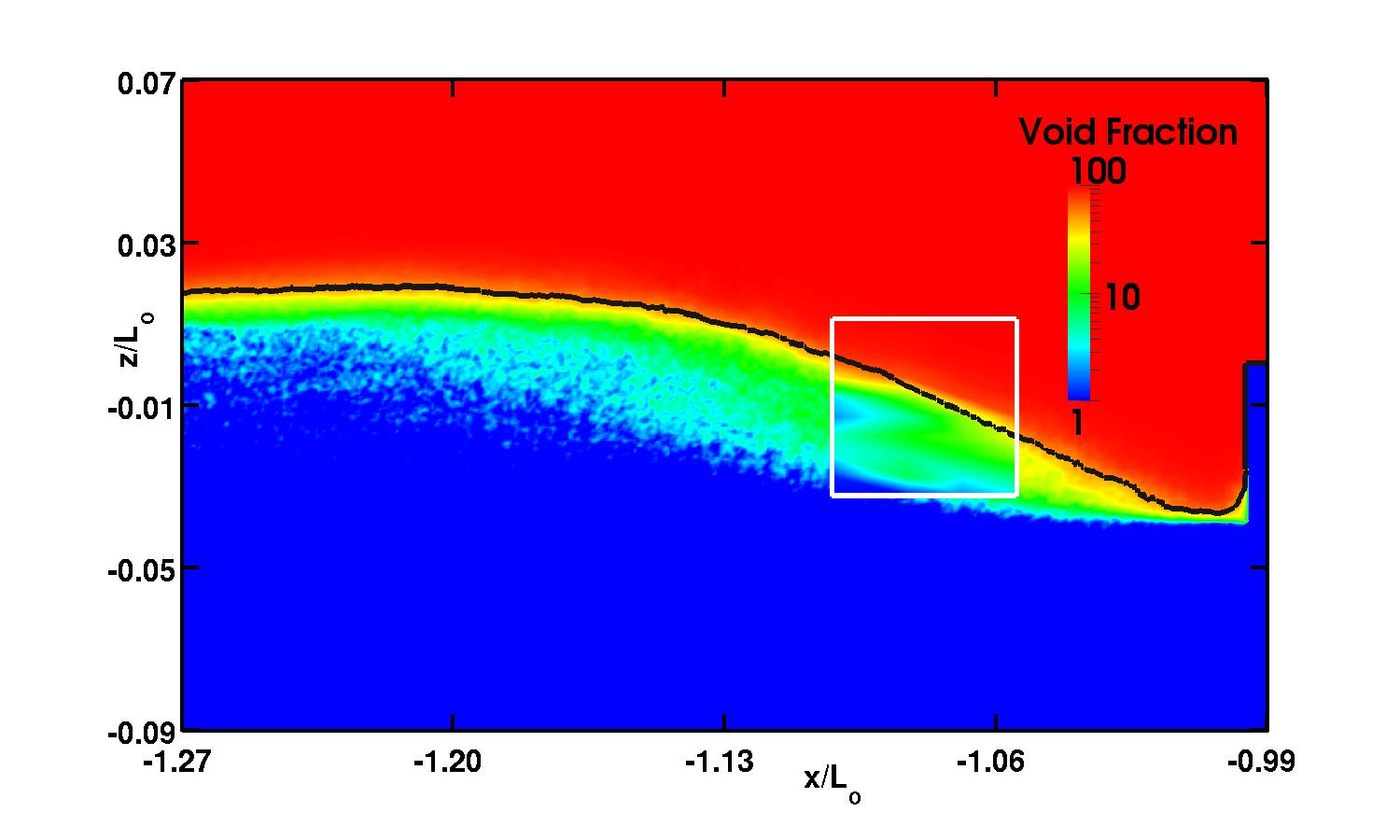} 
& & \includegraphics[trim = 12mm 12mm 12mm 15mm, clip,width=0.4\linewidth]{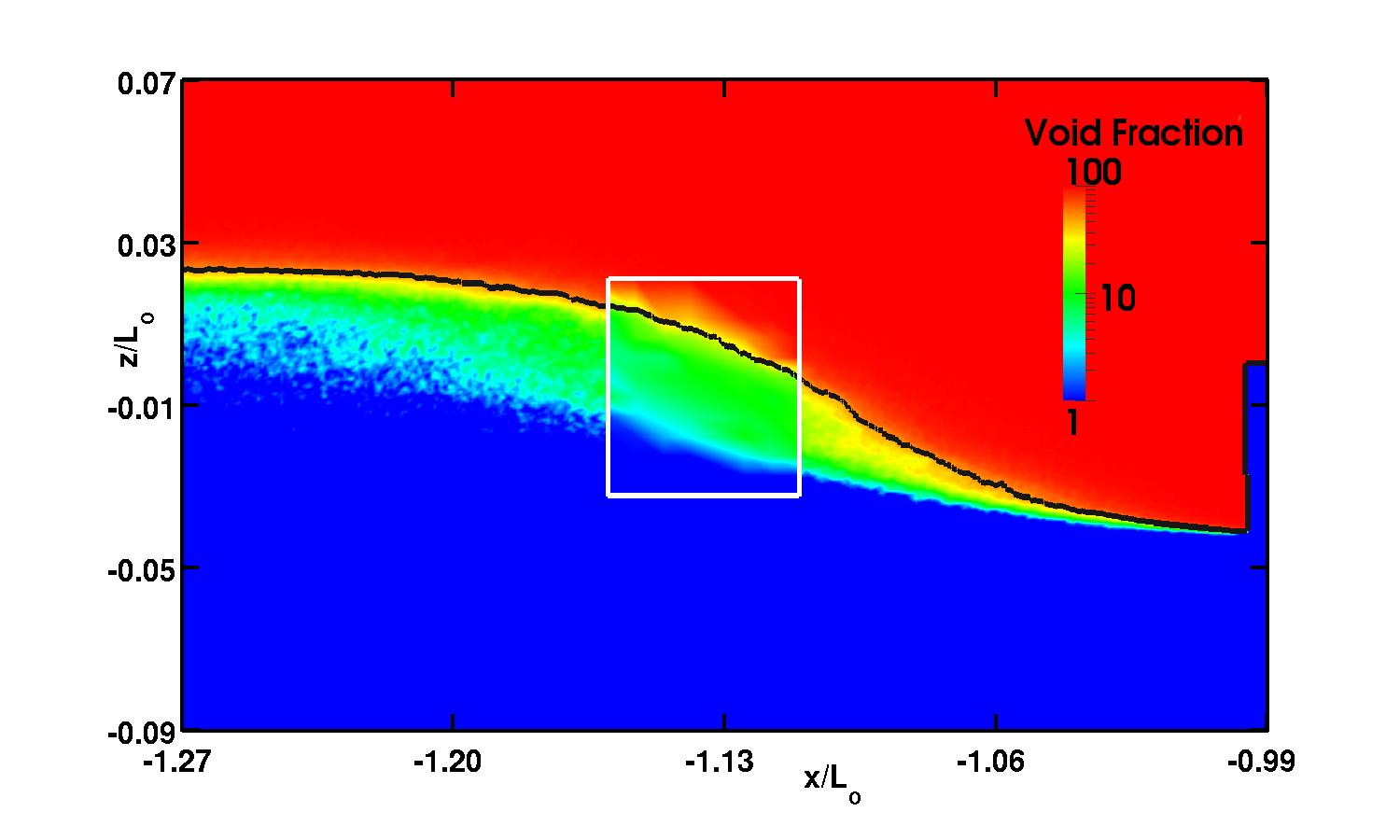}
\end{tabular} 
\end{center} 
\caption{\label{nfa_void} Void fraction predictions compared to experimental measurements on centerplane.  (a) NFA, 3.60 m/s (7 knots).  (b) NFA, 4.12 m/s (8 knots).  (c) NFA with void fraction measurements inserted, 3.60 m/s (7 knots).  (d) NFA with void fraction measurements inserted, 4.12 m/s (8 knots).  Transom is on right edge of plots.  Black lines denote 0.5 isosurface.   Measurements are denoted by white framing.  The numerical results have been time-averaged over the last 10,000 time steps of the numerical simulation.} 
\end{figure*}

In Figure~\ref{nfa_spectra}, top-down processing and LiDAR all have a higher
noise floor than height-function processing due to the effects of spray and
wave overturning.  The noise floor is much lower using height-function
processing.    Height-function processing shows a $-5/3$ power-law behavior for
3.60 m/s (7 knot) and 4.12 m/s (8 knot) results.  QViz also shows a $-5/3$
power-law behavior at the station closest to the the transom for both speeds.
There are 2 Hz features in the 3.60 m/s (7 knots) cases, and there are 2.5 Hz
features in the 4.12 m/s (8 knots) cases.   The 2.5 Hz features become more
prominent as the distance downstream increases.  The shear layer is thinner for
the 4.12 m/s (8 knot) case than the 3.60 m/s (7 knot) case, which may explain why the spectral
peak occurs at a higher frequency.   As the distance downstream increases, the
frequencies of the spectral peaks increase perhaps because the layers get
thinner with less entrained air and/or the effects of turbulent decay.   As a
possible source of the spectral peaks, a vortex shedding mechanism off the
trailing edge of the transom is not supported by the results in
Figure~\ref{nfa_spectra} because  both the 3.60 m/s (7 knot) case with a partially wet
transom and the 4.12 m/s (8 knot) case with a dry transom have spectral peaks.

Figure \ref{nfa_void} shows a centerplane cut behind the transom of NFA
predictions of the void fraction compared to measurements.   The top row of
plots shows NFA predictions.  The bottom row of plots shows NFA predictions
with void-fraction measurements inserted.   The plots are from 2.13m aft of the
transom to the transom, and from 0.6m above the mean waterline to -0.72m below.
Predictions are in good agreement with measurements.  An unsteady multiphase
shear layer forms in the rooster-tail region.   The air is primarily entrained
at toe of the spilling region and degasses over the top of the rooster tail.
For the 3.60 m/s (7 knot) case, the toe moves forward and wets the transom with foam.
For the 4.12 m/s (8 knot) case, the toe is slightly aft of the transom.  For
the 3.60 m/s (7 knot) and 4.12 m/s (8 knot) cases, the primary entrainment of air occurs aft of the transom.
The shear layer is thicker and longer for the 3.60 m/s (7 knot) case than for the 4.12
m/s (8 knot) case.   The difference in thicknesses and amount of air entrainment affects
the temporal and spatial structure of the shear layer, which is evident in the
spectra in Figure~\ref{nfa_spectra}.   The foam that wets the transom in the
3.60 m/s (7 knot) case does not appear to have enough momentum to affect the drag of the
ship.   At present, there is no evidence of large-scale vortical structures
being shed from the transom in the NFA simulations.  If coherent structures are
being shed from the backside of the transom, they must be dynamically weak due
to the entrained air.   The results of the NFA simulations suggest that the
spatial and temporal structure that is observed in the rooster-tail region for
3.60 m/s (7 knot) and for 4.12 m/s (8 knot) at shorter scales and higher frequencies is due to
the effects of unsteady multiphase shear layers. 

\begin{figure*}
\begin{center} 
\begin{tabular}{llll} 
(a1) & & (b1) & \vspace{-15pt} \\ 
& \includegraphics[trim = 12mm 12mm 12mm 15mm, clip,width=0.4\linewidth]{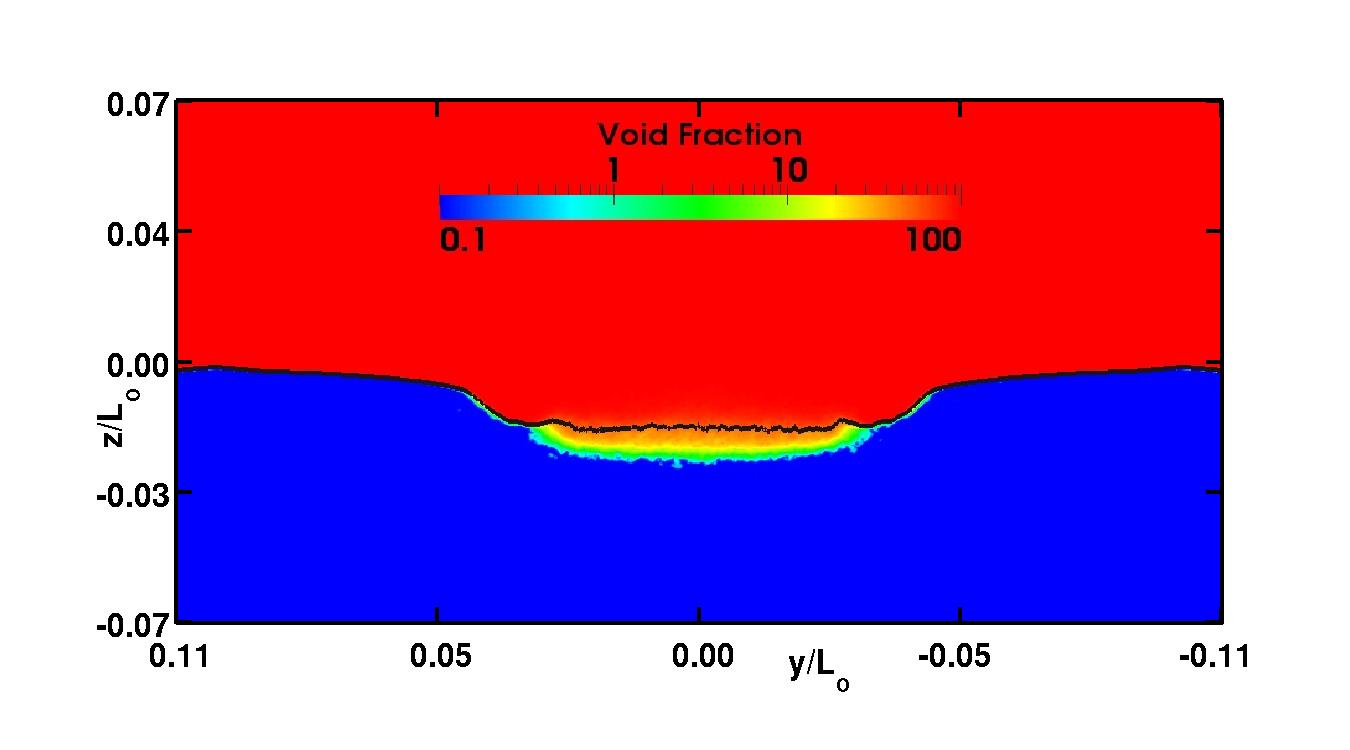}
& & \includegraphics[trim = 12mm 12mm 12mm 15mm, clip,width=0.4\linewidth]{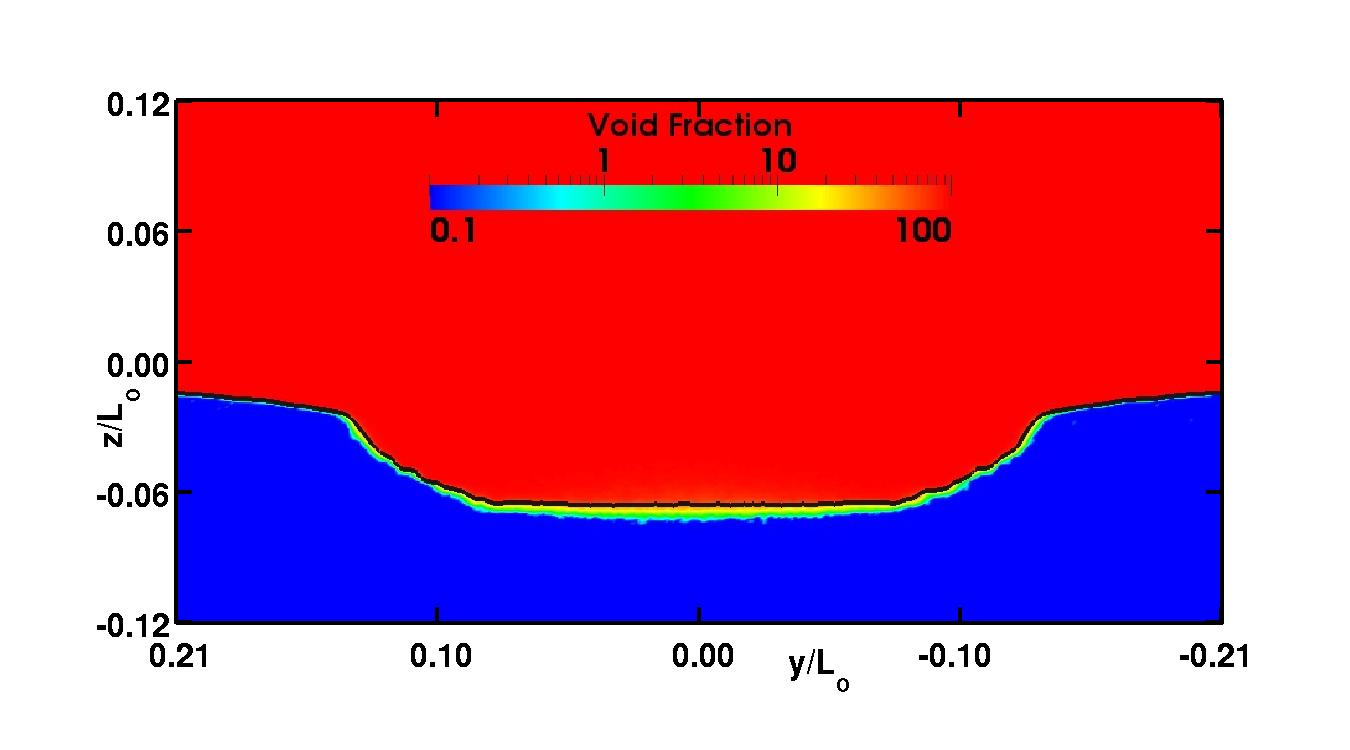} \\
(a2) & & (b2) & \vspace{-15pt} \\ 
& \includegraphics[trim = 12mm 12mm 12mm 15mm, clip,width=0.4\linewidth]{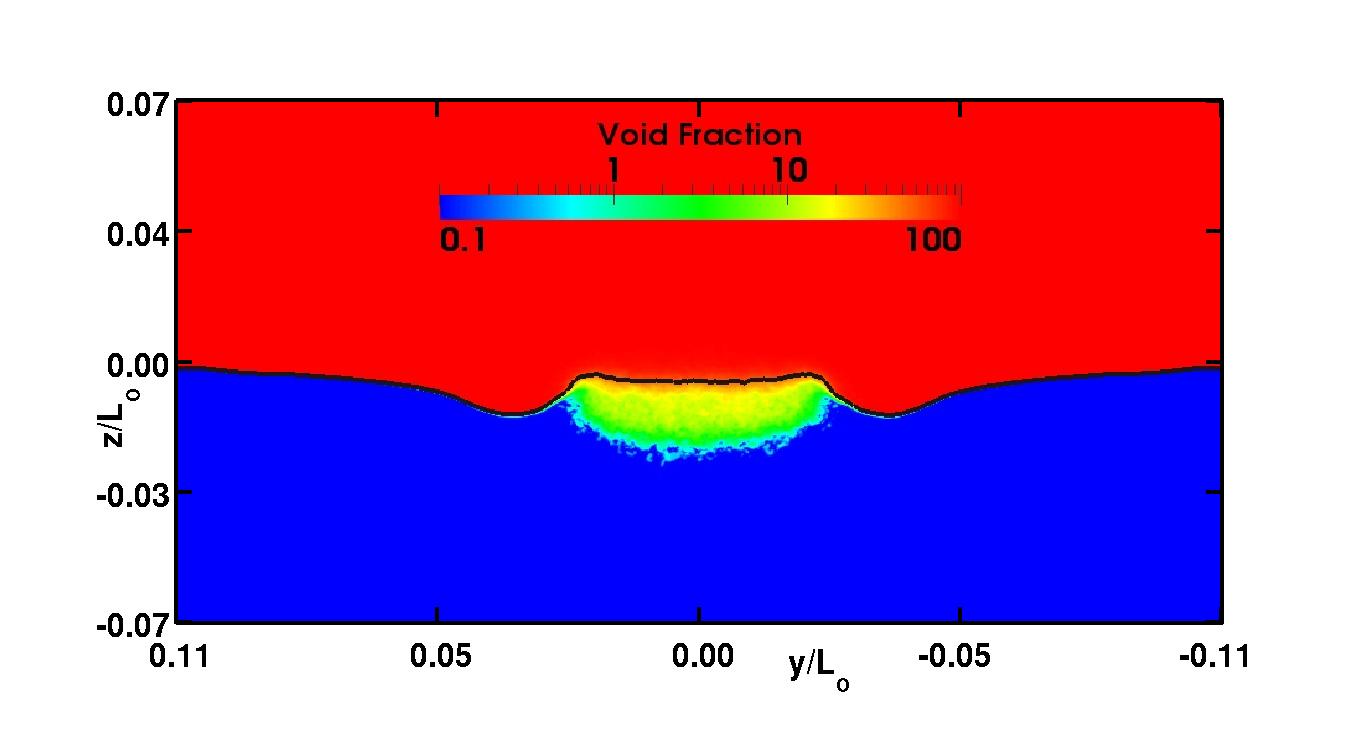}
& & \includegraphics[trim = 12mm 12mm 12mm 15mm, clip,width=0.4\linewidth]{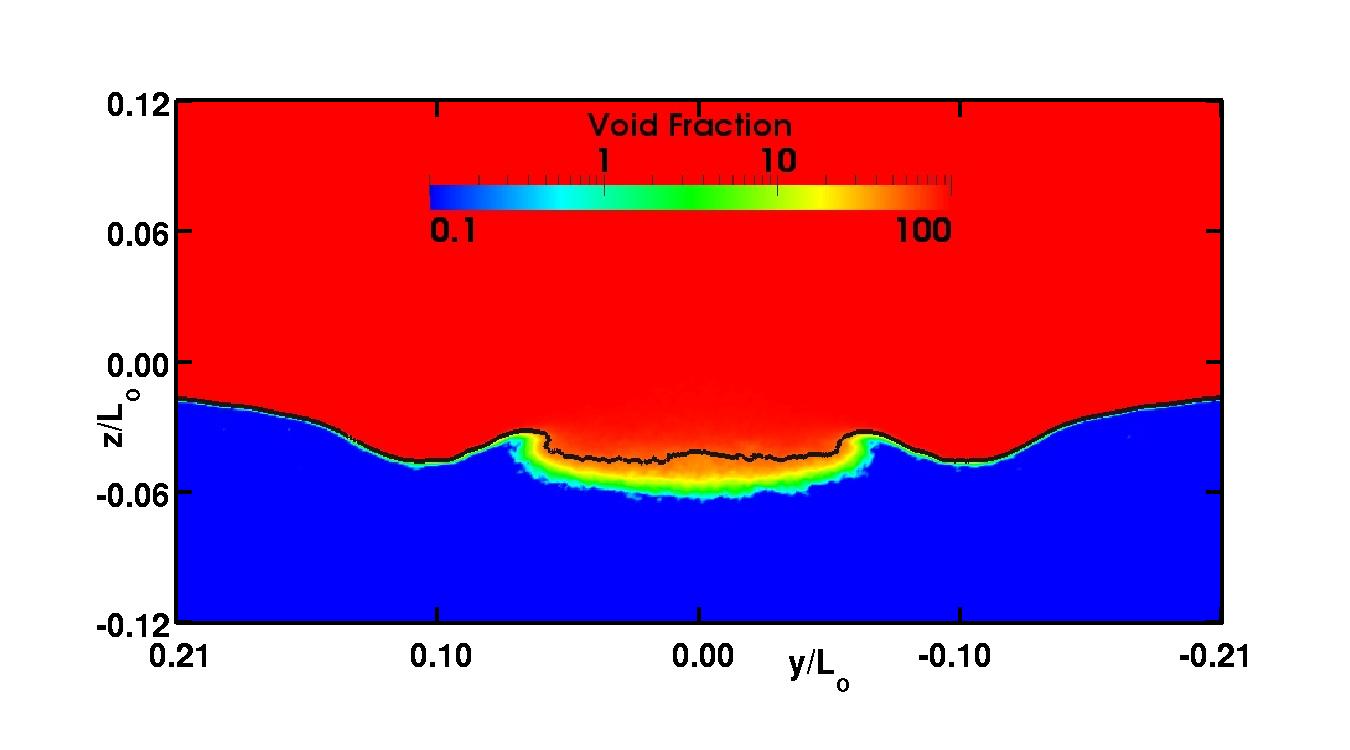} \\
(a3) & & (b3) & \vspace{-15pt} \\ 
& \includegraphics[trim = 12mm 12mm 12mm 15mm, clip,width=0.4\linewidth]{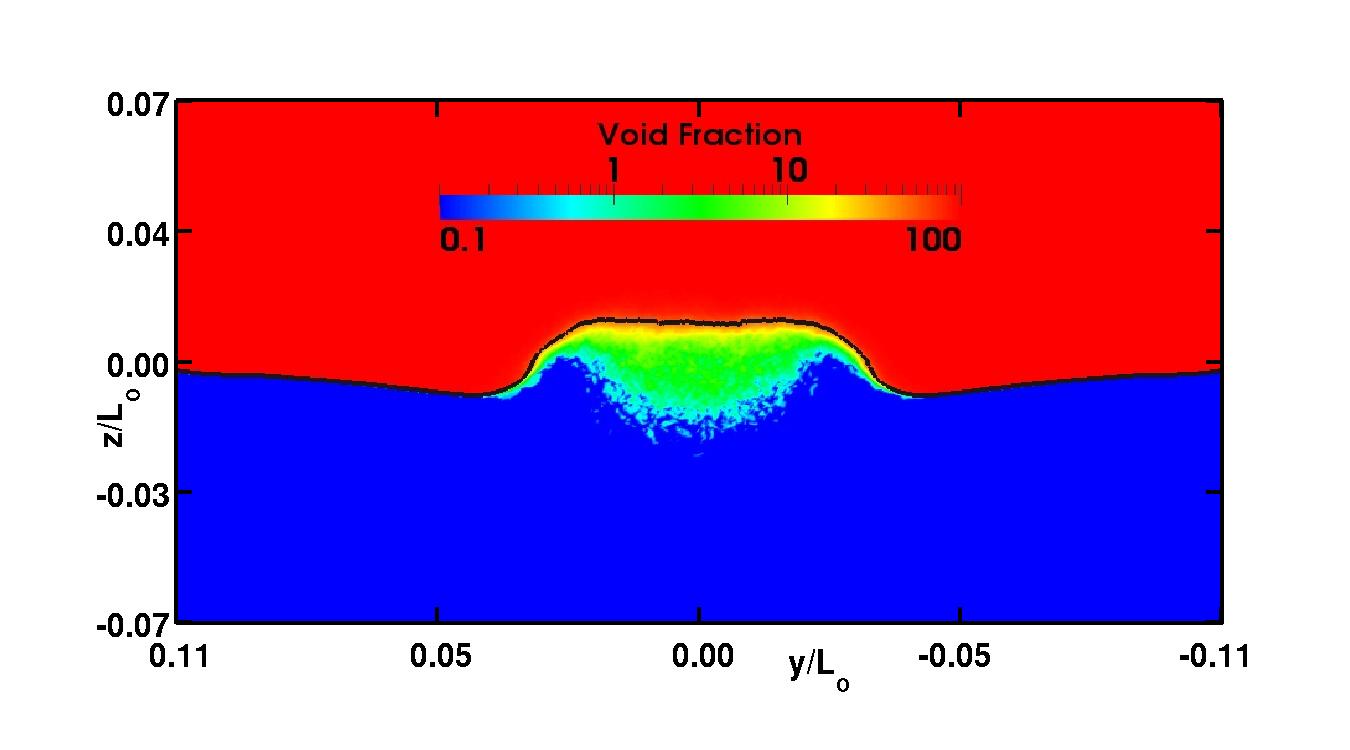}
& & \includegraphics[trim = 12mm 12mm 12mm 15mm, clip,width=0.4\linewidth]{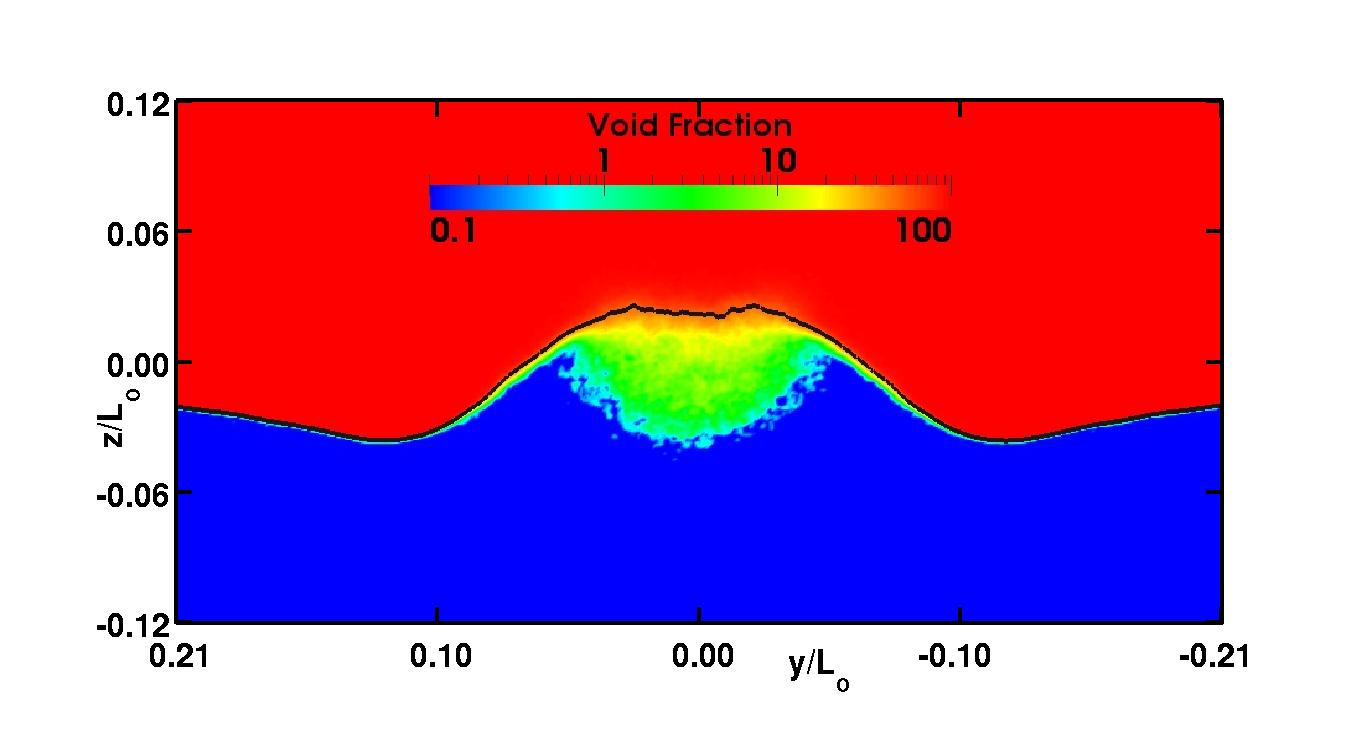} \\
(a4) & & (b4) & \vspace{-15pt} \\ 
& \includegraphics[trim = 12mm 12mm 12mm 15mm, clip,width=0.4\linewidth]{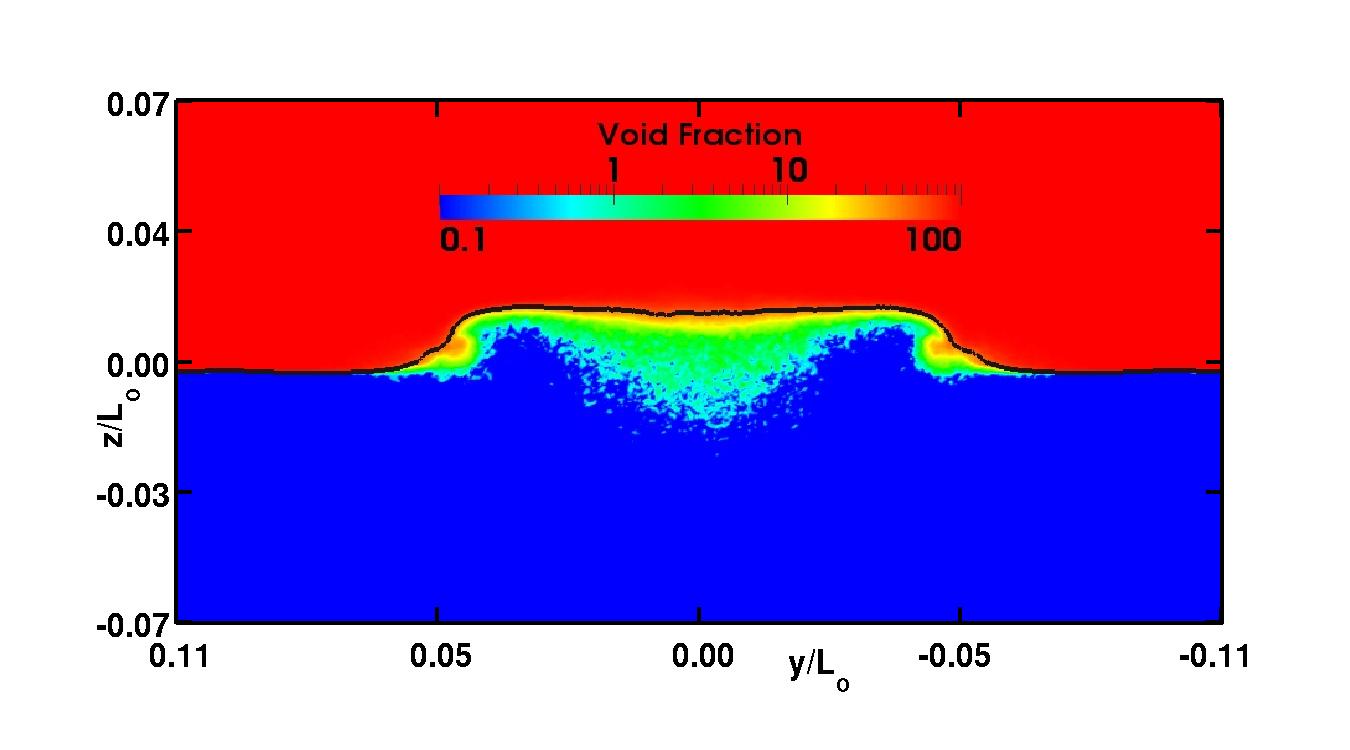}
& & \includegraphics[trim = 12mm 12mm 12mm 15mm, clip,width=0.4\linewidth]{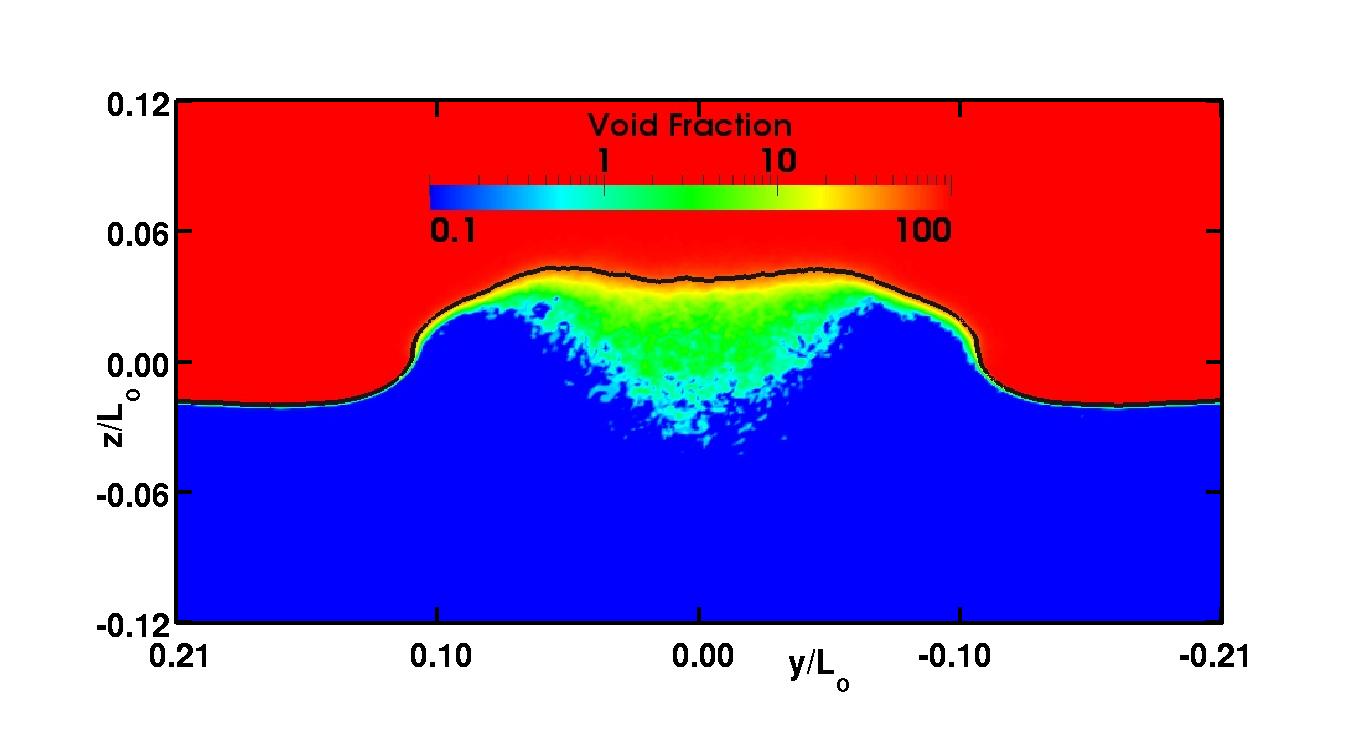} \\
(a5) & & (b5) & \vspace{-15pt} \\ 
& \includegraphics[trim = 12mm 12mm 12mm 15mm, clip,width=0.4\linewidth]{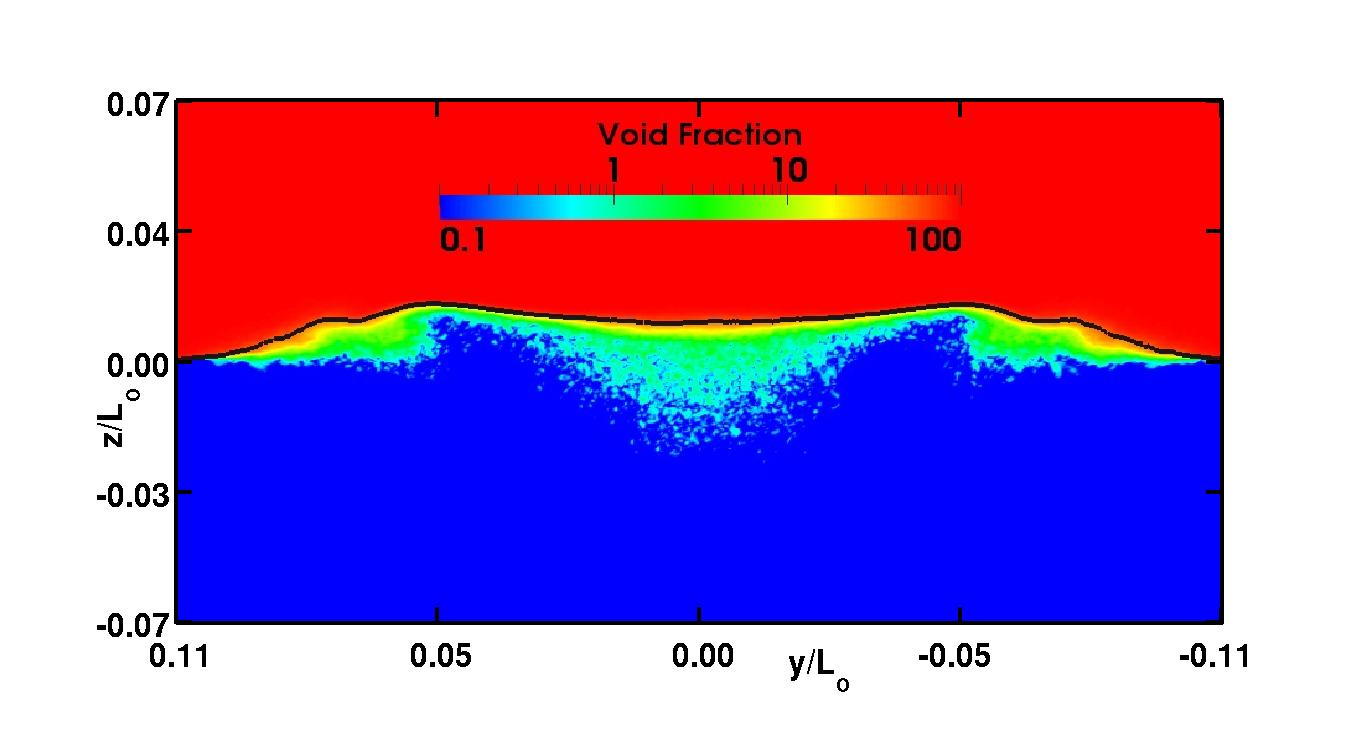}
& & \includegraphics[trim = 12mm 12mm 12mm 15mm, clip,width=0.4\linewidth]{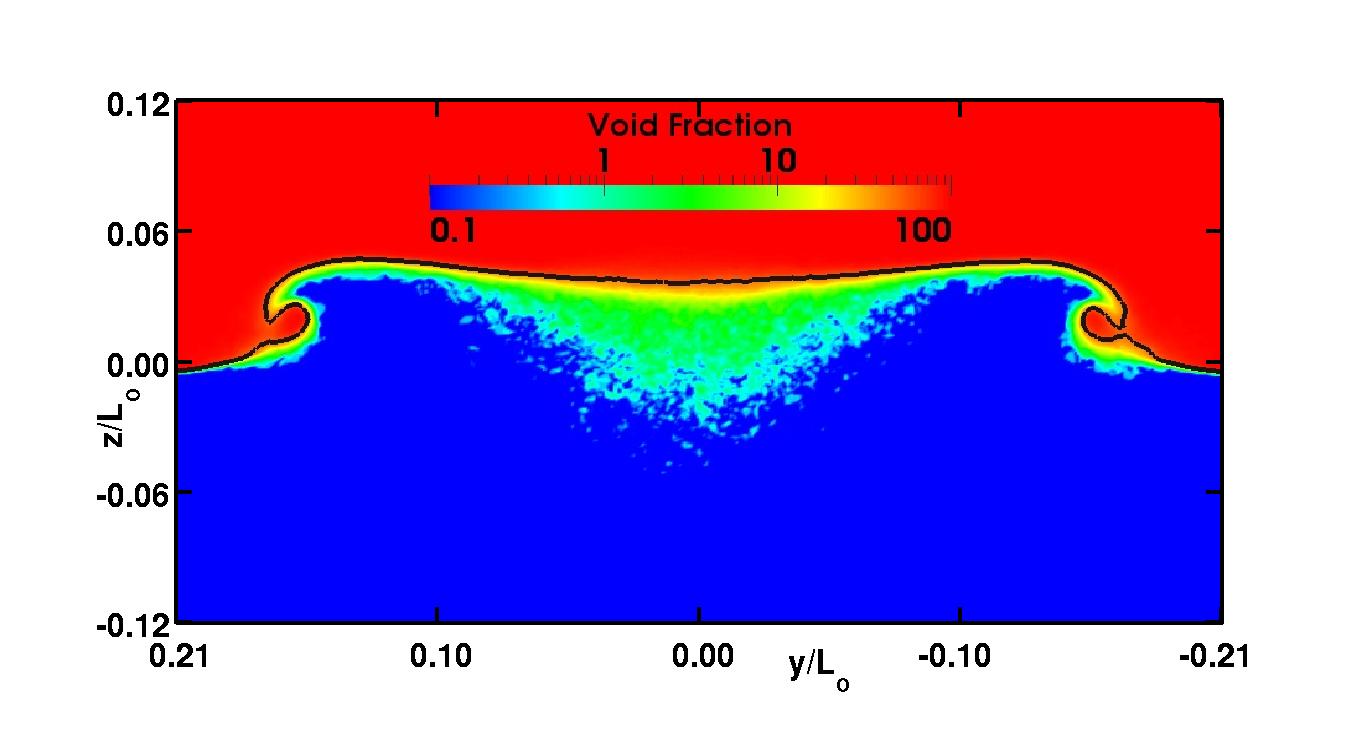}
\end{tabular}
\end{center} \caption{\label{nfa_foam} Air entrainment due to
wave breaking.  Color contours of the time-averaged volume fraction are plotted
for various transverse cuts.  The black lines denote the 0.5 isosurface of the
volume fraction.   (a) 3.60 m/s (7 knots)  and (b) 4.12 m/s (8 knots).   (1) $x/L_o=0.04$.  (2) $x/L_o=0.08$, (3) $x/L_o=0.16$.  (4) $x/L_o=0.24$.   (5) $x/L=0.32$.  Here, $x$ and $L_o$ are respectively the distance aft of the transom and the length of the model. The numerical results have been time-averaged over the last 10,000 time steps of the numerical simulation.}
\end{figure*}

Figure \ref{nfa_foam} shows transverse cuts of the time-averaged volume
fraction for various distances aft of the transom.    The 3.60 m/s (7 knot) plots
extend from -1m to 1m in transverse direction and from -.62m to .64m in the vertical.
The 4.12 m/s (8 knot) plots extend from -1.9m to 1.9m in transverse direction and from
-1.14m to 1.14m in the vertical. Most of the air entrainment occurs in the
rooster-tail region and along the edges of the stern breaking wave.   The
structure of the void fraction wake has three dominant structures corresponding
to the centerline entrainment in the rooster-tail region and the
spilling-breaking entrainment that occurs along the cusp line.

    %
    %

\subsection{CFDShip-IOWA Predictions and Assessments}

CFDShip-Iowa V4 mean and unsteady wave elevation predictions using
detached eddy simulation (DES) for the transom stern model for
a wet ($F_{r}$ = 0.38) and a dry ($F_{r}$ = 0.43) transom is
assessed using experimental data, and the dominant wetted
transom flow frequency is explained as a Karman-like vortex shedding.

\subsubsection{Computational Method}

The simulations are performed using a single phase solver in absolute
inertial earth-fixed coordinates \cite{Carrica07}. The turbulence modeling is
performed using DES and the interface modeling using level-set methods. A
multi-block overset grid approach is used to allow grid refinement in the
regions of interest. The governing equations are discretized using
cell-centered finite difference schemes on body-fitted curvilinear grids and
solved using a predictor-corrector method. The time marching is done using the
2$^{nd}$ order backward difference scheme. The convection terms and level-set
equations are discretized using a hybrid 2$^{nd}$/4$^{th}$ order Total
Variation Diminishing (TVD) scheme.
The pressure Poisson equation is solved using the Portable Extensible Toolkit
for Scientific computing (PETSc) using a
projection algorithm to satisfy continuity. MPI-based
domain decomposition is used, where each decomposed block is mapped to a single
processor.

\subsubsection{Domain, Grids, Boundary and Simulation Conditions}

\begin{figure} \centering
\includegraphics[width=.8\columnwidth]{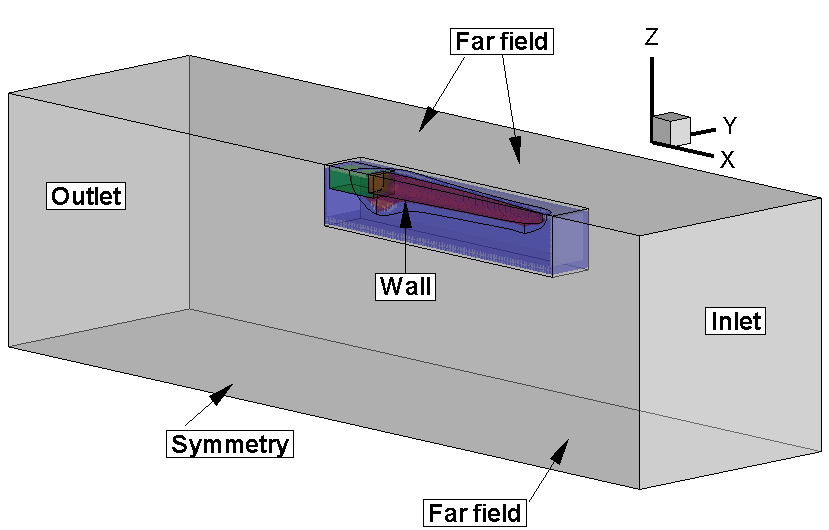}
\caption{\label{CFDShip_Fig_1} Domain and boundary conditions for CFDShip-Iowa
V4.} \end{figure}

The simulations are performed for half of a domain only as shown in Figure
\ref{CFDShip_Fig_1}. The grid consists of a background block $X/L_o$ = [-0.5,
3.0], $Y/L_o$ = [0,1.0], and $Z/L_o$ = [-1.0, 0.035], where $X$, $Y$, $Z$ and
$L_o$ are the streamwise, spanwise, normal directions and ship length,
respectively. Refinement blocks are used near the free-surface and in the
transom region to accurately resolve the transom flow features. The grid
consists of a total of 16.5M points, which is partitioned  into 98 blocks for
parallel computing. The uniform inlet velocity $U$, zero-gradient exit,
far-field at $Z$-Min, $Z$-Max and $Y$-Max planes, and $Y= 0$ symmetric boundary
conditions are applied to the background grid. The no-slip boundary condition
is applied at the boundary layer grid J=1 plane. The averaged $y^+$ = 0.8 to
1.0 for the near-wall resolution. The boundary layer grid is translated and
rotated to match the sinkage and trim from the experimental test.

%
%

Fixed sinkage and trim simulations are performed for the speeds of 3.60 m/s (7
knots) ($Re$ = 1.093x10$^7$, $F_{r}$ = 0.3803) and 4.12 m/s (8 knots) ($Re$ =
1.25x10$^7$, $F_{r}$ = 0.4346). The $Re$ values are computed assuming
20$^{\circ}$ C water temperature. Simulations are performed using a time step
size $\Delta t$ = 2.5x10$^{-3}$. The 3.60 m/s (7 knots) simulation is performed
for a time equivalent to 8.5 ship lengths (8.5$L_o/U$), and every fourth time-step solution for the last $6L_o/U$ data
is used for averaging.  The 4.12 m/s (8
knots) simulation is performed for only $6.5L_o/U$ as flow reaches a steady
state sooner and averaging is performed for the last $2L_o/U$ data.

\subsubsection{Prediction Assessments}

The wave elevation mean, RMS, and elevation spectra results were compared against QViz
close to the transom and LiDAR away from the transom. Void
fraction data are not compared, as they cannot be quantified by the level-set
interface modeling.

\begin{figure*} [t] \begin{center} \begin{tabular}{llll} (a) & & & \vspace{-15pt}
\\ & \includegraphics[width=0.4\linewidth]{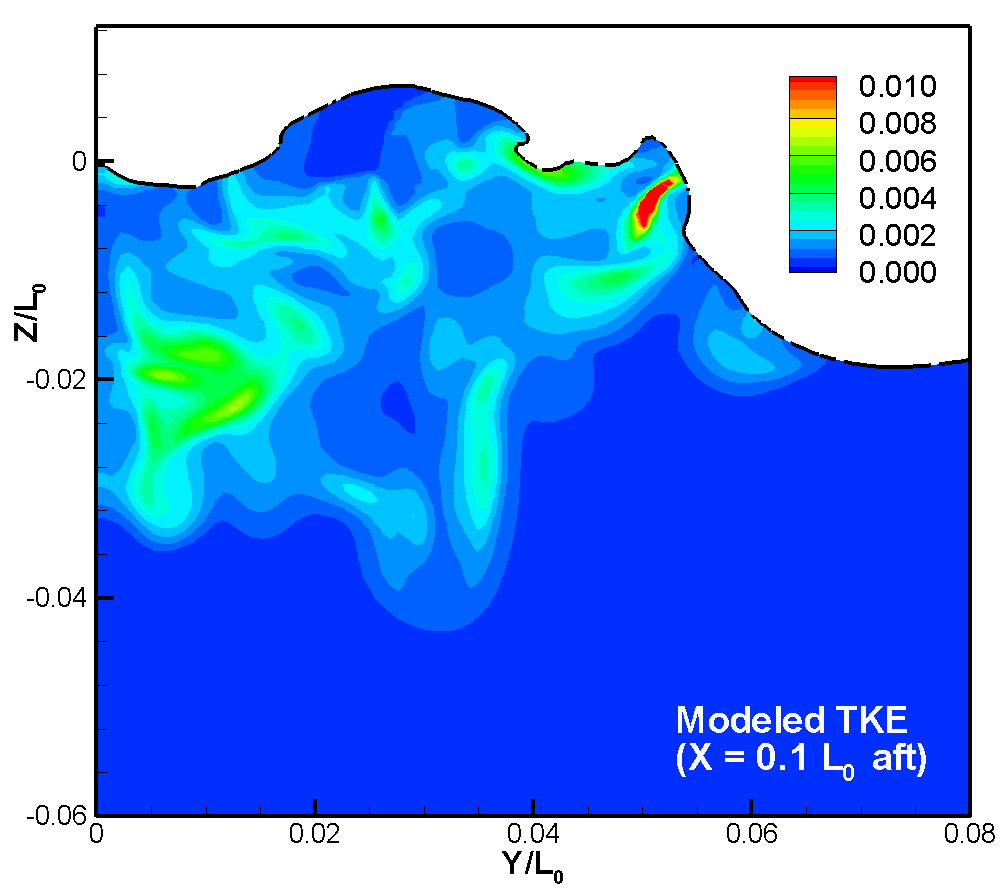}
& & \includegraphics[width=0.4\linewidth]{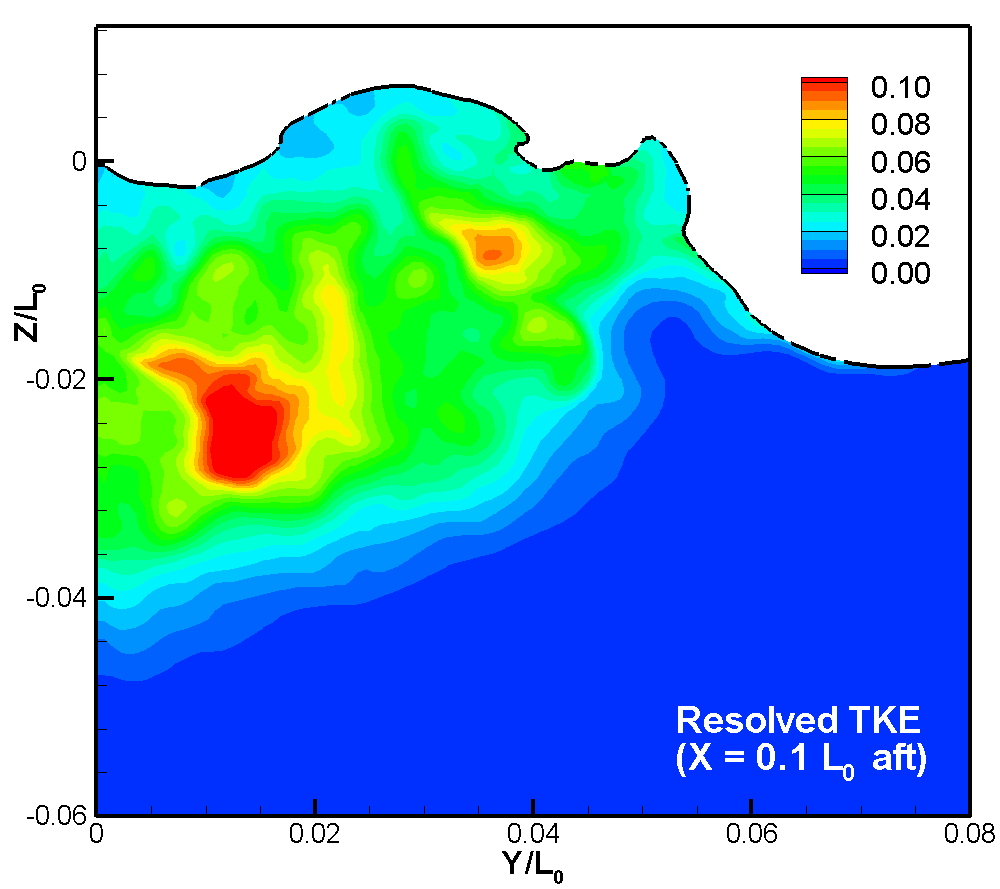}
\\ (b) & & & \vspace{-15pt} \\ &
\includegraphics[width=0.4\linewidth]{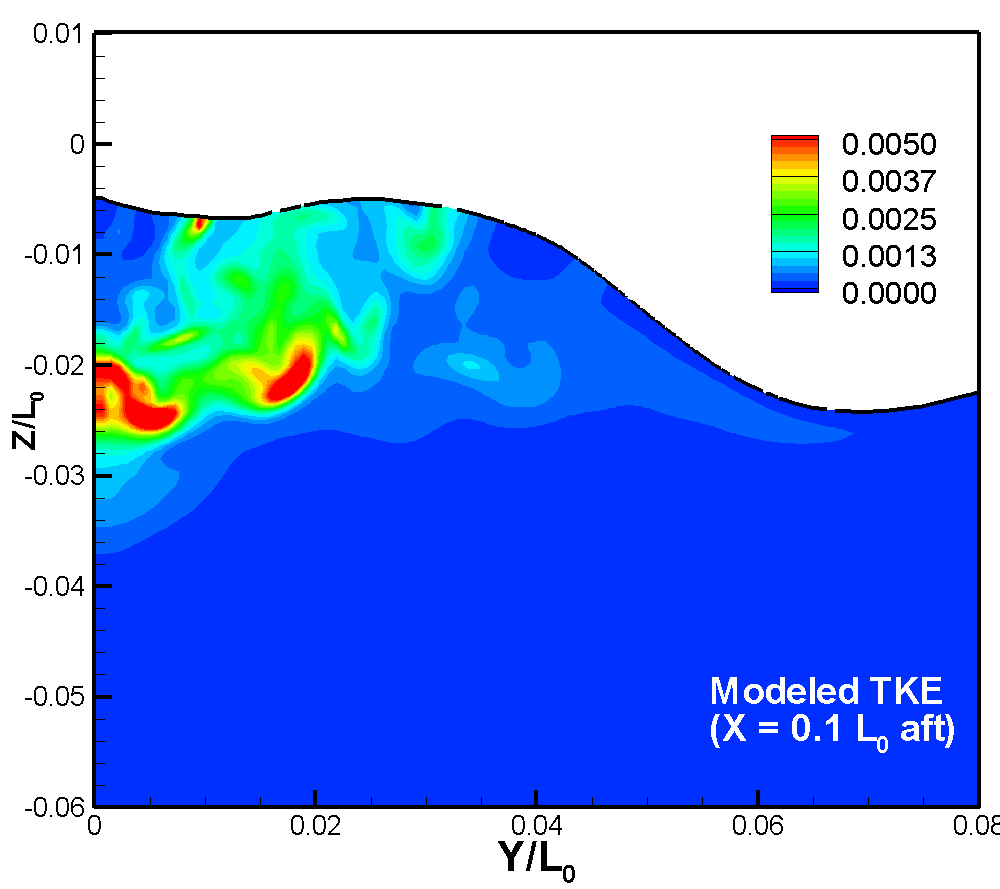} & &
\includegraphics[width=0.4\linewidth]{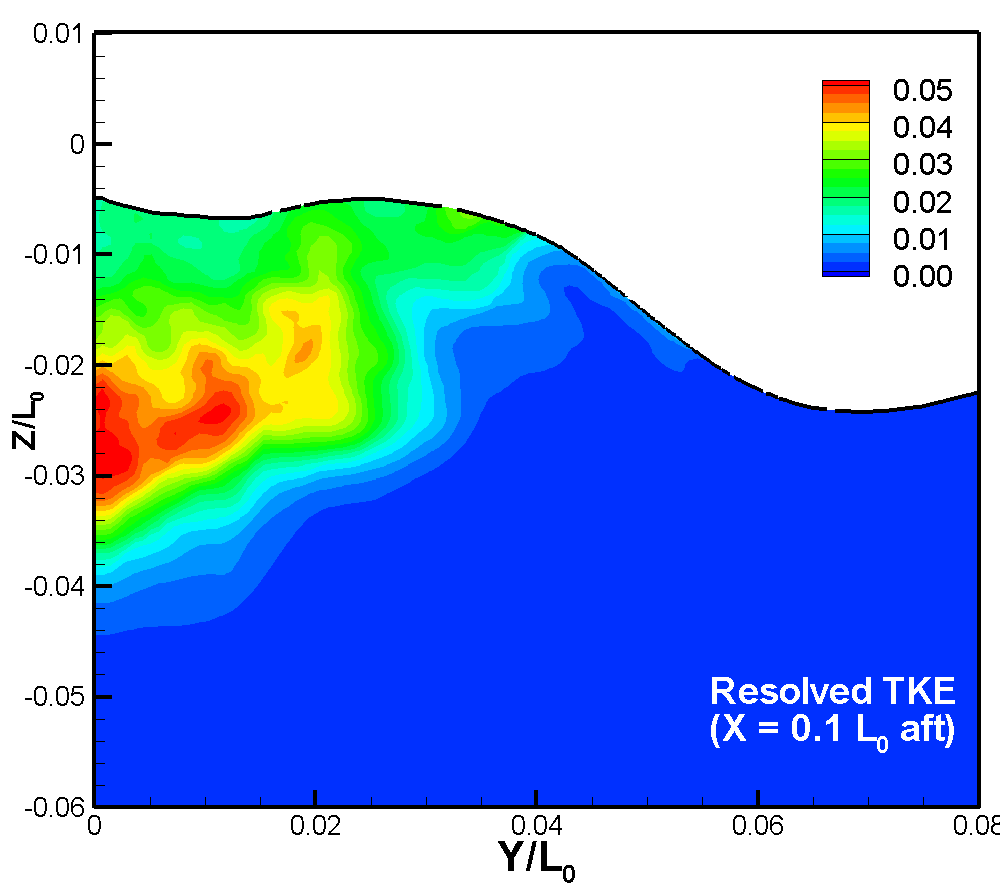} \\
\end{tabular} \end{center} \caption{\label{CFDShip_Fig_2} Modeled (left) and
resolved (right) TKE predictions using CFDShip-Iowa V4 for (a)3.60 m/s (7
knots), and (b) 4.12 m/s (8 knots) simulations at $X/L_o$ = -1.1.} \end{figure*}

The validation study focuses only on the comparison error $E$ between the
experimental
and CFDShip-Iowa V4 results, as a grid verification study was not performed.
Thus the grid uncertainty $U_G$ or validation uncertainty intervals cannot be estimated.
$U_I$ = 0.7\% based on the resistance prediction $C_t$ are similar to that
predicted for the CFDShip-Iowa V4 appended Athena study \cite{Bhushan10}. As shown in Figure \ref{CFDShip_Fig_2}, the resolved turbulent
kinetic energy (TKE) levels are greater than 92\% of the total TKE (modeled +
resolved) in the
transom stern region for both flow conditions. This result suggests that the grid is
sufficiently fine to properly resolve the turbulent fluctuation in the Large Eddy Simulation (LES)
region. Herein, only half domain simulations are performed, which may result in
under prediction of RMS as observed by \citeasnoun{Bhushan10}.

%
%

%
%

\begin{figure*} \begin{center} \begin{tabular}{llll} (a) & & (b) & \vspace{-15pt}
\\ & \includegraphics[width=.8\columnwidth]{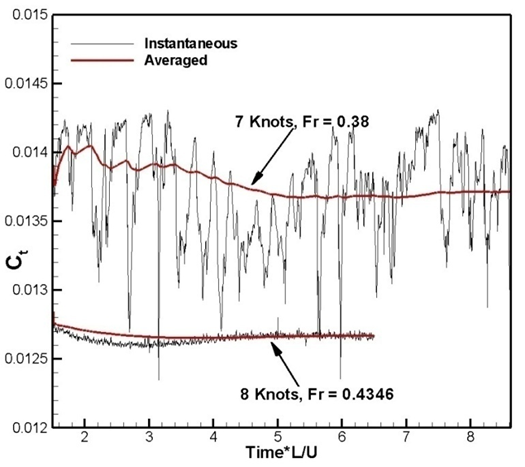}
& & \includegraphics[width=.8\columnwidth]{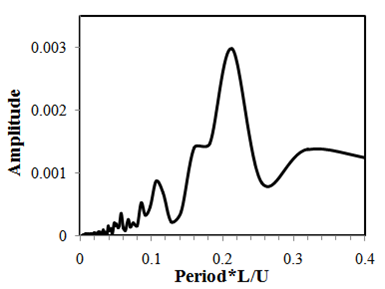}
\end{tabular} \end{center} \caption{\label{CFDShip_Fig_3} CFDShip-Iowa V4 Ct prediction (a)
time history, and (b) spectra for 3.60 m/s (7 knots).} \end{figure*}


%
%

\begin{table*} \caption{\label{CFDShip_Table1} Resistance, mean and unsteady
wave elevation validation for transom-stern model, and bare hull and appended
Athena } \begin{center}
\includegraphics[width=.9\textwidth]{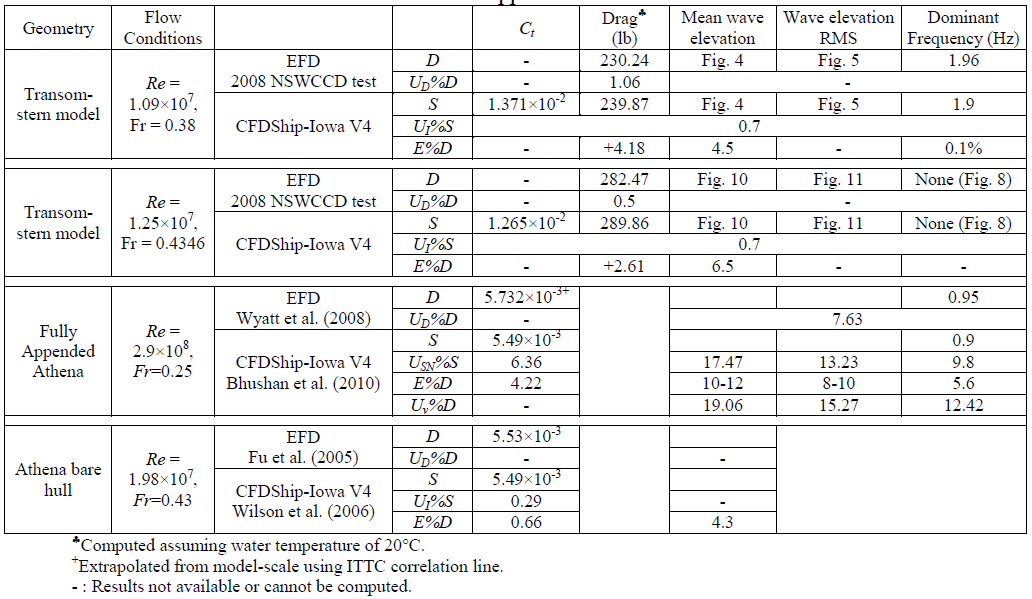}
\end{center} \end{table*}

An unsteady $C_t$ is predicted for 3.60 m/s (7 knots), whereas steady values
are predicted for 4.12 m/s (8 knots) as shown in Figure \ref{CFDShip_Fig_3}.
The time period of the $C_t$ unsteadiness, $\tau_p$ = 0.21$L_o/U$, is
due to the transom vortex shedding and will be discussed later. As summarized in Table
\ref{CFDShip_Table1}, the total drag predictions are 4.18\% and 2.61\% higher
than the experimental results for 3.60 m/s (7 knots) and 4.12 m/s (8 knots), respectively.

\begin{figure*} [t] \begin{center} \begin{tabular}{llll} (a) & & (b) &
\vspace{-15pt} \\ &
\includegraphics[width=0.4\linewidth]{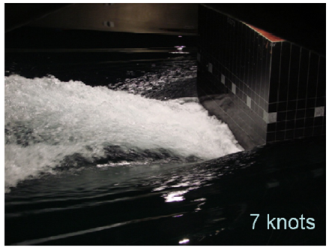} & &
\includegraphics[width=0.4\linewidth]{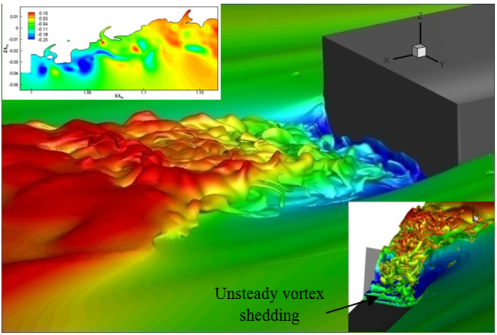} \\ (c) & &
(d) & \vspace{-15pt} \\ &
\includegraphics[width=0.4\linewidth]{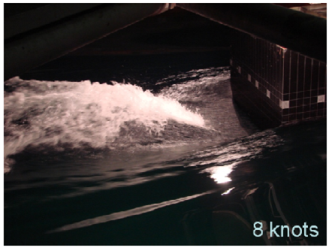} & &
\includegraphics[width=0.4\linewidth]{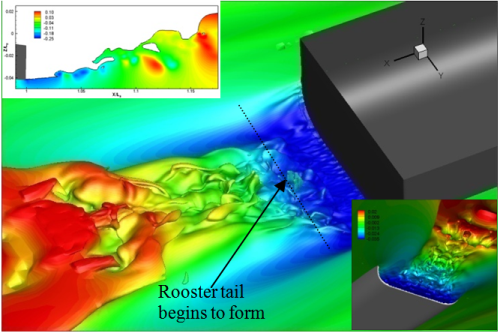} \\
\end{tabular} \end{center} \caption{\label{CFDShip_Fig_4} Images of the instantaneous transom
wave elevation during NSWCCD testing (left panel) and CFDShip-Iowa V4 (right panel)
for (a,b) 3.60 m/s (7 knots), $F_r$ = 0.38, and (c,d) 4.12 m/s (8 knots), $F_r$ =
0.43. Inset shows wave elevation at $Y/L_o$ = 0.01 colored using piezometric
pressure, and wave elevation and isosurface of $Q$-criterion = 3000
\protect\cite{Hunt88} as viewed from the
bottom } \end{figure*}

The experimental results show a wetted transom flow for the 3.60 m/s (7 knots) case with
a recirculation region on either side of the model centerline. As shown in Figure
\ref{CFDShip_Fig_4}, the wake spans the entire transom width and is dominated
by a large amount of entrained air. For the 4.12 m/s (8 knots) case, a dry
transom was observed with a rooster tail which begins to form at $X/L_o$ =
0.07-0.1 aft of the transom. The wake in this case is well defined, narrow, and
quickly steepens to a defined peak where it begins to spill out and widen.
CFDShip-Iowa V4 predicts a wetted transom for 3.60 m/s (7 knots) with unsteady vortex shedding
from the transom bottom corner. Wave breaking and air entrainment is predicted
up to the transom edge. For the 4.12 m/s (8 knots) case, a rooster tail begins
to form at $X/L_o$ = 0.086 aft of the transom, a narrow wake is observed up to
$X/L_o$ =0.15 aft, and starts to widen thereafter. The breaking wave and air
entrainment occurs mostly in the rooster tail region.

%
%

\begin{figure*} \begin{center} \begin{tabular}{llll} (a) & & (b) &
\vspace{-15pt} \\ &
\includegraphics[width=0.4\linewidth]{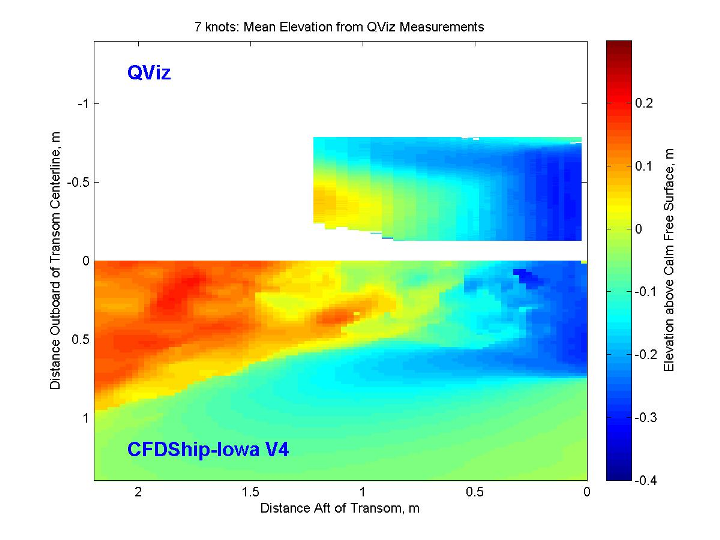} & &
\includegraphics[width=0.4\linewidth]{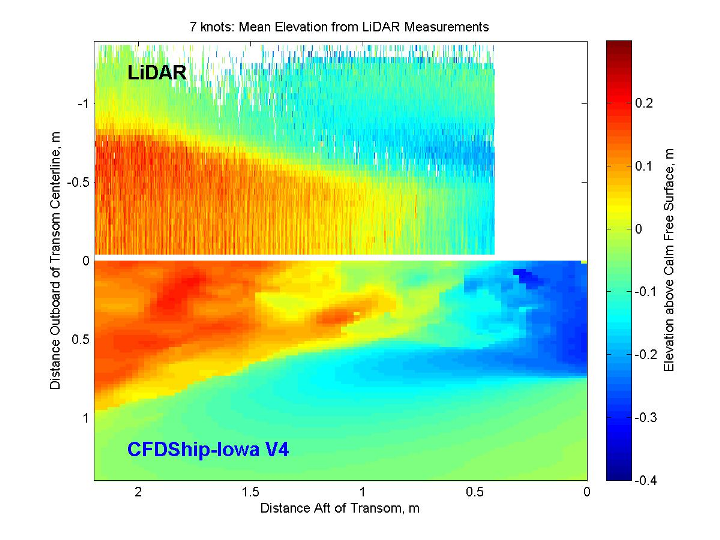} \\
(c) & & (d) & \vspace{-15pt} \\ &
\includegraphics[width=0.4\linewidth]{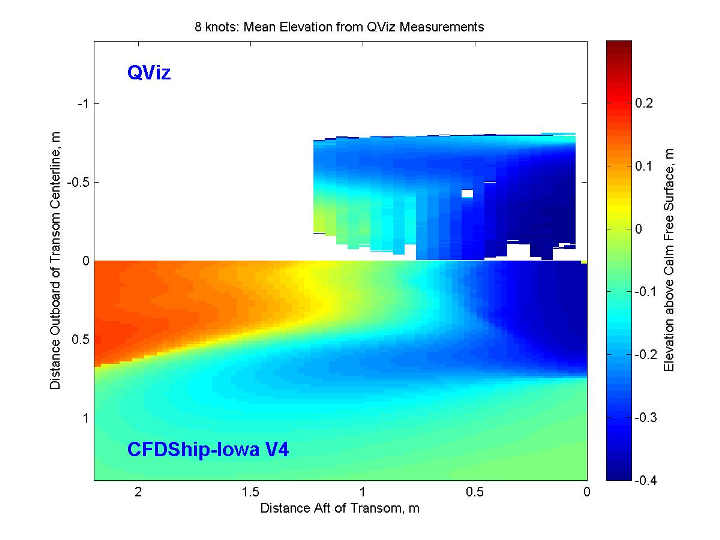} & &
\includegraphics[width=0.4\linewidth]{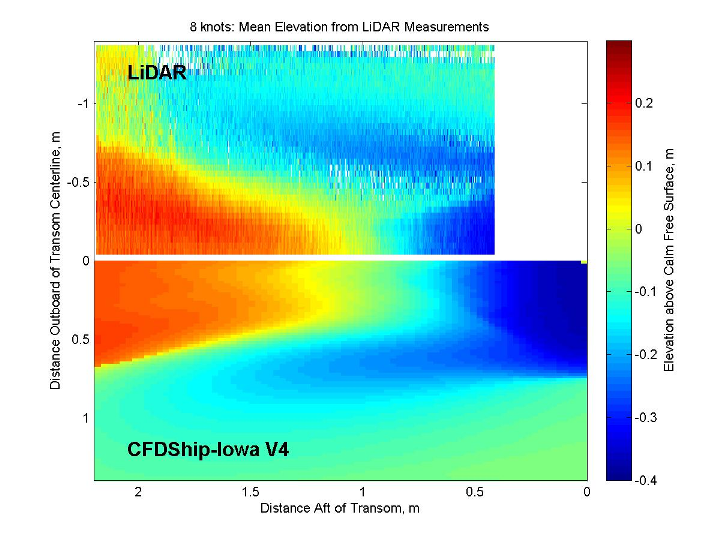} \\
\end{tabular} \end{center} \caption{\label{CFDShip_Fig_5} CFDShip-Iowa V4 (top)
mean transom wave elevation predictions for 3.60 m/s (7 knots) are compared
with (a) QViz, (b) LiDAR measurements (top), and for 4.12 m/s (8 knots) (c)
QViz, and (d) LiDAR measurements. The transom edge is at $X/L_o$ = 1.0. }
\end{figure*}

The experimentally measured mean wave elevation in Figure \ref{CFDShip_Fig_5} shows good agreement
between the QViz and LiDAR datasets in the small overlap region for both 3.60
m/s (7 knots) and 4.12 m/s (8 knots). The datasets
show a low wave elevation near the transom and a diverging wave trough emerging
from the transom edge. The diverging wave angle is 20.2$^\circ$ and 26$^\circ$ at 3.60 m/s
(7 knots) and 4.12 m/s (8 knots), respectively. A transverse wave is observed between the centerline and the diverging wave at both speeds, with
the wave forming further back at a speed of 4.12 m/s (8 knots) than the 3.60
m/s (7 knot) case.

CFDShip-Iowa V4 predictions are within 3-4\% of the experimental results both near the transom and in the diverging
waves trough at 3.60 m/s (7 knots) and 4.12 m/s (8 knots). The diverging wave
peak is predicted to be outside the LiDAR's measurement window and thus will not be compared.
The diverging wave angle is 12\% and 10\% higher than that found during the NSWCCD testing for 3.60 m/s (7 knots) and 4.12 m/s (8 knot), respectively.

A shoulder wave is seen in the wake elevation cross-sections for both the 3.60
m/s (7-knot) and 4.12 m/s (8-knot) cases. There are slight indications of a
shoulder wave in the LiDAR results, likely due to the large RMS at those
locations and the uncertainty inherent in the LiDAR measurements ($\pm$ 2.54
cm). CFDShip-Iowa V4 transverse wave peak elevation predictions
are within 2-4\% of the experimental results for both the 3.60 m/s (7 knot) and
4.12 m/s (8 knot) cases for the location closest to the transom. At 4.12 m/s (8 knots) the peak values reported by
CFDShip-Iowa V4 are lower by 25\% for the locations further from the
transom. For the farthest cross section, almost uniform
elevations are observed in the wake at 3.60 m/s (7 knots), which agrees within 2\% of the
experimental results.

%
%

%
%

\begin{figure*} \begin{center} \begin{tabular}{cccc} (a) & (b) & (c) & (d) \\
\includegraphics[width=0.22\linewidth]{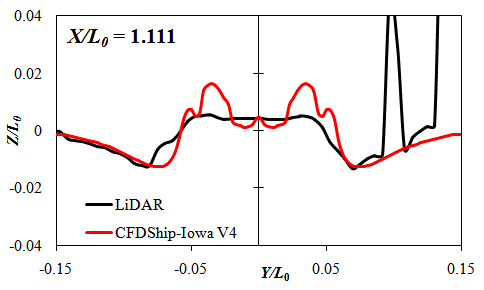} &
\includegraphics[width=0.22\linewidth]{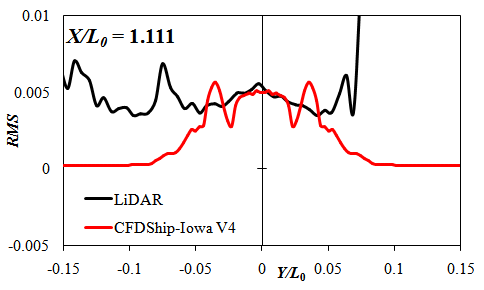} &
\includegraphics[width=0.22\linewidth]{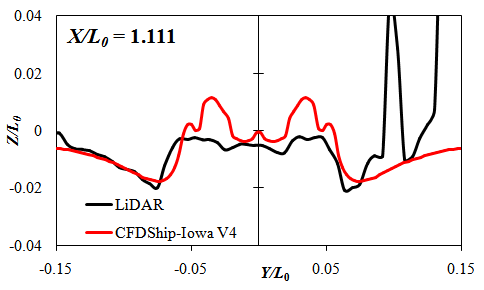} &
\includegraphics[width=0.22\linewidth]{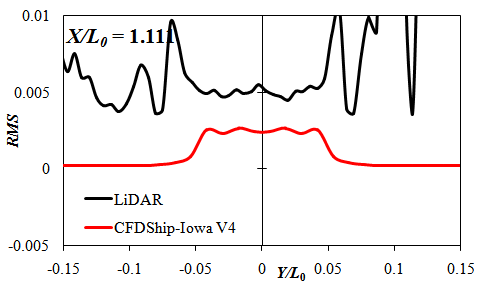}  \\
\includegraphics[width=0.22\linewidth]{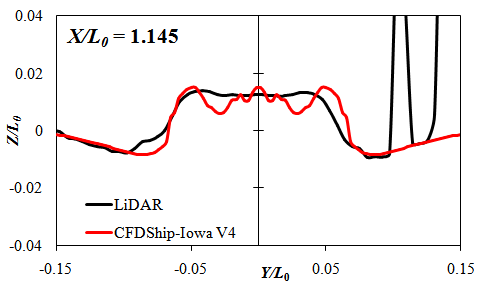} &
\includegraphics[width=0.22\linewidth]{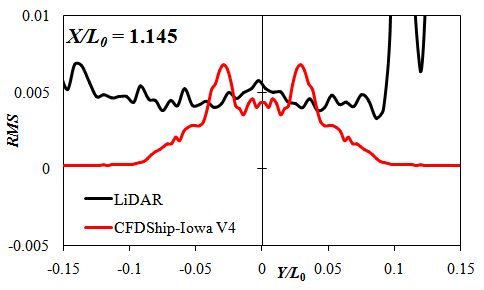} &
\includegraphics[width=0.22\linewidth]{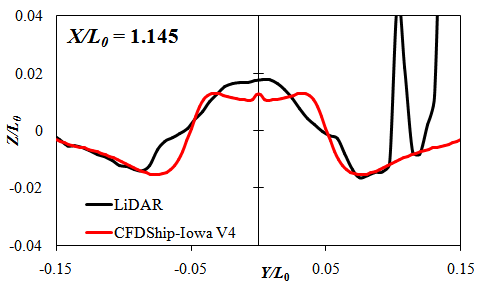} &
\includegraphics[width=0.22\linewidth]{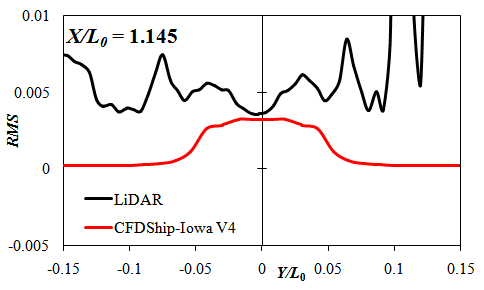}  \\
\includegraphics[width=0.22\linewidth]{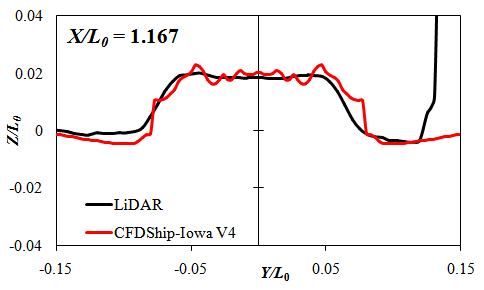} &
\includegraphics[width=0.22\linewidth]{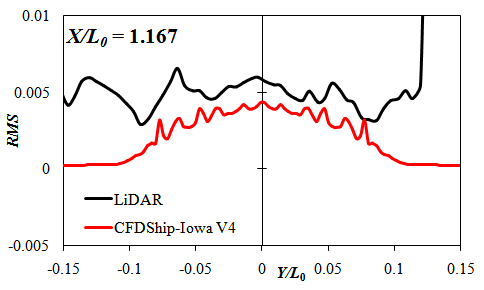} &
\includegraphics[width=0.22\linewidth]{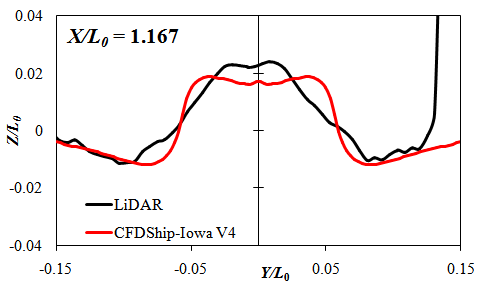} &
\includegraphics[width=0.22\linewidth]{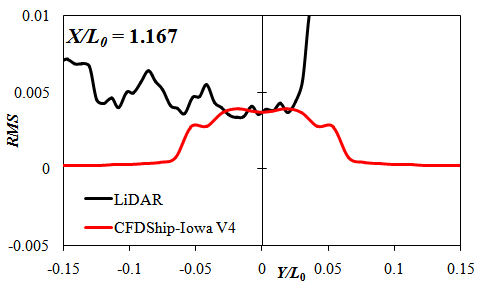}  \\
\end{tabular} \end{center} \caption{\label{CFDShip_Fig_6} Free-surface
transverse cuts.  $x$-axis denotes distance from centerline in cm. The $y$-axis
represents the elevation from the calm free-surface in cm.  Red, black lines
denote CFDShip-Iowa V4 and LiDAR results, respectively.  Column (a) 3.60 m/s (7 knots), mean height
0.254m, 0.508m aft and 0.762m aft. Column (b) 3.60 m/s (7 knots), RMS 0.254m,
0.508m aft and 0.762m aft. Column (c) 4.12 m/s (8 knots), mean height 0.254m,
0.508m aft and 0.762m aft. Column (d) 4.12 m/s (8 knots), RMS 0.254m, 0.508m
aft and 0.762m aft.} \end{figure*}

%
%

\begin{figure*} \begin{center} \begin{tabular}{llll} (a) & & (b) \vspace{-12pt}
\\ & \includegraphics[width=0.4\linewidth]{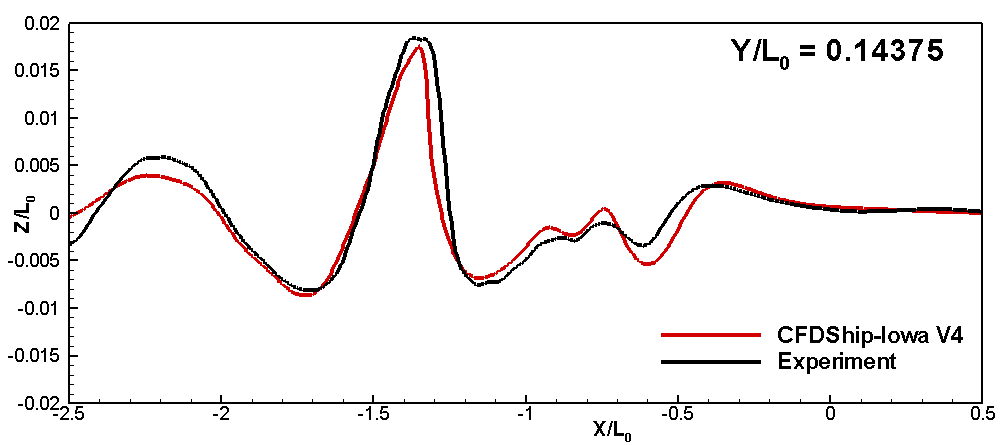} & &
\includegraphics[width=0.4\linewidth]{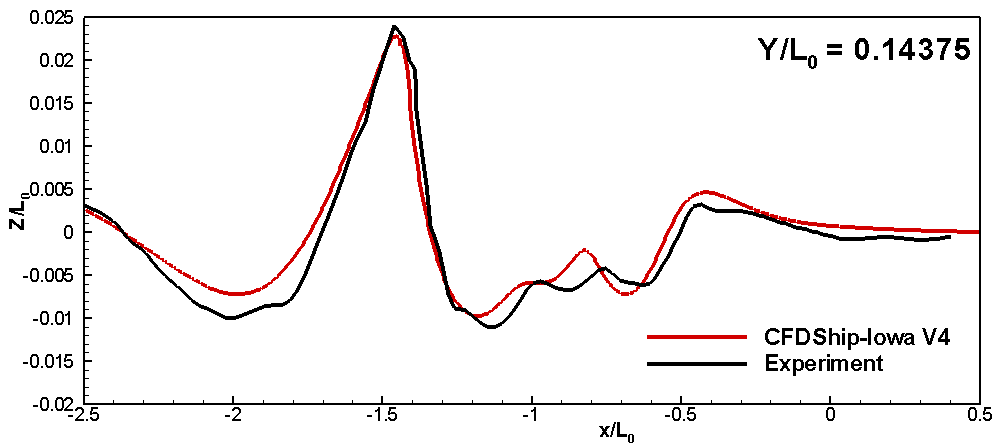} \\ & & &
\vspace{-12pt} \\ &
\includegraphics[width=0.4\linewidth]{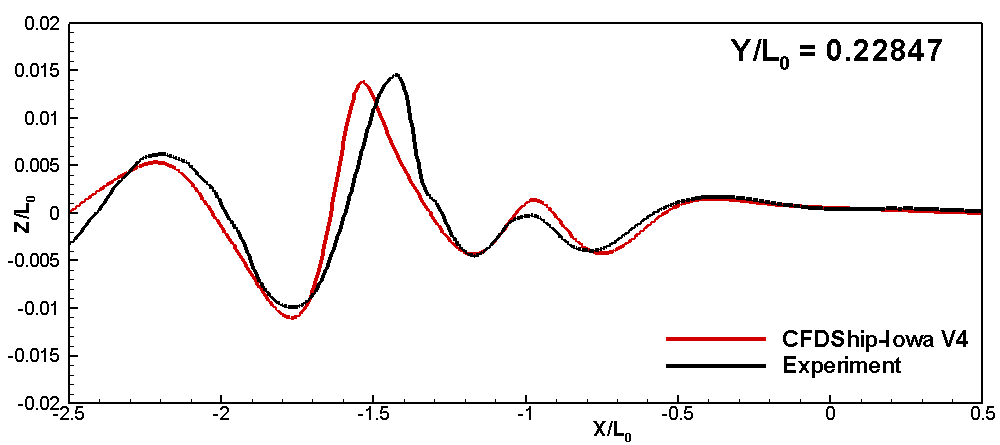} & &
\includegraphics[width=0.4\linewidth]{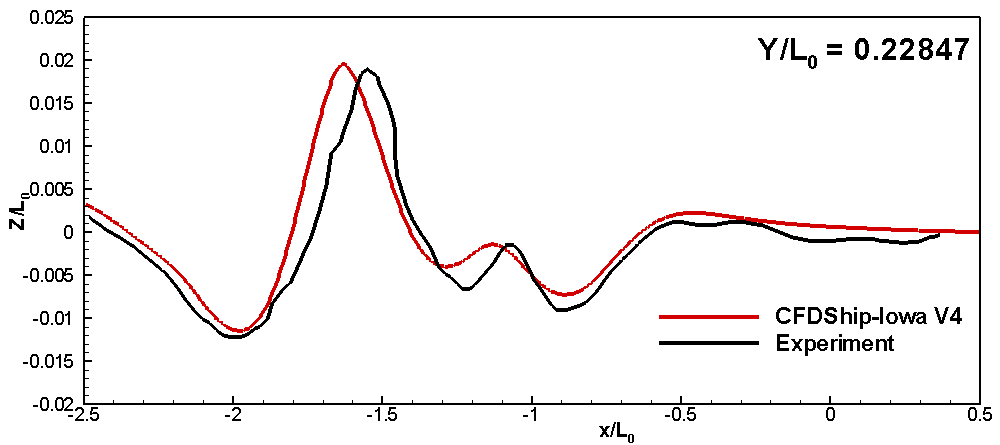} \\ & & &
\vspace{-12pt} \\ &
\includegraphics[width=0.4\linewidth]{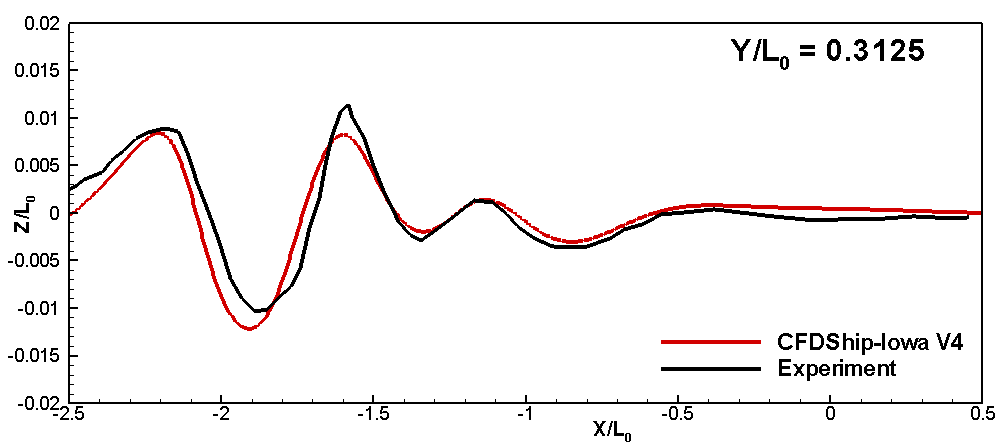} & &
\includegraphics[width=0.4\linewidth]{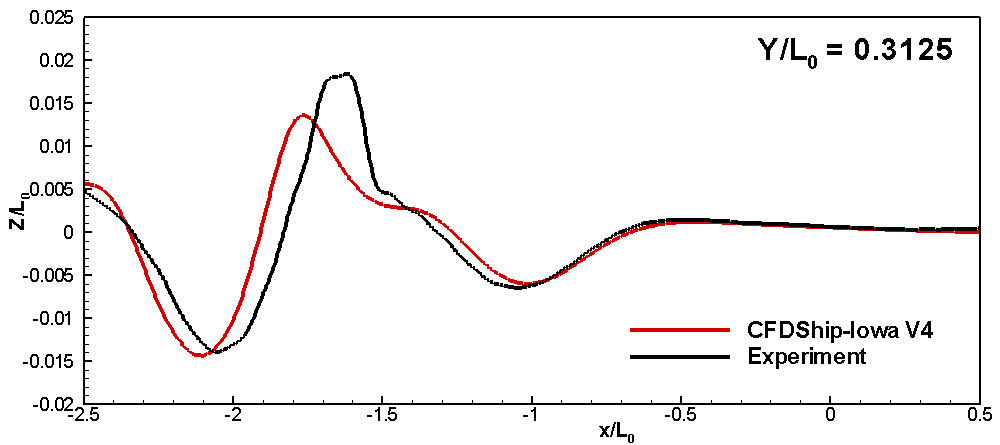} \\ & & &
\vspace{-12pt} \\ &
\includegraphics[width=0.4\linewidth]{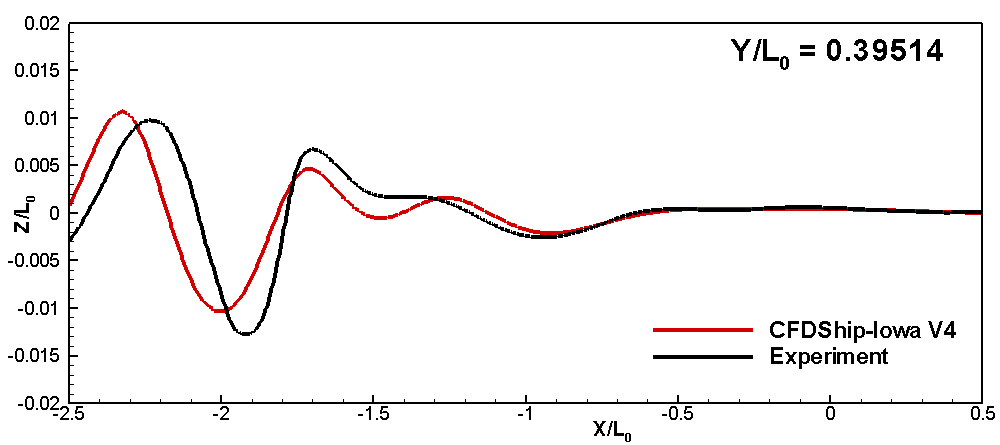} & &
\includegraphics[width=0.4\linewidth]{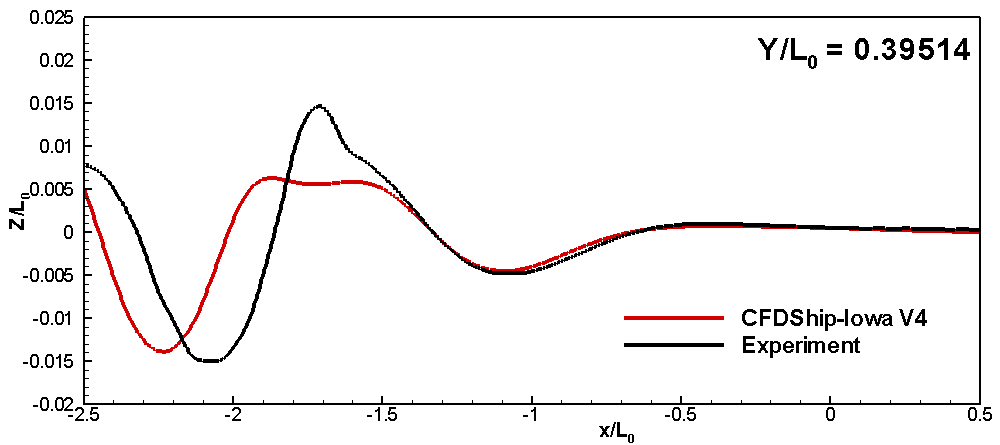}
\end{tabular} \end{center} \caption{\label{CFDShip_Fig_7} Transom-stern model
wave cuts at $Y/L_o$ = 0.14375, 0.22847, 0.03125 and 0.39514 predicted by
CFDShip-Iowa V4 is compared with Senix Ultrasonic Sensors measurements for (a)
3.60 m/s (7 knots) and (b) 4.12 m/s (8 knots). The X coordinates are oriented to
match the experiment's coordinates, i.e., X = 0 corresponds to bow and X = -1 the transom stern.} \end{figure*}

%
%

\begin{table*} \caption{\label{CFDShip_Table2} Characteristics of the
Karman-like vortex shedding for different geometries } \begin{center}
\includegraphics[width=.9\textwidth]{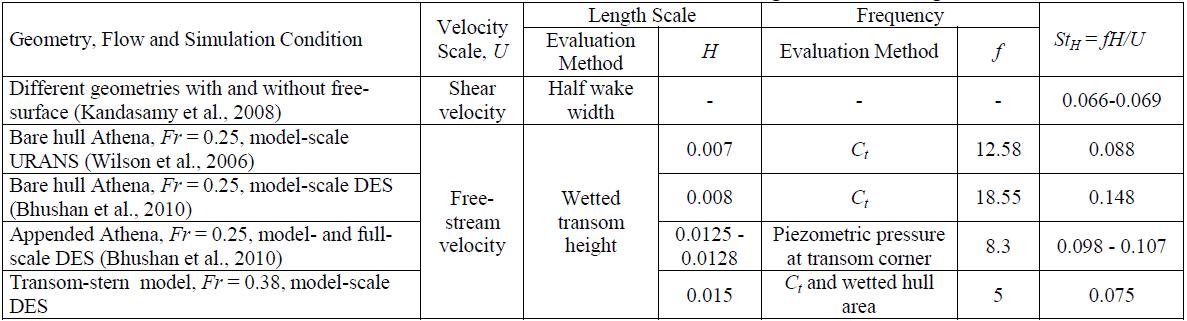}
\end{center} \end{table*}

%
%

\begin{figure*} \begin{center} \begin{tabular}{llll} (a) & & (b) \vspace{-12pt}
\\ & \includegraphics[width=0.4\linewidth]{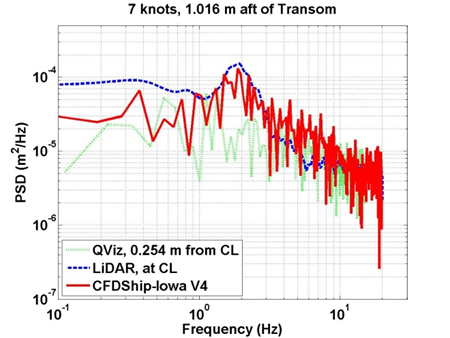} & &
\includegraphics[width=0.4\linewidth]{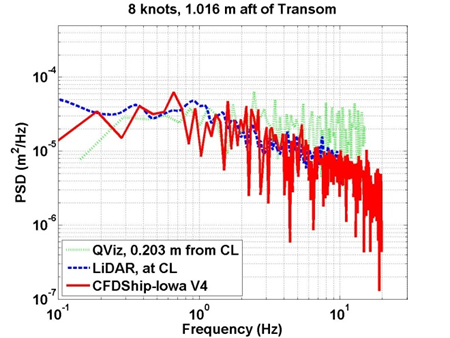} \\ & & &
\vspace{-12pt} \\ &
\includegraphics[width=0.4\linewidth]{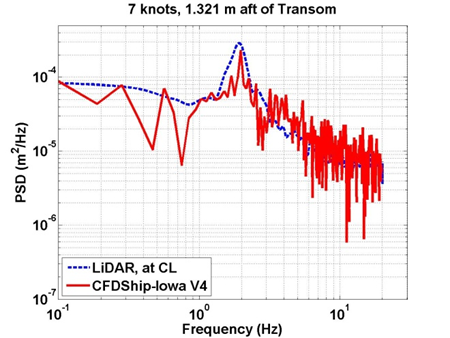} & &
\includegraphics[width=0.4\linewidth]{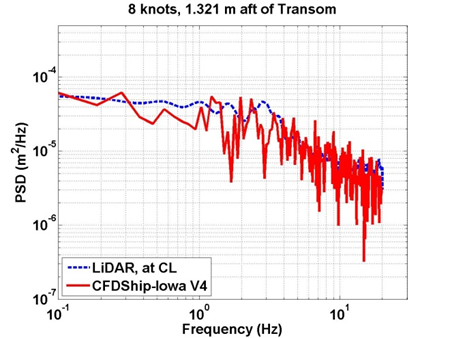} \\ & & &
\vspace{-12pt} \\ &
\includegraphics[width=0.4\linewidth]{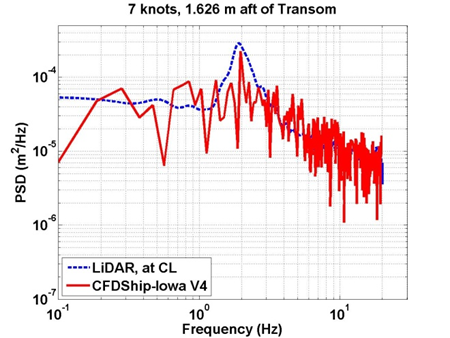} & &
\includegraphics[width=0.4\linewidth]{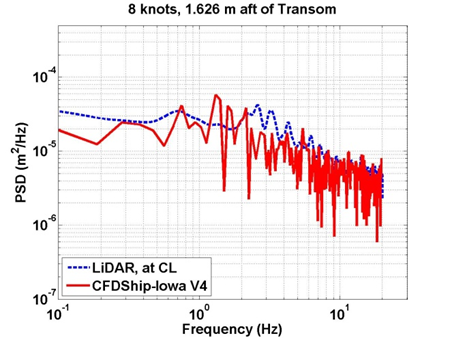}
\end{tabular} \end{center} \caption{\label{CFDShip_Fig_8} Spectra of wave
elevation unsteadiness at $X/L_o$ = 1.016, 1.321 and 1.626 predicted by
CFDShip-Iowa V4 is compared with experimental data for (a) 3.60 m/s (7 knots) (a) 3.60 m/s (7
knots) and (b) 4.12 m/s (8 knots). } \end{figure*}

Both LiDAR and CFDShip-Iowa V4 indicate high RMS values in the transverse wave
region with a peak at the centerline at both 3.60 m/s (7 knots) and 4.12 m/s (8
knots). Values taper off outside the wake region due to breaking as shown in Figure
\ref{CFDShip_Fig_6}(b).

CFDShip-Iowa V4 mean wave cut predictions in Figure
\ref{CFDShip_Fig_7}(a) compare within 4-5\% of the experiments for the first three
locations at a speed of 3.60 m/s (7 knots). An averaged error of 7-9\% is obtained for the farthest transverse
location. At 4.12 m/s (8 knots) the two closest tranverse wave cut locations
agree within 2-4\% of the experimental results. For the two outer locations,
the peak values are 20-30\% lower than the experimental data. A lag is expected
between the CFDShip-Iowa V4 and experimental results due ot the differences in
the diverging wave angles for both model speeds.

The wave elevation spectra from the LiDAR measurements in Figure \ref{CFDShip_Fig_8}(a)
show a dominant frequency of 1.96Hz at all the locations at a speed of 3.60
m/s (7 knots). CFDShip-Iowa V4 predicts the
dominant frequency within 0.1\% of the experiments, but the peak amplitude is under
predicted by 10\%. CFDShip-Iowa V4 results show large amplitude oscillations for high
frequencies, which could be due to the iterative error in the level-set function.
At a speed of 4.12 m/s (8 knots), the results between CFDShip-Iowa V4 and the
experimental results agree well. No spectral peak is seen at
this speed. The flow streamlines exit parallel to the hull bottom for this
case, and vortical structures are not predicted near the transom region. Any
remaining unsteadiness is likely due to wave breaking.

Overall, the experimental datasets are reasonably good for the drag, mean wave elevation and
wave-cuts. RMS datasets do not show any coherent pattern, but the spectra do show a
dominant frequency at 3.60 m/s (7 knots) and the dominant frequency compares
within 0.1\% of the experiments. CFDShip-Iowa V4 drag predictions are 4.18\% higher than the experimental
measurements at 3.60 m/s (7 knots) and 2.61\% higher at 4.12 m/s (8 knots). CFDShip-Iowa V4
predictions show a peak in the transom wave elevation cross-section close
to the transom in the shoulder region, which was not observed in the LiDAR
data. The mean wave elevation and
wave-cut predictions are 4\% higher when compared to the experimental results
at 3.60 m/s (7 knots) and 7\% higher at 4.12 m/s (8 knots). The limited validation
here supports the credibility of CFDShip-Iowa V4 simulation results.

\begin{figure} [t] \centering
\includegraphics[width=.8\columnwidth]{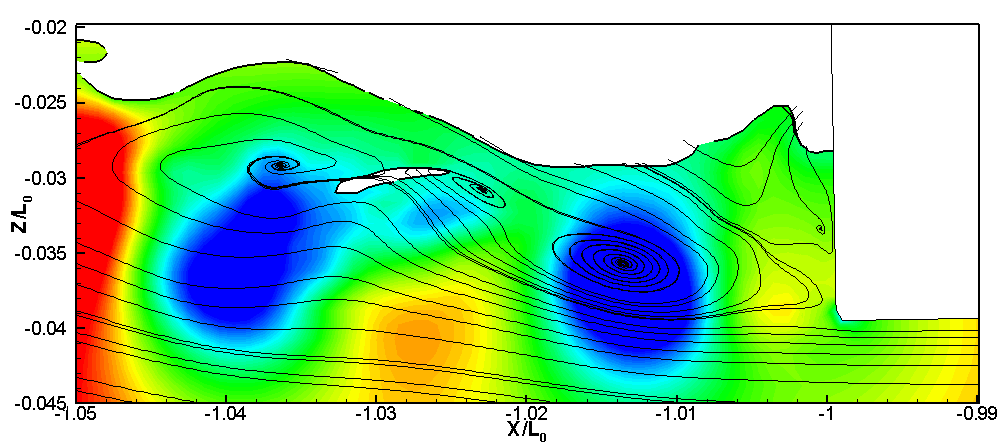} \\
\includegraphics[width=.8\columnwidth]{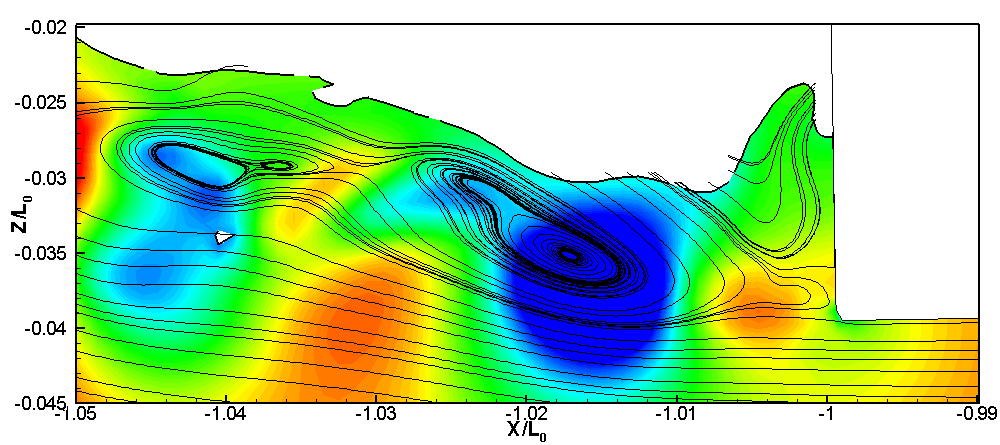} \\
\includegraphics[width=.8\columnwidth]{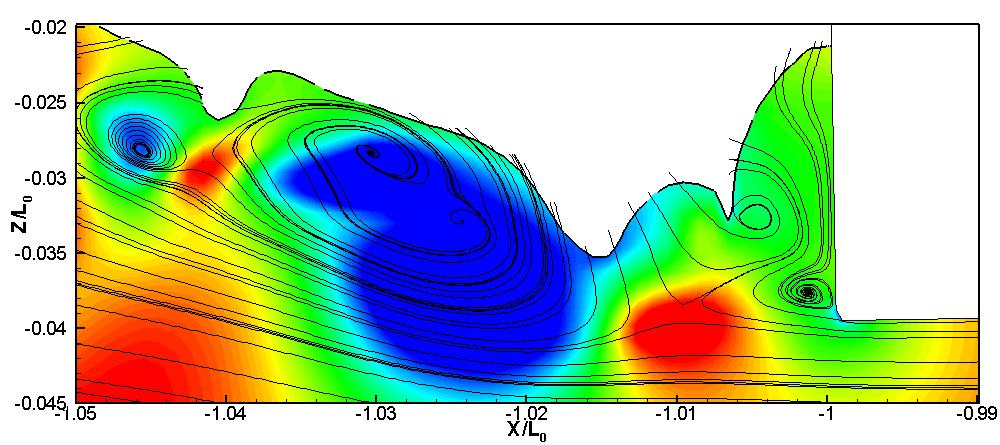} \\
\includegraphics[width=.8\columnwidth]{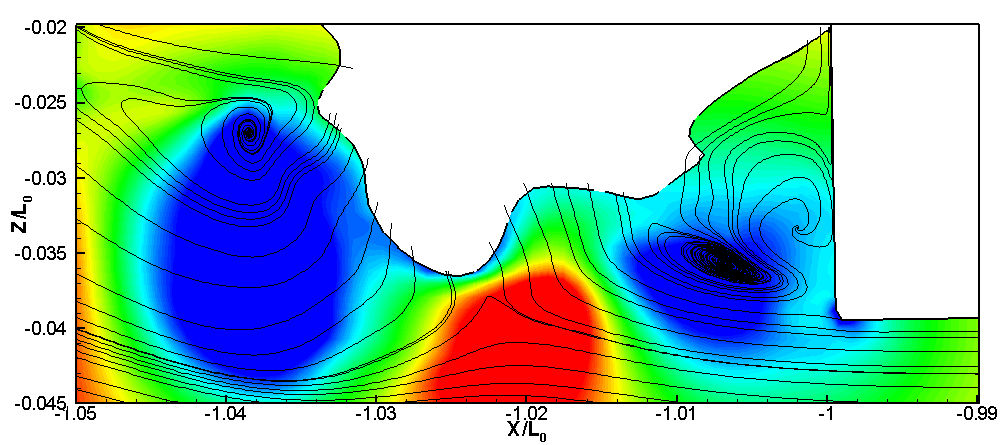}
\caption{\label{CFDShip_Fig_9} Analysis of unsteady transom vortex shedding
predicted by CFDShip-Iowa V4 for 3.60 m/s (7 knots). (a) Quarter phases of
transom vortex shedding are shown by streamlines and contours of piezometric
pressure at $Y/L_o$ = 0.01 plane. } \end{figure}

The transom vortex shedding is analyzed from the volume solution as shown in
Figure \ref{CFDShip_Fig_9}. The wave elevation at the transom increases when
the vortex forms at the transom corner and decreases as the vortex is shed,
whereas the wave elevation elsewhere on the hull remains the same. This
mechanism induces
an unsteady hull-wetted area with a frequency that is the same as that of transom
vortex shedding. Similarly, the $C_t$ decreases as the transom wave elevation
increases due to the excess pressure acting on the transom, and vice versa.
Thus either $C_t$ or hull-wetted area can be used to evaluate transom vortex
shedding frequency, and the former was previously used for the bare hull Athena
study \cite{Wilson06b}. As expected, both hull-wetted area and $C_t$ give
the same  $\tau_p$ = 0.21$L_o/U$, i.e., frequency= 1.9Hz, which is confirmed
from the volume solution analysis. The transom vortex shedding frequency is
within 3\% of the transom wave elevation unsteadiness, thus is identified to be
the main cause of transom wave elevation unsteadiness. Similar results were
obtained for the bare hull and appended Athena wetted transom flow.

%
%

The transom vortex shedding is identified as the Karman-like shedding as in the
Athena studies. As summarized in Table \ref{CFDShip_Table2}, transom-model
$\tau_p$ is 70\% and 36\% higher than those predicted for bare hull and
appended Athena using CFDShip-Iowa V4 DES at $F_{r}$ = 0.25, respectively. The averaged
wetted transom height H = 0.0145 is 51\% and 13\% higher than the values
predicted for bare hull and appended Athena studies, respectively. This give
$St_H$ = $fH/U$ = 0.075, which is half of that predicted for bare hull and 33\%
lower than that for appended Athena. $St_H$ is 8.5\% higher than the upper
limit of the Karman-like shedding $St_H$ range of 0.066-0.069 \cite{Kandasamy08}
for different geometries with and without a free-surface.

%
%

\section{CONCLUSIONS}

\subsubsection{Experimental Measurements}

We have presented a summary of data obtained from a series of experiments on a
model transom stern. The aim of the experimental work was to obtain an
understanding of the gross transom wake properties as well as detailed
statistical measurements across the wet/dry transom condition. A variety of
instrumentation was used and there is generally good agreement between the
results, despite an uncertainty in the registration of the LiDAR.

The experimental work described in this paper demonstrates the utility of such measurements compared to
larger-scale field experiments. While full-scale experiments allow for in-situ
measurements of the transom stern wake, accurate and repeatable measurements in
the field are difficult to make. Laboratory experiments allow for more detailed
measurements on specific aspects of the flow that would not be possible in the
field. The ability to obtain highly repeatable wake elevation measurements from
instruments such as the LiDAR are essential to our understanding of the
full-scale breaking transom wake as well as providing validation for numerical tools.

\subsubsection{NFA}

The agreement between NFA predictions and laboratory measurements is good.
Since NFA uses free-slip conditions on the hull, the good agreement suggests
that the wall boundary layer does not affect wave breaking and air entrainment in the transom
region when the Reynolds number is sufficiently high.  Analysis of the air entrainment suggests that the
peaks that are observed in the spectra of the free-surface elevation in the transom region are due to the
effects of a multi-phase shear layer that forms beneath the rooster tail and continues into the stern breaking wave.
 In terms of processing data, one advantage of the numerical simulations over laboratory measurements
is that all of the data is readily available.    NFA predictions of drag,
free-surface elevations, and air entrainment are all within experimental error.
Without the experiments, there is no basis to validate the numerical simulations,
and the development of computer codes such as NFA would not be possible.
NFA's ease of use, numerical stability, rapid turn around, and high accuracy
provide a robust framework for simulating complex flows around naval combatants.
Future improvements to the NFA algorithm are only possible under the guidance of
high-fidelity laboratory and field experiments such as those reported in this paper.

\subsubsection{CFDShip-Iowa V4}

CFDShip-Iowa V4 mean and unsteady wave elevation predictions using DES are validated for a
transom-stern model for 3.60 m/s (7 knots), $F_{r}$=0.38 and 4.12 m/s (8 knots), $F_{r}$=0.43. The
drag predictions are within 4.2\% of the experimental results for both flow conditions. The
grids are found to be sufficiently fine to properly activate LES in the transom
region as resolved turbulent kinetic energy is greater than 92\% of the total. The transom flow pattern
compares well with those from the experiments, i.e., a wetted transom is
predicted for 3.60 m/s (7 knots) with
wide wake dominated by wave breaking up to the transom. The 4.12 m/s (8 knot) simulation
predicts a dry transom with well-defined narrow wake that forms a rooster tail
with breaking waves.

Comparison with the experimental results are reasonably good for the drag, mean wave elevation, wave-cuts
and FFT, but coherent patterns are not observed for the RMS. CFDShip-Iowa V4 mean wave
elevation and wave-cut predictions are within 4.5\% and 6.5\% of the
experimental results for the wetted and dry transom case, respectively. The larger errors for the dry
transom case were due to the 18\% under prediction of the diverging wave angle.
CFDShip-Iowa V4 predictions show a shoulder wave close to the transom for both speeds, which
are not observed in the LiDAR wave elevation cross-sectional profiles. The transom wave elevation dominant frequency for the wetted transom
is predicted within 0.1\% of that found from the experimental work, whereas
neither results predict any
dominant frequency for the dry transom. The comparison errors are reasonable
and supports the credibility of CFDShip-Iowa V4 simulations.

The dominant wetted transom flow frequency is explained as a Karman-like
vortex shedding from the transom bottom corner, as the frequencies are within
3\%. The Karman-like shedding Strouhal number, $S_t$= 0.075, is 33\% lower than that for
appended Athena and 8.5\% higher than the upper limit of the Karman-like
shedding range based on the Strouhal number, $S_t$= 0.066-0.069.

%
%
\section{ACKNOWLEDGEMENTS}

The Office of Naval Research supports this research under multiple Office of
Naval Research contract vehicles as directed by Dr. L. Patrick Purtell (ONR
Code 331), program manager.  The authors would like to acknowledge the efforts
of Susan Brewton, Connor Bruns, Michael Capitain, Lisa Minnick, Toby Ratcliffe,
James Rice, Lauren Russell, and Don Walker from NSWCCD - Code 50, the NSWCCD
Media Lab, Eric Terrill and Genivieve Lada from the Scrips Institute of
Oceanography, UCSD, and David Jeon, Daegyoum Kim and Mory Gharib from the
California Institute of Technology.  The support of the Data Analysis and
Assessment Center (DAAC) is gratefully acknowledged. DAAC members include Dr.
Michael Stephens, Randall Hand, Paul Adams, Miguel Valenciano, Kevin George, Tom
Biddlecome, and Richard Walters. Animations of NFA simulations are available at
\href{http://www.youtube.com/waveanimations}{http://www.youtube.com/waveanimations}. Science Applications International Corporation IR\&D supported recent upgrades to the NFA algorithm.  NFA predictions are supported in part by a grant of computer time from the Department of Defense High Performance Computing Modernization Program (\href{http://www.hpcmo.hpc.mil/}{http://www.hpcmo.hpc.mil/}).  NFA simulations have been performed on the SGI Altix ICE at the U.S. Army Engineering Research and Development Center
(ERDC).

%
%

\bibliography{28onr}
\bibliographystyle{28onr}

%
%

\end{document}